\documentclass[12pt]{article}

\pdfoutput=1

\usepackage[dvipsnames]{xcolor}
\usepackage{amsfonts}
\usepackage{amsmath}
\usepackage{epsfig}
\usepackage{latexsym}
\usepackage{graphicx}
\usepackage{color}
\usepackage{amssymb}
\usepackage{epstopdf,mathtools,braket}
\numberwithin{equation}{section}
\usepackage{tikz}
\usetikzlibrary{intersections, calc}
\usetikzlibrary {arrows.meta,bending,positioning}
\usetikzlibrary{decorations.pathreplacing,decorations.markings}
\usepackage[hang,small,bf]{caption}
\usepackage[subrefformat=parens]{subcaption}
\allowdisplaybreaks
\usepackage{afterpage}
\usepackage{slashed}
\usepackage{url}
\usepackage{bm}
\usepackage{cancel}
\usepackage{hyperref}
\usepackage{titling}
\hypersetup{colorlinks,linkcolor=blue,citecolor=blue ,urlcolor=blue!80}
\usepackage{cite}
\usepackage{comment}
\usepackage{tikz}
\usetikzlibrary{intersections, calc,positioning,decorations.pathreplacing,decorations.pathmorphing}
\tikzset{>=latex}

\usepackage{comment}

\usepackage{centernot}
\usepackage{mathtools}
\usepackage{stmaryrd}

\def\i{\text{i}}

\def\be{\begin{equation}}
\def\ee{\end{equation}}
\def\bea{\begin{eqnarray}}
\def\eea{\end{eqnarray}}
\def\bequ{\begin{equation}}
\def\eequ{\end{equation}}

\def\Tr{\mbox{Tr}}

\def\i{\text{i}}
\renewcommand{\thefootnote}{\fnsymbol{footnote}}

\def\Tr{\mathrm{Tr}}

\def\CR{\mathcal{R}}

\usepackage{tikz}
\usetikzlibrary{intersections, calc,positioning,decorations.pathreplacing,decorations.pathmorphing}
\tikzset{>=latex}

\def\i{{\rm i}}

\def\CH{{\cal H}}

\def\CR{{\cal R}}

\def\CO{{\cal O}}

\def\b0{\bm{0}_\perp}

\usepackage{centernot}
\usepackage{mathtools}
\usepackage{stmaryrd}

\usepackage{simplewick}

\usepackage{framed}
 {\endMakeFramed}

\pretitle{%
  \begin{flushright}
    \small OU-HET 1284
  \end{flushright}
  \begin{center}
  \LARGE }
\posttitle{\end{center} }
\preauthor{\begin{center} \large}
\postauthor{\end{center}}
\predate{\begin{center} \small}
\postdate{\end{center}}

\title{\hspace{-3mm}Entanglement Asymmetry and Quantum Mpemba Effect for Non-Abelian Global Symmetry\hspace{3mm} \vspace{2cm}}

\author{Harunobu Fujimura and Soichiro Shimamori \\ 
\vspace{5mm}
{\small\it Department of Physics, The University of Osaka, Machikaneyama-Cho 1-1, Toyonaka 560-0043, Japan}}

\date{\small{September 2025}}

\begin{document}
\maketitle
\thispagestyle{empty}

\begin{abstract}
Entanglement asymmetry is a measure that quantifies the degree of symmetry breaking at the level of a subsystem. In this work, we investigate the entanglement asymmetry in $\widehat{su}(N)_k$ Wess-Zumino-Witten model and discuss the quantum Mpemba effect for $\text{SU}(N)$ symmetry, the phenomenon that the more symmetry is initially broken, the faster it is restored. Due to the Coleman–Mermin–Wagner theorem, spontaneous breaking of continuous global symmetries is forbidden in $1+1$ dimensions. To circumvent this no-go theorem, we consider excited initial states which explicitly break non-Abelian global symmetry. We particularly focus on the initial states built from primary operators in the fundamental and adjoint representations. In both cases, we study the real-time dynamics of the R\'enyi entanglement asymmetry and provide clear evidence of quantum Mpemba effect for $\text{SU}(N)$ symmetry. Furthermore, we find a new type of quantum Mpemba effect for the primary operator in the fundamental representation: increasing the rank $N$ leads to stronger initial symmetry breaking but faster symmetry restoration. Also, increasing the level $k$ leads to weaker initial symmetry breaking but slower symmetry restoration. On the other hand, no such behavior is observed for adjoint case, which may suggest that this new type of quantum Mpemba effect is not universal.
\end{abstract}

\renewcommand{\thefootnote}{\arabic{footnote}}
\setcounter{footnote}{0}

\newpage
\pagenumbering{arabic}
\setcounter{page}{1}
\tableofcontents


\section{Introduction}
Symmetry breaking is one of the most fundamental concepts in modern physics, ranging from high energy physics to condensed matter physics. 
Traditionally, the symmetry breaking is characterized through order parameters such as magnetization and chiral condensation. 
However, formulating the symmetry breaking in non-equilibrium systems is much more subtle, and it is necessary to introduce more suitable quantity to capture and quantify it. 

As one way to characterize the symmetry breaking for non-equilibrium systems, {\it entanglement asymmetry} was proposed in \cite{Ares:2022koq} as a measure to quantify the degree of symmetry breaking at the level of subsystem. Let $G$ be the symmetry of interest and $\rho_A$ the reduced density matrix of a subsystem $A$. 
As reviewed in the main text, one can define the symmetrized reduced density matrix, denoted by $\rho_{A,G}$, obtained by projecting $\rho_A$ onto the $G$-invariant subspace. 
The entanglement asymmetry is then defined as the relative entropy between $\rho_A$ and $\rho_{A,G}$:
\begin{align}\label{eq:EA}
    \Delta S_{A}
    &=
    \Tr_{A} \left[ \rho_A ( \log \rho_A  -\log \rho_{A,G}) \right].
\end{align}
By construction, the entanglement asymmetry is non-negative and vanishes if and only if the subsystem $A$ preserves the symmetry $G$ \cite{Kullback:1951zyt}. The entanglement asymmetry has been studied in various contexts e.g., \cite{Chen:2023gql,Fossati:2024xtn,Capizzi:2023yka,Lastres:2024ohf,Fossati:2024ekt,Kusuki:2024gss}.

One of the most mysterious phenomena in non-equilibrium systems is the {\it quantum Mpemba effect} through symmetry restoration. This is a phenomenon that the more symmetry is initially broken, the faster it is restored.\footnote{As discussed in \cite{Summer:2025wsa}, the framework of resource theories provides a way to relate the symmetry Mpemba effect to the original thermal Mpemba effect.} While originally studied from a purely theoretical perspective, the quantum Mpemba effect has recently been experimentally observed in trapped-ion quantum simulators \cite{Joshi:2024sup,Zhang_2025,AharonyShapira:2024nrt,Xu:2025wml}. In recent studies, it has been revealed that the entanglement asymmetry can be used to describe this counterintuitive phenomenon in various quantum systems, e.g., one-dimensional spin chains \cite{Ares:2023kcz,Ares:2024nkh,Ferro:2023sbn,Rylands:2024fio,Chalas:2024wjz,Liu:2024uqf,Bhore:2025nko,Ares:2025vjw,Yamashika:2025vpd,Gibbins:2025bdd}, two-dimensional spin systems \cite{Yamashika:2024hpr,Yamashika:2024mut}, and conformal field theories (CFTs) \cite{Benini:2024xjv}.\footnote{In the context of quantum information theory, the quantum Mpemba effect has also been investigated in random quantum circuits \cite{Liu:2024kzv,Turkeshi:2024juo,Ares:2025ljj,Yu:2025lku,Klobas:2024png,Russotto:2024pqg,Chen:2024lxe} and in periodically driven quantum systems \cite{Banerjee:2024zqb,Klobas:2024mlb}.} For more detail, we refer the readers to recent reviews \cite{Ares:2025onj,Yu:2025vth}.

While the studies mentioned in the previous paragraph focus mainly on the restoration of Abelian symmetries including U(1) or $\mathbb{Z}_N$, the quantum Mpemba effect for non-Abelian symmetries remains essentially elusive.\footnote{Although the entanglement asymmetries for non-Abelian symmetries are discussed in \cite{Capizzi:2023xaf,Lastres:2024ohf,Fossati:2024ekt,Kusuki:2024gss}, the quantum Mpemba effects are not discussed. Also, while the quantum Mpemba effects for non-Abelian global symmetries are discussed for random states \cite{Liu:2024kzv,Russotto:2024pqg}, the continuum examples have not been explored so far.} In this work, we present the detailed analysis on the symmetry restoration structures for non-Abelian symmetries by using the $\widehat{su}(N)_k$ Wess-Zumino-Witten (WZW) models through the lends of entanglement asymmetry. 
Due to the Coleman–Mermin–Wagner theorem \cite{Mermin:1966fe, Coleman:1973ci}, spontaneous breaking of continuous global symmetries in $1+1$ dimensions is forbidden. To bypass this no-go theorem, we consider an excited initial state $\ket{\mathcal{O}} = \mathcal{O}\ket{0}$, created by a local primary operator $\mathcal{O}$ carrying non-trivial charges under the non-Abelian symmetry \cite{Nozaki:2014hna}. For concreteness, we focus on WZW primaries in the fundamental and adjoint representations.

Given the difficulty of computing entanglement asymmetry directly, we instead evaluate the $n$-th Rényi entanglement asymmetry $\Delta S_{A}^{(n)}$ which yields the entanglement asymmetry in the limit $n \to 1$.
Crucially, conformal symmetry enables us to express these Rényi entanglement asymmetries in terms of $2n$-point correlation functions, which are tractable via CFT techniques. 
In particular, we compute the second Rényi entanglement asymmetry (i.e., $n=2$) using exact four-point functions, which are obtained analytically by solving the Knizhnik–Zamolodchikov equations \cite{Knizhnik:1984nr} and are explicitly known in \cite{Christe:1986cy, Zamolodchikov:1986bd}.

With these exact results, we analyze the R\'enyi entanglement asymmetry in WZW model and explore various asymptotic regimes, such as the long-time limit and the large-interval limit.
We also investigate the real-time dynamics of the R\'enyi entanglement asymmetry and provide clear evidence of the quantum Mpemba effect for $\text{SU}(N)$ symmetry in both cases of fundamental and adjoint primaries. 
More interestingly, for the initial state built from the primary in the fundamental representation, we uncover a new type of quantum Mpemba effect: increasing the rank $N$ amplifies the initial symmetry breaking, while accelerating the symmetry restoration.\footnote{This phenomenon differs from most existing studies of entanglement asymmetry and the quantum Mpemba effect in CFTs. Previous works (e.g. \cite{Chen:2023gql}) typically fix a single theory and vary the initial state, for instance by changing the representation of the excitation, to study how the asymmetry evolves. In contrast, we fix the initial state (e.g. in the fundamental or adjoint representation) and vary the rank $N$, which corresponds to moving across a family of distinct CFTs rather than changing parameters within a single theory. From this perspective, the observed Mpemba-like behavior reflects a nontrivial dependence on the underlying theory itself. To our knowledge, realizing quantum Mpemba behavior through varying $N$ at fixed representation has not been explored previously, and thus constitutes a new direction in the study of entanglement asymmetry.}
Conversely, increasing the level $k$ weakens the initial symmetry breaking, while decelerating the symmetry restoration. Such a realization of quantum Mpemba effect though symmetry restoration has not been reported in previous works. On the other hand, in the case of adjoint primary, we are not able to find any evidence for this new type of quantum Mpemba effect. This may suggest that this quantum Mpemba effect is not a universal phenomenon although we have not explored other representations. We leave further analysis about this new type of quamtum Mpemba effect to future studies.

This paper is organized as follows.
In Section \ref{sec:EA_in_CFT}, we review the definition and basic properties of the entanglement asymmetry in quantum field theories and explain how it can be expressed in terms of correlation functions.
In Section \ref{sec:EA_in_WZW},  we derive the exact result of second R\'enyi entanglement asymmetry for the initial state build from the fundamental primary, and analyze its asymptotic behavior as well as the quantum Mpemba effect.
In Section \ref{sec:current_EA}, we analyze the symmetry restoration structures for the initial state constructed from the WZW current in the adjoint representation.
Finally, in Section \ref{sec:conclusion_and_discussion}, we summarize this paper and list some future directions. 
In Appendix \ref{sec:4pt_function_in_WZW}, we review the four-point function of two fundamental and two anti-fundamental primaries.
In Appendix \ref{sec:Asymptotic_derivation}, we give detailed analysis on asymptotic behaviors of the second R\'enyi entanglement asymmetry.

\section{Entanglement asymmetry in quantum field theories}
\label{sec:EA_in_CFT}
In this section, we provide a brief review of entanglement asymmetry and R\'enyi entanglement asymmetry introduced in \cite{Ares:2022koq} and discuss how to analyze the R\'enyi entanglement asymmetry in conformal field theories (CFTs). 
From the perspective of the replica trick \cite{Holzhey:1994we, Calabrese:2004eu, Casini:2009sr} and the symmetry operators \cite{Gaiotto:2014kfa}, the R\'enyi entanglement asymmetry can be expressed in terms of correlation functions.

\subsection{Symmetry operators}
Before introducing the entanglement asymmetry,  we briefly review the symmetry operators \cite{Gaiotto:2014kfa}, which will be necessary later.
For a more comprehensive introduction to symmetry operators, we refer the readers to the recent review on generalized symmetries \cite{Bhardwaj:2023kri}.
For simplicity, let us consider quantum field theories with group-like global symmetry $G$ on the two-dimensional Euclidean spacetime. Every group element $g\in G$ is associated to the corresponding symmetry operator $U_{\Sigma }(g)$, where $\Sigma$ is one-dimensional manifold in two-dimensional spacetime. The symmetry operator $U_{\Sigma }(g)$ is topological in a sense that the correlation function with $U_{\Sigma }(g)$ inserted is invariant under continuous and smooth deformation of $\Sigma \to \Sigma'$. This topological property can be rephrased by the following operator equation:
\begin{align}
    U_{\Sigma }(g) 
    &=
    U_{\Sigma' }(g)\ .
    \label{eq:topologicality_symmetry_op}
\end{align}
Moreover, this symmetry operator inherits the group structure of $G$, namely,
\begin{align}
    U_{\Sigma }(g_1) U_{\Sigma }(g_2)
    &=
    U_{\Sigma }(g_1 g_2) \ , \quad \forall g_1 , g_2 \in G \ .
    \label{eq:fusion_symmetry_op}
\end{align}
which is schematically expressed as\\
\begin{align}
    \begin{tikzpicture}
    [scale=1,baseline={([yshift=-.5ex]current bounding box.center)}]
        \coordinate (left_down) at (0,0);
        \coordinate (left_up) at (0,2.3);
        \coordinate (right_down) at (2,0);
        \coordinate (right_up) at (2,2.3);
        \draw[line width=1] (left_down)--(right_down)--(right_up)--(left_up)--cycle;
        \draw [blue, thick,decoration={markings, mark=at position 0.6 with {\arrow[scale=1.2]{>}}}, postaction={decorate}] 
        ($(left_down)!0.7!(left_up)$)--($(right_down)!0.7!(right_up)$)
        node[midway, above] {$g_1$};
        \draw [blue, thick,decoration={markings, mark=at position 0.6 with {\arrow[scale=1.2]{>}}}, postaction={decorate}] 
        ($(left_down)!0.3!(left_up)$)--($(right_down)!0.3!(right_up)$)
        node[midway, below] {$g_2$};
    \end{tikzpicture}
    \quad
    =
    \quad
    \begin{tikzpicture}[scale=1,baseline={([yshift=-.5ex]current bounding box.center)}]
        \coordinate (left_down) at (0,0);
        \coordinate (left_up) at (0,2.3);
        \coordinate (right_down) at (2,0);
        \coordinate (right_up) at (2,2.3);
        \draw[line width=1] (left_down)--(right_down)--(right_up)--(left_up)--cycle;
        \draw [blue, thick,decoration={markings, mark=at position 0.6 with {\arrow[scale=1.2]{>}}}, postaction={decorate}] 
        ($(left_down)!0.5!(left_up)$)--($(right_down)!0.5!(right_up)$)
        node[midway, above] {$g_1 g_2$};
    \end{tikzpicture}\ ,
    \label{eq:fusion_rule_graphical_rep}
\end{align}\\
\noindent where the blue line labeled by $g$ represents the symmetry operator $U_{\Sigma}(g)$, and the box means the entire spacetime. The transformation of an operator $\CO (x)$ under a group element $g\in G$ is described by the symmetry operator as
\begin{align}
    U_{\Sigma}^{\dagger}(g) \CO (x) U_{\Sigma}(g)
    =
    R(g) \CO (x)\ ,\quad x\in \Sigma\ ,
    \label{eq:operator_trsf_group}
\end{align}
where $R(g)$ denotes a representation of $g$ of $\CO$. 
Diagrammatically, this rule is depicted as follows:
\begin{align}
\begin{tikzpicture}[scale=1,baseline={([yshift=-.5ex]current bounding box.center)}]
    \draw[line width=1,->,blue!80] (0,0) arc(-270:100:1);
    \draw[blue] (0,0.4) node {$g$};
    \draw (0,-1) node {$\times$};
    \draw (0,-0.6) node {$\CO(x)$};
    \draw (2,-1) node {$=$};
    \draw (4,-1) node {$\times$};
    \draw (4,-0.6) node {$R(g) \CO(x)$};
\end{tikzpicture} \ .
\label{eq:operator_trsf_symmetry_op}
\end{align}

\subsection{Entanglement asymmetry}

\begin{figure}
    \centering
    \begin{tikzpicture}
        \coordinate(left_down)  at (0,0);
        \coordinate(right_down) at (5,0);
        \coordinate(left_up)  at (0,3);
        \coordinate(right_up) at (5,3);
        \coordinate(u) at ($($(left_down)!0.5!(left_up)$)!0.3!($(right_down)!0.5!(right_up)$)$);
        \coordinate(v) at ($($(left_down)!0.5!(left_up)$)!0.7!($(right_down)!0.5!(right_up)$)$);
        \draw[line width=1] (left_down)--(right_down)--(right_up)--(left_up)--(left_down)--cycle;
        \draw[line width=1.2, dashed] ($(left_down)!0.5!(left_up)$)--(u);
        \draw[line width=1.2, dashed] ($(right_down)!0.5!(right_up)$)--(v);
        \draw[line width=1.2, red] (u)--(v);
        \draw[red] ($($(u)!0.5!(v)$)+(0,0.5)$) node {\scalebox{1.2}{$A$}};
        \draw ($($($(left_down)!0.5!(left_up)$)!0.5!(u)$)+(0,0.5)$) node {\scalebox{1.2}{$B$}};
        \draw ($($(v)!0.5!($(right_down)!0.5!(right_up)$)$)+(0,0.5)$) node {\scalebox{1.2}{$B$}};
        \draw[->,line width=1] ($(left_down)+(-2,0)$)--($(left_down)+(-2,1)$) node[above] {$\tau$};
        \draw[->,line width=1] ($(left_down)+(-2,0)$)--($(left_down)+(-1,0)$) node[right] {$x$};
    \end{tikzpicture}
    \caption{The sketch of subregions $A$ and $B$ in two-dimensional Euclidean spacetime. 
    Here subregion $B$ is defined as the total system excluding the subregion $A$.
    }
    \label{fig:region_A_B}
\end{figure}
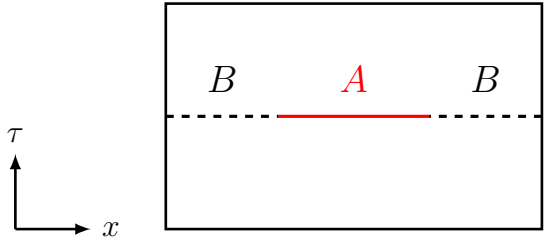

In this subsection, we introduce the concept of entanglement asymmetry \cite{Ares:2022koq}. 
Let the total spatial region be an infinite line\footnote{One can extend our analysis to the case where the total spatial region is a finite line by employing the conformal transformation used in \cite{Cardy:2016fqc}. However, in this situation finite-size effects arise, and the entanglement asymmetry becomes a periodic function of the system size. In this paper, we focus on the infinite line as the total spacial region to clearly see the symmetry restoration and quantum Mpemba effect.} and we divide it into two subregions $A$ and $B$. The subregion $A = [0,\ell]$ is a single interval, as illustrated in Fig.\,\ref{fig:region_A_B}.
We assume that the total Hilbert space $\CH_{\text{tot}}$ decomposes into a tensor product as
\begin{align}
    \CH_{\text{tot}}=\CH_{A} \otimes \CH_{B}\ ,
    \label{eq:Hilbert_space_separation}
\end{align}
where $\CH_{A/B}$ denote the Hilbert spaces of subsystems $A/B$, respectively. With this decomposition, the symmetry operator $U_{\Sigma} (g)$ factorizes as
\begin{align}
    U_{\Sigma} (g)
    =
    U_{A} (g) \otimes U_{B} (g)\ ,
\end{align}
where $\Sigma$ is chosen to be the spatial slice at fixed time, $\Sigma=A \cup B$, and $U_{A/B} (g)$ denote the symmetry operators acting on the Hilbert spaces $\CH_{A/B}$, respectively. The reduced density matrix $\rho_A$ of the subsystem $A$ is defined by tracing out the degrees of freedom on subsystem $B$ from the total density matrix $\rho_{\text{tot}}$:
\begin{align}
    \rho_{A} 
    \equiv
    \Tr_{B}\left[\rho_{\text{tot}} \right]\ .
\end{align}
Under the action of a group element $g$, the reduced density matrix $\rho_A$ transforms as $\rho_A \mapsto U_{A} (g) \rho_A U_{A}^{\dagger} (g)$. This means that when the global symmetry $G$ is preserved in the subsystem~$A$, the reduced density matrix has to commute with the symmetry operator $U_{\Sigma}(g)$ for $\forall g \in G$:
\begin{align}
    \left[\rho_A,  U_{A}(g)\right] 
    &=
    0\quad \text{(symmetry preserved)} \ .
\end{align}
Otherwise, the symmetry $G$ is said to be broken in the subsystem~$A$:
\begin{align}
    \left[\rho_A,  U_{A}(g)\right] 
    &\not =
    0\quad \text{(symmetry broken)} \ . 
\end{align}
To define the entanglement asymmetry,  we introduce the symmetrized reduced density matrix \cite{Fossati:2024xtn}:
\begin{align}
    \rho_{A,G} 
    &\equiv
    \int_G \text{d}g \  U_{A}(g) \rho_A U_{A}^{\dagger}(g)\ ,
    \label{eq:rho_A_G_def}
\end{align}
where $\text{d}g$ denotes the Haar measure on the group $G$, normalized as $\int_G \text{d}g =1$\footnote{For a discrete group, the symmetrized reduced density matrix can be defined in the same way, with replacement $\int_G \text{d}g \to \frac{1}{|G|}\sum_{g\in G} $.}. By construction, the symmetrized reduced density matrix $\rho_{A,G}$ always commutes with the symmetry operator $U_{A}(g)$. With this definition, the criterion for symmetry preservation/breaking in $A$ described above can be expressed as
\begin{align}
\left\{\quad
\begin{aligned}
    \rho_A
    &=\rho_{A,G} \quad \text{(Symmetry preserved)}\\
    \rho_A
    &\neq
    \rho_{A,G}\quad \text{(Symmetry broken)}
\end{aligned}
\right.
\label{eq:symmetry_breaking_def}
\end{align}
In quantum information theory, the distinguishability between two density matrices $\rho $ and $\sigma $ can be quantified by relative entropy \cite{Nielsen:2012yss}. The entanglement asymmetry is thus naturally defined as the relative entropy between $\rho_A$ and $\rho_{A,G}$:
\begin{align}
\begin{aligned}
    \Delta S_{A}
    &\equiv
    \Tr_{A} \left[ \rho_A ( \log \rho_A  -\log \rho_{A,G}) \right]  \ .
\end{aligned}
\label{eq:EA_def}
\end{align}
Importantly, the entanglement asymmetry satisfies the following property \cite{Kullback:1951zyt,Nielsen:2012yss}:
\begin{align}
    \left\{\ 
    \begin{aligned}
        \Delta S_{A} = 0\  &\iff \  \rho_A = \rho_{A,G}\\
        \Delta S_{A} > 0\  &\iff \  \rho_A \neq \rho_{A,G}
    \end{aligned}
    \right.\ .
    \label{eq:EA_property}
\end{align}
By combining this with \eqref{eq:symmetry_breaking_def}, we can regard the entanglement asymmetry as a quantitative measure of symmetry breaking at the level of the subsystem $A$. In general, analyzing the entanglement asymmetry is challenging due to the presence of $\log \rho$ in its definition. It is therefore convenient to introduce the R\'enyi entanglement asymmetry:
\begin{align}
    \Delta S_{A}^{(n)}
    \equiv
    \frac{1}{1-n} 
    \left( 
    \log 
    \Tr_{A} \left[ \rho_{A,G}^n \right]
    -
    \log 
    \Tr_{A} \left[ \rho_{A}^n \right]
    \right)\ ,
    \label{eq:REA_def}
\end{align}
where $n$ is a positive integer. The entanglement asymmetry in \eqref{eq:EA_def} is recovered by analytic continuation of the R\'enyi entanglement asymmetry and taking the following limit:
\begin{align}
    \lim_{n\to1} \Delta S_{A}^{(n)} 
    =
    \Delta S_{A}\ .
\end{align}

Let us rewrite the R\'enyi entanglement asymmetry in terms of the symmetry operators. Substituting the definition of the symmetrized reduced density matrix in \eqref{eq:rho_A_G_def} into $\Tr_{A} \left[ \rho_{A,G}^n \right]$, we obtain the following:
\begin{align}
    \begin{aligned}
    \Tr_{A} \left[ \rho_{A,G}^n \right]
    &=
    \int_G \text{d}h_1 \cdots \text{d}h_n \
    \Tr_A 
    \left[
    U_{A}(h_1) \rho_A U_{A}^{\dagger}(h_1) \cdot
    U_{A}(h_2) \rho_A U_{A}^{\dagger}(h_2) \cdots
    U_{A}(h_n) \rho_A U_{A}^{\dagger}(h_n)
    \right]\\
    &=
    \int_G \text{d}h_1 \cdots \text{d}h_n \
    \Tr_A 
    \left[
    \rho_A U_{A} \left(h_1^{-1}h_2\right)  
    \rho_A U_{A} \left(h_2^{-1}h_3\right)
    \cdots 
    \rho_A U_{A} \left(h_{n}^{-1}h_1\right) 
    \right],
    \end{aligned}
\end{align}
where we used the cyclic property of trace and the fusion rule \eqref{eq:fusion_symmetry_op}. By changing integral variables as $g_i^{-1}=h_{i}^{-1} h_{i+1}$ for $i=1,\cdots n\ ,(h_{n+1}\equiv h_1)$, we arrive at\footnote{Note that the $g_n$ does not appear in the integrand, and its integration simply yields $\int_G \text{d}g_n = 1$.}
\begin{align}
    \Tr_{A} \left[ \rho_{A,G}^n \right]
    &=
    \int_G \text{d}g_1 \cdots \text{d}g_{n-1}  \
    \Tr_A 
    \left[
    \rho_A U_{A} \left(g_1^{-1}\right) \cdots 
    \rho_A U_{A} \left(g_{n-1}^{-1}\right) \cdot
    \rho_A U_{A} \left(g_{n-1}\cdots g_{1}\right)
    \right]\ .
\end{align}
We now define the charged moments $Z(g_1,\cdots ,g_{n-1})$ as 
\begin{align}
    Z(g_1,\cdots ,g_{n-1})
    \equiv
    \frac{
    \Tr_A 
    \left[
    \rho_A U_{A} \left(g_1^{-1}\right) \cdots 
    \rho_A U_{A} \left(g_{n-1}^{-1}\right) \cdot
    \rho_A U_{A} \left(g_{n-1}\cdots g_{1} \right)
    \right]
    }{
    \Tr_A\left[\rho_A^n \right]
    }\ .
    \label{eq:charged_moment}
\end{align}
In terms of the charged moment, the R\'enyi entanglement asymmetry can be expressed as
\begin{align}
    \Delta S_{A}^{(n)}
    =
    \frac{1}{1-n} 
    \log 
    \left[
    \int_G \text{d}g_1 \cdots \text{d}g_{n-1} \
    Z(g_1,\cdots ,g_{n-1})
    \right]\ .
    \label{eq:REA_charged_moment}
\end{align}
To evaluate the charged moments $Z(g_1,\cdots ,g_{n-1})$, we employ the path integral formulation, as explained in the next subsection.

\subsection{Replica trick and operator insertion}

Let us now consider the path integral representation of the charged moments. In the case where the quantum state of the total system is vacuum, i.e., $\rho_{\text{tot}} = \ket{0} \bra{0}$, the quantity $\Tr_A[\rho_A^n]$ is given by the partition function on the $n$-sheeted manifold $\CR_n$ \cite{Calabrese:2004eu, Casini:2009sr}. In this case, a charged moment is identified with the partition function on the $n$-sheeted manifold $\CR_n$ with the insertion of symmetry operators \cite{Goldstein:2017bua, Cornfeld:2018wbg}. 

\begin{figure}
    \centering
    \begin{tikzpicture}[scale=0.6,baseline={([yshift=-.5ex]current bounding box.center)}]
        \coordinate(left_lower_edge_of_plane)  at (0,0);
        \coordinate(right_lower_edge_of_plane) at (10,0);
        \coordinate(left_upper_edge_of_plane)  at (0+2,2);
        \coordinate(right_upper_edge_of_plane) at (10+2,2);
        \coordinate(a)  at (0,2.5);
        \coordinate(upper_u1) at (5+0.1,1+0.1);
        \coordinate(lower_u1) at (5-0.1,1-0.1);
        \coordinate(upper_v1) at (9+0.1,1+0.1);
        \coordinate(lower_v1) at (9-0.1,1-0.1);
        \coordinate(epsilon) at (0.04,0.24);
        \coordinate(o_up) at ($($(left_lower_edge_of_plane)!0.5!(left_upper_edge_of_plane)$)+(epsilon)$);
        \coordinate(o_down) at ($($(left_lower_edge_of_plane)!0.5!(left_upper_edge_of_plane)$)-(epsilon)$);
        \draw[line width=1, name path=lower_plane]
        (left_lower_edge_of_plane)--
        (right_lower_edge_of_plane)--
        (right_upper_edge_of_plane)--
        (left_upper_edge_of_plane)--cycle;
        \draw[line width=1, name path=upper_plane]
        ($(left_lower_edge_of_plane)  + (a)$)--
        ($(right_lower_edge_of_plane) + (a)$)--
        ($(right_upper_edge_of_plane) + (a)$)--
        ($(left_upper_edge_of_plane)  + (a)$)--cycle;
        \draw[line width=1,red!80](lower_u1)--(lower_v1)--(upper_v1)--(upper_u1)--cycle;
        \draw[line width=1,red!80]
        ($(lower_u1) + (a)$)--
        ($(lower_v1) + (a)$)--
        ($(upper_v1) + (a)$)--
        ($(upper_u1) + (a)$)--cycle;
        \fill[lightgray!50]
        (upper_u1)--
        (upper_v1)--
        ($(upper_v1)!0.59!($(lower_v1) + (a)$)$)--
        ($(upper_u1)!0.59!($(lower_u1) + (a)$)$)--cycle;
        \fill[lightgray!20]
        ($(upper_v1)!0.63!($(lower_v1) + (a)$)$)--
        ($(upper_u1)!0.63!($(lower_u1) + (a)$)$)--
        ($(lower_u1) + (a) -(0,0.05)$)--
        ($(lower_v1) + (a) -(0,0.05)$)--cycle;
        \fill[lightgray!50]
        ($(upper_u1)+ (a)$)--
        ($(upper_v1)+ (a)$)--
        ($($(upper_v1)+ (a)$)!0.59!($(lower_v1) + 2*(a)$)$)--
        ($($(upper_u1)+ (a)$)!0.59!($(lower_u1) + 2*(a)$)$)--cycle;
        \fill[lightgray!20]
        ($(lower_u1)-(0,0.05)$)--
        ($(lower_v1)-(0,0.05)$)--
        ($(lower_v1) - 0.2*(a) + (+0.2,0)$)--
        ($(lower_u1) - 0.2*(a) + (+0.2,0)$)--cycle;
        \draw($($(upper_u1)+ (a)$)!0.5!($(lower_v1)+ 2.5*(a)$)$) node[above=-3mm] {\scalebox{1}{$\vdots $}};
        \draw ($(left_lower_edge_of_plane)+(1,0.5)$) node {\scalebox{1}{\rotatebox{0}{$1$}}};
        \draw ($(left_lower_edge_of_plane)+(a)+(1,0.5)$) node {\scalebox{1}{\rotatebox{0}{$2$}}};
        \draw ($($(left_upper_edge_of_plane)!0.2!(right_upper_edge_of_plane)+(a)$)!0.3!($(left_lower_edge_of_plane)!0.2!(right_lower_edge_of_plane)+(a)$)$) node {\scalebox{1}{$\times$}};
        \draw ($($(left_upper_edge_of_plane)!0.2!(right_upper_edge_of_plane)+(a)$)!0.7!($(left_lower_edge_of_plane)!0.2!(right_lower_edge_of_plane)+(a)$)$) node {\scalebox{1}{$\times$}};
        \draw ($($(left_upper_edge_of_plane)!0.2!(right_upper_edge_of_plane)$)!0.3!($(left_lower_edge_of_plane)!0.2!(right_lower_edge_of_plane)$)$) node {\scalebox{1}{$\times$}};
        \draw ($($(left_upper_edge_of_plane)!0.2!(right_upper_edge_of_plane)$)!0.7!($(left_lower_edge_of_plane)!0.2!(right_lower_edge_of_plane)$)$) node {\scalebox{1}{$\times$}};
        \draw [blue, thick,decoration={markings, mark=at position 0.5 with {\arrow[scale=1.2]{<}}}, postaction={decorate}] 
        (upper_u1) -- (upper_v1)
        node[midway, above] {$g_1$};
        \draw [blue, thick,decoration={markings, mark=at position 0.5 with {\arrow[scale=1.2]{<}}}, postaction={decorate}] 
        ($(upper_u1) + (a)$) -- ($(upper_v1) + (a)$)
        node[midway, above] {$g_2$};
        \draw ($($(right_upper_edge_of_plane)!0.5!($(right_lower_edge_of_plane)+(a)$)$)+(2,0)$) node{\scalebox{1.5}{$=$}};
    \end{tikzpicture}
    \begin{tikzpicture}[scale=0.6,baseline={([yshift=-.5ex]current bounding box.center)}]
        \coordinate(left_lower_edge_of_plane)  at (0,0);
        \coordinate(right_lower_edge_of_plane) at (10,0);
        \coordinate(left_upper_edge_of_plane)  at (0+2,2);
        \coordinate(right_upper_edge_of_plane) at (10+2,2);
        \coordinate(a)  at (0,2.5);
        \coordinate(upper_u1) at (5+0.1,1+0.1);
        \coordinate(lower_u1) at (5-0.1,1-0.1);
        \coordinate(upper_v1) at (9+0.1,1+0.1);
        \coordinate(lower_v1) at (9-0.1,1-0.1);
        \coordinate(epsilon) at (0.04,0.24);
        \coordinate(o_up) at ($($(left_lower_edge_of_plane)!0.5!(left_upper_edge_of_plane)$)+(epsilon)$);
        \coordinate(o_down) at ($($(left_lower_edge_of_plane)!0.5!(left_upper_edge_of_plane)$)-(epsilon)$);
        \draw[line width=1, name path=lower_plane]
        (left_lower_edge_of_plane)--
        (right_lower_edge_of_plane)--
        (right_upper_edge_of_plane)--
        (left_upper_edge_of_plane)--cycle;
        \draw[line width=1, name path=upper_plane]
        ($(left_lower_edge_of_plane)  + (a)$)--
        ($(right_lower_edge_of_plane) + (a)$)--
        ($(right_upper_edge_of_plane) + (a)$)--
        ($(left_upper_edge_of_plane)  + (a)$)--cycle;
        \draw[line width=1,red!80](lower_u1)--(lower_v1)--(upper_v1)--(upper_u1)--cycle;
        \draw[line width=1,red!80]
        ($(lower_u1) + (a)$)--
        ($(lower_v1) + (a)$)--
        ($(upper_v1) + (a)$)--
        ($(upper_u1) + (a)$)--cycle;
        \fill[lightgray!50]
        (upper_u1)--
        (upper_v1)--
        ($(upper_v1)!0.59!($(lower_v1) + (a)$)$)--
        ($(upper_u1)!0.59!($(lower_u1) + (a)$)$)--cycle;
        \fill[lightgray!20]
        ($(upper_v1)!0.63!($(lower_v1) + (a)$)$)--
        ($(upper_u1)!0.63!($(lower_u1) + (a)$)$)--
        ($(lower_u1) + (a) -(0,0.05)$)--
        ($(lower_v1) + (a) -(0,0.05)$)--cycle;
        \fill[lightgray!50]
        ($(upper_u1)+ (a)$)--
        ($(upper_v1)+ (a)$)--
        ($($(upper_v1)+ (a)$)!0.59!($(lower_v1) + 2*(a)$)$)--
        ($($(upper_u1)+ (a)$)!0.59!($(lower_u1) + 2*(a)$)$)--cycle;
        \fill[lightgray!20]
        ($(lower_u1)-(0,0.05)$)--
        ($(lower_v1)-(0,0.05)$)--
        ($(lower_v1) - 0.2*(a) + (+0.2,0)$)--
        ($(lower_u1) - 0.2*(a) + (+0.2,0)$)--cycle;
        \draw($($(upper_u1)+ (a)$)!0.5!($(lower_v1)+ 2.5*(a)$)$) node[above=-3mm] {\scalebox{1}{$\vdots $}};
        \draw ($(left_lower_edge_of_plane)+(1,0.5)$) node {\scalebox{1}{\rotatebox{0}{$1$}}};
        \draw ($(left_lower_edge_of_plane)+(a)+(1,0.5)$) node {\scalebox{1}{\rotatebox{0}{$2$}}};
        \draw ($($(left_upper_edge_of_plane)!0.2!(right_upper_edge_of_plane)+(a)$)!0.3!($(left_lower_edge_of_plane)!0.2!(right_lower_edge_of_plane)+(a)$)$) node {\scalebox{1}{$\times$}};
        \draw ($($(left_upper_edge_of_plane)!0.2!(right_upper_edge_of_plane)+(a)$)!0.7!($(left_lower_edge_of_plane)!0.2!(right_lower_edge_of_plane)+(a)$)$) node {\scalebox{1}{$\times$}};
        \draw ($($(left_upper_edge_of_plane)!0.2!(right_upper_edge_of_plane)$)!0.3!($(left_lower_edge_of_plane)!0.2!(right_lower_edge_of_plane)$)$) node {\scalebox{1}{$\times$}};
        \draw ($($(left_upper_edge_of_plane)!0.2!(right_upper_edge_of_plane)$)!0.7!($(left_lower_edge_of_plane)!0.2!(right_lower_edge_of_plane)$)$) node {\scalebox{1}{$\times$}};
        \draw [blue, thick,decoration={markings, mark=at position 0.5 with {\arrow[scale=1.2]{<}}}, postaction={decorate}] 
        ($(upper_u1) + (a)$) -- ($(upper_v1) + (a)$)
        node[midway, above] {$g_2 g_1$};
        \draw[line width=1,->,blue!80] 
        ($($($(left_upper_edge_of_plane)!0.2!(right_upper_edge_of_plane)+1*(a)$)!0.3!($(left_lower_edge_of_plane)!0.2!(right_lower_edge_of_plane)+1*(a)$)$)+(0,0.4)$)
        arc(-270:118:0.4);
        \draw[line width=1,->,blue!80] 
        ($($($(left_upper_edge_of_plane)!0.2!(right_upper_edge_of_plane)+1*(a)$)!0.7!($(left_lower_edge_of_plane)!0.2!(right_lower_edge_of_plane)+1*(a)$)$)+(0,0.4)$)
        arc(-270:118:0.4);
        \draw[blue!80]
        ($($($(left_upper_edge_of_plane)!0.2!(right_upper_edge_of_plane)+1*(a)$)!0.3!($(left_lower_edge_of_plane)!0.2!(right_lower_edge_of_plane)+1*(a)$)$)+(0.5,-0.8)$)
        node {\scalebox{1}{$g_1$}};
    \end{tikzpicture}
    \caption{The symmetry operators on an $n$-sheeted manifold can be topologically deformed. For example, the symmetry operator $U_A(g_1)$ on the first sheet can be moved to the second sheet and fuse with the symmetry operator $U_A(g_2)$ there. As a result, the operators on the second sheet transform as $\CO_2 \to U_{\Sigma}^{\dagger}(g_1) \CO_2 U_{\Sigma}(g_1)$ The same applies to the remaining sheets.}
    \label{fig:fusion_symmetry_op}
\end{figure}
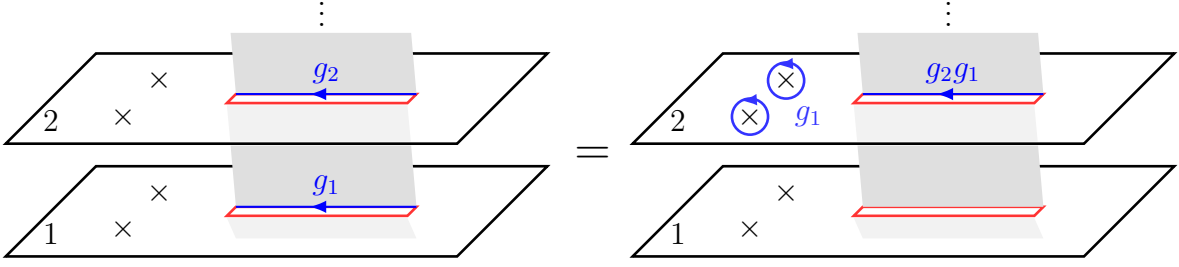

To investigate the dynamics of symmetry restoration, we have to consider an initially symmetry breaking state. However, in $1+1$ dimensions, the Coleman-Mermin-Wagner theorem \cite{Mermin:1966fe, Coleman:1973ci} forbids {\it spontaneous} breaking of continuous symmetries. Therefore, we focus on an initial quantum state which {\it explicitly} breaks the global symmetry $G$. As such a state, we prepare an initial state $\ket{\psi}$ by inserting a local operator $\mathcal{O}$ \cite{Nozaki:2014hna, He:2014mwa}:
\begin{align}
    \ket{\psi} = \frac{1}{\sqrt{Z_{\CO}}} \CO\left(z_{-}, \overline{z}_{-}\right) \ket{0} \ , 
\end{align}
where $Z_\mathcal{O}$ is the normalization constant fixed by $\braket{\psi|\psi}=1$ and $z_{-}=x_0 - \i \tau_{0}$ denotes the insertion point in complex coordinates. The real-time evolution of the density matrix is obtained by analytical continuation from Euclidean time\cite{Nozaki:2014hna, He:2014mwa}:
\begin{align}
    \rho_{\text{tot}}(t)
    &=
    e^{-\i H t} \ket{\psi}\bra{\psi} e^{\i H t}\nonumber\\
    &=
    \frac{1}{Z_{\CO}}\CO\left(z_{-}(t), \overline{z}_{-}(t)\right) \ket{0}\bra{0} \CO^{\dagger} \left(z_{+}(t), \overline{z}_{+}(t)\right)\ ,
    \label{eq:density_matrix_op_insertion}
\end{align}
where $z_{\pm}(t) = x_0  \pm \i \tau_{0} +t,\ \overline{z}_{\pm}(t) = x_0 \mp \i \tau_{0} - t $.\footnote{In our notation, $\overline{z}$ is not complex conjugate to $z$ in Minkowski spacetime.\label{foodnote:z}}
The reduced density matrix $\rho_A = \Tr_B \left[\ket{\psi} \bra{\psi} \right]$ can then be represented as a path integral form:\vspace{3mm}
\begin{align}
    \rho_A 
    =
    \frac{1}{Z_{\CO}}\ 
    \begin{tikzpicture}[scale=1,baseline={([yshift=-.5ex]current bounding box.center)}]
        \coordinate(left_down)  at (-0.5,-3/2);
        \coordinate(right_down) at (3.5,-3/2);
        \coordinate(left_up)  at (-0.5,3/2);
        \coordinate(right_up) at (3.5,3/2);
        \coordinate(epsilon) at (0,0.1);
        \coordinate(u_up) at ($(1,0)+(epsilon)$);
        \coordinate(v_up) at ($(3,0)+(epsilon)$);
        \coordinate(u_down) at ($(1,0)-(epsilon)$);
        \coordinate(v_down) at ($(3,0)-(epsilon)$);
        \draw[line width=1] (left_up)--(left_down)--(right_down)--(right_up)--cycle;
        \draw[line width=1,red!80] (u_down)--(v_down);
        \draw[line width=1,red!80] (u_up)--(v_up);
        \draw ($(right_down)!0.5!(right_up) + 2*(epsilon)$) node[right=2mm] {$\tau = +0$};
        \draw ($(right_down)!0.5!(right_up) - 2*(epsilon)$) node[right=2mm] {$\tau = -0$};
        \draw ($($(left_up)!0.2!(right_up)$)!0.4!($(left_down)!0.2!(right_down)$)$) node {\scalebox{1}{$\times$}}; 
        \draw ($($(left_up)!0.2!(right_up)$)!0.6!($(left_down)!0.2!(right_down)$)$) node {\scalebox{1}{$\times$}}; 
        \draw ($($(left_up)!0.2!(right_up)$)!0.4!($(left_down)!0.2!(right_down)$)$) node[above=0.1] {$\CO^{\dagger}$}; 
        \draw ($($(left_up)!0.2!(right_up)$)!0.6!($(left_down)!0.2!(right_down)$)$) node[below=0.1] {$\CO$}; 
    \end{tikzpicture}\ .
        \label{eq:rho_A_path_int}
\end{align}
From this representation, the trace $\Tr_A \left[ \rho_A^n \right]$ can be expressed as a path integral over the $n$-sheeted manifold $\CR_n$ \cite{Calabrese:2004eu, Casini:2009sr}.
\begin{align}
    \Tr_{A} \left[\rho_A^n \right]
    &=
    \frac{1}{Z_{\CO}^n} \times\ 
    \begin{tikzpicture}[scale=0.6,baseline={([yshift=-.5ex]current bounding box.center)}]
        \coordinate(left_lower_edge_of_plane)  at (0,0);
        \coordinate(right_lower_edge_of_plane) at (10,0);
        \coordinate(left_upper_edge_of_plane)  at (0+2,2);
        \coordinate(right_upper_edge_of_plane) at (10+2,2);
        \coordinate(a)  at (0,2.5);
        \coordinate(upper_u1) at (5+0.1,1+0.1);
        \coordinate(lower_u1) at (5-0.1,1-0.1);
        \coordinate(upper_v1) at (9+0.1,1+0.1);
        \coordinate(lower_v1) at (9-0.1,1-0.1);
        \coordinate(epsilon) at (0.04,0.24);
        \coordinate(o_up) at ($($(left_lower_edge_of_plane)!0.5!(left_upper_edge_of_plane)$)+(epsilon)$);
        \coordinate(o_down) at ($($(left_lower_edge_of_plane)!0.5!(left_upper_edge_of_plane)$)-(epsilon)$);
        \draw[line width=1, name path=lower_plane]
        (left_lower_edge_of_plane)--
        (right_lower_edge_of_plane)--
        (right_upper_edge_of_plane)--
        (left_upper_edge_of_plane)--cycle;
        \draw[line width=1, name path=upper_plane]
        ($(left_lower_edge_of_plane)  + (a)$)--
        ($(right_lower_edge_of_plane) + (a)$)--
        ($(right_upper_edge_of_plane) + (a)$)--
        ($(left_upper_edge_of_plane)  + (a)$)--cycle;
        \draw[line width=1, name path=upper_plane]
        ($(left_lower_edge_of_plane)  + 2.5*(a)$)--
        ($(right_lower_edge_of_plane) + 2.5*(a)$)--
        ($(right_upper_edge_of_plane) + 2.5*(a)$)--
        ($(left_upper_edge_of_plane)  + 2.5*(a)$)--cycle;
        \draw[line width=1,red!80](lower_u1)--(lower_v1)--(upper_v1)--(upper_u1)--cycle;
        \draw[line width=1,red!80]
        ($(lower_u1) + (a)$)--
        ($(lower_v1) + (a)$)--
        ($(upper_v1) + (a)$)--
        ($(upper_u1) + (a)$)--cycle;
        \draw[line width=1,red!80]
        ($(lower_u1) + 2.5*(a)$)--
        ($(lower_v1) + 2.5*(a)$)--
        ($(upper_v1) + 2.5*(a)$)--
        ($(upper_u1) + 2.5*(a)$)--cycle;
        \fill[lightgray!50]
        (upper_u1)--
        (upper_v1)--
        ($(upper_v1)!0.59!($(lower_v1) + (a)$)$)--
        ($(upper_u1)!0.59!($(lower_u1) + (a)$)$)--cycle;
        \fill[lightgray!20]
        ($(upper_v1)!0.63!($(lower_v1) + (a)$)$)--
        ($(upper_u1)!0.63!($(lower_u1) + (a)$)$)--
        ($(lower_u1) + (a) -(0,0.05)$)--
        ($(lower_v1) + (a) -(0,0.05)$)--cycle;
        \fill[lightgray!50]
        ($(upper_u1)+ (a)$)--
        ($(upper_v1)+ (a)$)--
        ($($(upper_v1)+ (a)$)!0.59!($(lower_v1) + 2*(a)$)$)--
        ($($(upper_u1)+ (a)$)!0.59!($(lower_u1) + 2*(a)$)$)--cycle;
        \fill[lightgray!20]
        ($($(upper_v1)+ (a)$)!0.8!($(lower_v1) + 2.5*(a)$)$)--
        ($($(upper_u1)+ (a)$)!0.8!($(lower_u1) + 2.5*(a)$)$)--
        ($(lower_u1) + 2.5*(a) -(0,0.05)$)--
        ($(lower_v1) + 2.5*(a) -(0,0.05)$)--cycle;
        \filldraw[lightgray!50]
        ($(upper_u1) + 2.5*(a)$)--
        ($(upper_v1) + 2.5*(a)$)--
        ($(upper_v1) + 2.5*(a) + 0.5*(a) + (-0.2,0)$) arc (180:90:0.35)--
        ($(upper_u1) + 2.5*(a) + 0.5*(a) + (-0.2,0) + (0.35,0.35)$ ) arc (90:180:0.35)--
        ($(upper_u1) + 2.5*(a) + 0.5*(a) + (-0.2,0)$ )--cycle;
        \draw[line width=1.5,lightgray!50] 
        ($(upper_v1) + 2.5*(a) + 0.5*(a) + (-0.2,0) + (0.35,0.35)$ ) arc (90:0:0.35) [rounded corners]--
        ($(upper_v1) + 2.5*(a) + 0.5*(a) + (-0.2,0) +(0.7,0) + (0, -0.6)$);
        \draw[line width=1.5,dashed,lightgray!50]
        ($(upper_v1) + 2.5*(a) + 0.5*(a) + (-0.2,0) +(0.7,0) + (0, -0.7)$)--
        ($(upper_v1) + 2.5*(a) + 0.5*(a) + (-0.2,0) +(0.7,0) + (0, -2.3)$);
        \draw[line width=1.5,lightgray!50]
        ($(upper_v1) +2.5*(a) + 0.5*(a) + (-0.2,0) +(0.7,0) + (0, -2.4)$)--
        ($(upper_v1) + 2.5*(a) + 0.5*(a) + (-0.2,0) +(0.7,0) + (0, -4)$);
        \draw[line width=1.5,dashed, lightgray!50]
        ($(upper_v1) + 2.5*(a) + 0.5*(a) + (-0.2,0) +(0.7,0) + (0, -4.1)$)--
        ($(upper_v1) +2.5*(a) + 0.5*(a) + (-0.2,0) +(0.7,0) + (0, -6.1)$);
        \draw[line width=1.5,lightgray!50]
        ($(upper_v1) +2.5*(a) + 0.5*(a) + (-0.2,0) +(0.7,0) + (0, -6.15)$)--
        ($(upper_v1) +2.5*(a) + 0.5*(a) + (-0.2,0) +(0.7,0) + (0, -6.6)$);
        \draw[line width=1.5,dashed, lightgray!50]
        ($(upper_v1) + 2.5*(a) + 0.5*(a) + (-0.2,0) +(0.7,0) + (0, -6.7)$)--
        ($(upper_v1) +2.5*(a) + 0.5*(a) + (-0.2,0) +(0.7,0) + (0, -7)$)[rounded corners]
        to [out = -90, in = 0]
        ($(lower_v1)       - 0.2*(a) + (+0.2,0)$);
        \fill[lightgray!20]
        ($(lower_u1)-(0,0.05)$)--
        ($(lower_v1)-(0,0.05)$)--
        ($(lower_v1) - 0.2*(a) + (+0.2,0)$)--
        ($(lower_u1) - 0.2*(a) + (+0.2,0)$)--cycle;
        \draw($($(upper_u1)+ (a)$)!0.5!($(lower_v1)+ 2.5*(a)$)$) node[above=-3mm] {\scalebox{1}{$\vdots $}};
        \draw ($(left_lower_edge_of_plane)+(1,0.5)$) node {\scalebox{1}{\rotatebox{0}{$1$}}};
        \draw ($(left_lower_edge_of_plane)+(a)+(1,0.5)$) node {\scalebox{1}{\rotatebox{0}{$2$}}};
        \draw ($(left_lower_edge_of_plane)+2.5*(a)+(1,0.5)$) node {\scalebox{1}{\rotatebox{0}{$n$}}};
        \draw ($($(left_upper_edge_of_plane)!0.2!(right_upper_edge_of_plane)+2.5*(a)$)!0.3!($(left_lower_edge_of_plane)!0.2!(right_lower_edge_of_plane)+2.5*(a)$)$) node {\scalebox{1}{$\times$}};
        \draw ($($(left_upper_edge_of_plane)!0.2!(right_upper_edge_of_plane)+2.5*(a)$)!0.7!($(left_lower_edge_of_plane)!0.2!(right_lower_edge_of_plane)+2.5*(a)$)$) node {\scalebox{1}{$\times$}};
        \draw ($($(left_upper_edge_of_plane)!0.2!(right_upper_edge_of_plane)+(a)$)!0.3!($(left_lower_edge_of_plane)!0.2!(right_lower_edge_of_plane)+(a)$)$) node {\scalebox{1}{$\times$}};
        \draw ($($(left_upper_edge_of_plane)!0.2!(right_upper_edge_of_plane)+(a)$)!0.7!($(left_lower_edge_of_plane)!0.2!(right_lower_edge_of_plane)+(a)$)$) node {\scalebox{1}{$\times$}};
        \draw ($($(left_upper_edge_of_plane)!0.2!(right_upper_edge_of_plane)$)!0.3!($(left_lower_edge_of_plane)!0.2!(right_lower_edge_of_plane)$)$) node {\scalebox{1}{$\times$}};
        \draw ($($(left_upper_edge_of_plane)!0.2!(right_upper_edge_of_plane)$)!0.7!($(left_lower_edge_of_plane)!0.2!(right_lower_edge_of_plane)$)$) node {\scalebox{1}{$\times$}};
    \end{tikzpicture}
    =
    \frac{1}{Z_{\CO}^n Z_n}
    \left\langle
    \prod_{i=1}^{n} 
    \CO_i \CO^{\dagger}_i
    \right\rangle_{\CR_n}\ ,
    \label{eq:Tr_rho_A^n}
\end{align}
where $\CO_i = \CO \left(z_{i,-}, \overline{z}_{i,-}\right),\ \CO^{\dagger}_i =  \CO^{\dagger} \left(z_{i,+}, \overline{z}_{i,+}\right)$ and $z_{i,\pm}$ denote the operator insertion points on the $i$-th sheet. The constant $Z_n$ denotes the partition function of $n$-sheeted manifold $\CR_n$. The numerator in charged moments \eqref{eq:charged_moment} can be evaluated in a similar way.
\begin{align}
    (\text{Numerator in } \eqref{eq:charged_moment})
    &=
    \frac{1}{Z_{\CO}^n} \times\ 
    \begin{tikzpicture}[scale=0.6,baseline={([yshift=-.5ex]current bounding box.center)}]
        \coordinate(left_lower_edge_of_plane)  at (0,0);
        \coordinate(right_lower_edge_of_plane) at (10,0);
        \coordinate(left_upper_edge_of_plane)  at (0+2,2);
        \coordinate(right_upper_edge_of_plane) at (10+2,2);
        \coordinate(a)  at (0,2.5);
        \coordinate(upper_u1) at (5+0.1,1+0.1);
        \coordinate(lower_u1) at (5-0.1,1-0.1);
        \coordinate(upper_v1) at (9+0.1,1+0.1);
        \coordinate(lower_v1) at (9-0.1,1-0.1);
        \coordinate(epsilon) at (0.04,0.24);
        \coordinate(o_up) at ($($(left_lower_edge_of_plane)!0.5!(left_upper_edge_of_plane)$)+(epsilon)$);
        \coordinate(o_down) at ($($(left_lower_edge_of_plane)!0.5!(left_upper_edge_of_plane)$)-(epsilon)$);
        \draw[line width=1, name path=lower_plane]
        (left_lower_edge_of_plane)--
        (right_lower_edge_of_plane)--
        (right_upper_edge_of_plane)--
        (left_upper_edge_of_plane)--cycle;
        \draw[line width=1, name path=upper_plane]
        ($(left_lower_edge_of_plane)  + (a)$)--
        ($(right_lower_edge_of_plane) + (a)$)--
        ($(right_upper_edge_of_plane) + (a)$)--
        ($(left_upper_edge_of_plane)  + (a)$)--cycle;
        \draw[line width=1, name path=upper_plane]
        ($(left_lower_edge_of_plane)  + 2.5*(a)$)--
        ($(right_lower_edge_of_plane) + 2.5*(a)$)--
        ($(right_upper_edge_of_plane) + 2.5*(a)$)--
        ($(left_upper_edge_of_plane)  + 2.5*(a)$)--cycle;
        \draw[line width=1,red!80](lower_u1)--(lower_v1)--(upper_v1)--(upper_u1)--cycle;
        \draw[line width=1,red!80]
        ($(lower_u1) + (a)$)--
        ($(lower_v1) + (a)$)--
        ($(upper_v1) + (a)$)--
        ($(upper_u1) + (a)$)--cycle;
        \draw[line width=1,red!80]
        ($(lower_u1) + 2.5*(a)$)--
        ($(lower_v1) + 2.5*(a)$)--
        ($(upper_v1) + 2.5*(a)$)--
        ($(upper_u1) + 2.5*(a)$)--cycle;
        \fill[lightgray!50]
        (upper_u1)--
        (upper_v1)--
        ($(upper_v1)!0.59!($(lower_v1) + (a)$)$)--
        ($(upper_u1)!0.59!($(lower_u1) + (a)$)$)--cycle;
        \fill[lightgray!20]
        ($(upper_v1)!0.63!($(lower_v1) + (a)$)$)--
        ($(upper_u1)!0.63!($(lower_u1) + (a)$)$)--
        ($(lower_u1) + (a) -(0,0.05)$)--
        ($(lower_v1) + (a) -(0,0.05)$)--cycle;
        \fill[lightgray!50]
        ($(upper_u1)+ (a)$)--
        ($(upper_v1)+ (a)$)--
        ($($(upper_v1)+ (a)$)!0.59!($(lower_v1) + 2*(a)$)$)--
        ($($(upper_u1)+ (a)$)!0.59!($(lower_u1) + 2*(a)$)$)--cycle;
        \fill[lightgray!20]
        ($($(upper_v1)+ (a)$)!0.8!($(lower_v1) + 2.5*(a)$)$)--
        ($($(upper_u1)+ (a)$)!0.8!($(lower_u1) + 2.5*(a)$)$)--
        ($(lower_u1) + 2.5*(a) -(0,0.05)$)--
        ($(lower_v1) + 2.5*(a) -(0,0.05)$)--cycle;
        \filldraw[lightgray!50]
        ($(upper_u1) + 2.5*(a)$)--
        ($(upper_v1) + 2.5*(a)$)--
        ($(upper_v1) + 2.5*(a) + 0.5*(a) + (-0.2,0)$) arc (180:90:0.35)--
        ($(upper_u1) + 2.5*(a) + 0.5*(a) + (-0.2,0) + (0.35,0.35)$ ) arc (90:180:0.35)--
        ($(upper_u1) + 2.5*(a) + 0.5*(a) + (-0.2,0)$ )--cycle;
        \draw[line width=1.5,lightgray!50] 
        ($(upper_v1) + 2.5*(a) + 0.5*(a) + (-0.2,0) + (0.35,0.35)$ ) arc (90:0:0.35) [rounded corners]--
        ($(upper_v1) + 2.5*(a) + 0.5*(a) + (-0.2,0) +(0.7,0) + (0, -0.6)$);
        \draw[line width=1.5,dashed,lightgray!50]
        ($(upper_v1) + 2.5*(a) + 0.5*(a) + (-0.2,0) +(0.7,0) + (0, -0.7)$)--
        ($(upper_v1) + 2.5*(a) + 0.5*(a) + (-0.2,0) +(0.7,0) + (0, -2.3)$);
        \draw[line width=1.5,lightgray!50]
        ($(upper_v1) +2.5*(a) + 0.5*(a) + (-0.2,0) +(0.7,0) + (0, -2.4)$)--
        ($(upper_v1) + 2.5*(a) + 0.5*(a) + (-0.2,0) +(0.7,0) + (0, -4)$);
        \draw[line width=1.5,dashed, lightgray!50]
        ($(upper_v1) + 2.5*(a) + 0.5*(a) + (-0.2,0) +(0.7,0) + (0, -4.1)$)--
        ($(upper_v1) +2.5*(a) + 0.5*(a) + (-0.2,0) +(0.7,0) + (0, -6.1)$);
        \draw[line width=1.5,lightgray!50]
        ($(upper_v1) +2.5*(a) + 0.5*(a) + (-0.2,0) +(0.7,0) + (0, -6.15)$)--
        ($(upper_v1) +2.5*(a) + 0.5*(a) + (-0.2,0) +(0.7,0) + (0, -6.6)$);
        \draw[line width=1.5,dashed, lightgray!50]
        ($(upper_v1) + 2.5*(a) + 0.5*(a) + (-0.2,0) +(0.7,0) + (0, -6.7)$)--
        ($(upper_v1) +2.5*(a) + 0.5*(a) + (-0.2,0) +(0.7,0) + (0, -7)$)[rounded corners]
        to [out = -90, in = 0]
        ($(lower_v1)       - 0.2*(a) + (+0.2,0)$);
        \fill[lightgray!20]
        ($(lower_u1)-(0,0.05)$)--
        ($(lower_v1)-(0,0.05)$)--
        ($(lower_v1) - 0.2*(a) + (+0.2,0)$)--
        ($(lower_u1) - 0.2*(a) + (+0.2,0)$)--cycle;
        \draw($($(upper_u1)+ (a)$)!0.5!($(lower_v1)+ 2.5*(a)$)$) node[above=-3mm] {\scalebox{1}{$\vdots $}};
        \draw ($(left_lower_edge_of_plane)+(1,0.5)$) node {\scalebox{1}{\rotatebox{0}{$1$}}};
        \draw ($(left_lower_edge_of_plane)+(a)+(1,0.5)$) node {\scalebox{1}{\rotatebox{0}{$2$}}};
        \draw ($(left_lower_edge_of_plane)+2.5*(a)+(1,0.5)$) node {\scalebox{1}{\rotatebox{0}{$n$}}};
        \draw ($($(left_upper_edge_of_plane)!0.2!(right_upper_edge_of_plane)+2.5*(a)$)!0.3!($(left_lower_edge_of_plane)!0.2!(right_lower_edge_of_plane)+2.5*(a)$)$) node {\scalebox{1}{$\times$}};
        \draw ($($(left_upper_edge_of_plane)!0.2!(right_upper_edge_of_plane)+2.5*(a)$)!0.7!($(left_lower_edge_of_plane)!0.2!(right_lower_edge_of_plane)+2.5*(a)$)$) node {\scalebox{1}{$\times$}};
        \draw ($($(left_upper_edge_of_plane)!0.2!(right_upper_edge_of_plane)+(a)$)!0.3!($(left_lower_edge_of_plane)!0.2!(right_lower_edge_of_plane)+(a)$)$) node {\scalebox{1}{$\times$}};
        \draw ($($(left_upper_edge_of_plane)!0.2!(right_upper_edge_of_plane)+(a)$)!0.7!($(left_lower_edge_of_plane)!0.2!(right_lower_edge_of_plane)+(a)$)$) node {\scalebox{1}{$\times$}};
        \draw ($($(left_upper_edge_of_plane)!0.2!(right_upper_edge_of_plane)$)!0.3!($(left_lower_edge_of_plane)!0.2!(right_lower_edge_of_plane)$)$) node {\scalebox{1}{$\times$}};
        \draw ($($(left_upper_edge_of_plane)!0.2!(right_upper_edge_of_plane)$)!0.7!($(left_lower_edge_of_plane)!0.2!(right_lower_edge_of_plane)$)$) node {\scalebox{1}{$\times$}};
        \draw [blue, thick,decoration={markings, mark=at position 0.5 with {\arrow[scale=1.2]{<}}}, postaction={decorate}] 
        (upper_u1) -- (upper_v1)
        node[midway, above] {$g_1$};
        \draw [blue, thick,decoration={markings, mark=at position 0.5 with {\arrow[scale=1.2]{<}}}, postaction={decorate}] 
        ($(upper_u1)+(a)$) -- ($(upper_v1)+(a)$)
        node[midway, above] {$g_2$};
        \draw [blue, thick,decoration={markings, mark=at position 0.5 with {\arrow[scale=1.2]{>}}}, postaction={decorate}] 
        ($(upper_u1)+2.5*(a)$) -- ($(upper_v1)+2.5*(a)$)
        node[midway, above] {$g_{n-1}\cdots g_1$};
    \end{tikzpicture}\nonumber\\\nonumber
\end{align}
To simplify this expression, we make use of the topological property of symmetry operators\footnote{One might concern about the presence of the boundary of subregion $A$ \cite{Ohmori:2014eia}. Here we consider the symmetric boundary condition at the edge of entangling surface $\partial A$ and the symmetry operators are also topological there. This means that the inserted local operators are only sources for symmetry breaking.}
As depicted in Fig.\,\ref{fig:fusion_symmetry_op}, we can move the symmetry operator on the $i$-th sheet to the $(i+1)$-th sheet $(i=1,\cdots ,n-1)$\cite{Benini:2024xjv}.
During this deformation, the symmetry operator acts on the operators  $\CO_i$ and $\CO^{\dagger}_i$, which transform as \eqref{eq:operator_trsf_group}. By repeating these deformations $(n-1)$ times, we arrive at the following expressions:
\begin{align}
    (\text{Numerator in } \eqref{eq:charged_moment})
    &=
    \frac{1}{Z_{\CO}^n} \times\ 
    \begin{tikzpicture}[scale=0.6,baseline={([yshift=-.5ex]current bounding box.center)}]
        \coordinate(left_lower_edge_of_plane)  at (0,0);
        \coordinate(right_lower_edge_of_plane) at (10,0);
        \coordinate(left_upper_edge_of_plane)  at (0+2,2);
        \coordinate(right_upper_edge_of_plane) at (10+2,2);
        \coordinate(a)  at (0,2.5);
        \coordinate(upper_u1) at (5+0.1,1+0.1);
        \coordinate(lower_u1) at (5-0.1,1-0.1);
        \coordinate(upper_v1) at (9+0.1,1+0.1);
        \coordinate(lower_v1) at (9-0.1,1-0.1);
        \coordinate(epsilon) at (0.04,0.24);
        \coordinate(o_up) at ($($(left_lower_edge_of_plane)!0.5!(left_upper_edge_of_plane)$)+(epsilon)$);
        \coordinate(o_down) at ($($(left_lower_edge_of_plane)!0.5!(left_upper_edge_of_plane)$)-(epsilon)$);
        \draw[line width=1, name path=lower_plane]
        (left_lower_edge_of_plane)--
        (right_lower_edge_of_plane)--
        (right_upper_edge_of_plane)--
        (left_upper_edge_of_plane)--cycle;
        \draw[line width=1, name path=upper_plane]
        ($(left_lower_edge_of_plane)  + (a)$)--
        ($(right_lower_edge_of_plane) + (a)$)--
        ($(right_upper_edge_of_plane) + (a)$)--
        ($(left_upper_edge_of_plane)  + (a)$)--cycle;
        \draw[line width=1, name path=upper_plane]
        ($(left_lower_edge_of_plane)  + 2.5*(a)$)--
        ($(right_lower_edge_of_plane) + 2.5*(a)$)--
        ($(right_upper_edge_of_plane) + 2.5*(a)$)--
        ($(left_upper_edge_of_plane)  + 2.5*(a)$)--cycle;
        \draw[line width=1,red!80](lower_u1)--(lower_v1)--(upper_v1)--(upper_u1)--cycle;
        \draw[line width=1,red!80]
        ($(lower_u1) + (a)$)--
        ($(lower_v1) + (a)$)--
        ($(upper_v1) + (a)$)--
        ($(upper_u1) + (a)$)--cycle;
        \draw[line width=1,red!80]
        ($(lower_u1) + 2.5*(a)$)--
        ($(lower_v1) + 2.5*(a)$)--
        ($(upper_v1) + 2.5*(a)$)--
        ($(upper_u1) + 2.5*(a)$)--cycle;
        \fill[lightgray!50]
        (upper_u1)--
        (upper_v1)--
        ($(upper_v1)!0.59!($(lower_v1) + (a)$)$)--
        ($(upper_u1)!0.59!($(lower_u1) + (a)$)$)--cycle;
        \fill[lightgray!20]
        ($(upper_v1)!0.63!($(lower_v1) + (a)$)$)--
        ($(upper_u1)!0.63!($(lower_u1) + (a)$)$)--
        ($(lower_u1) + (a) -(0,0.05)$)--
        ($(lower_v1) + (a) -(0,0.05)$)--cycle;
        \fill[lightgray!50]
        ($(upper_u1)+ (a)$)--
        ($(upper_v1)+ (a)$)--
        ($($(upper_v1)+ (a)$)!0.59!($(lower_v1) + 2*(a)$)$)--
        ($($(upper_u1)+ (a)$)!0.59!($(lower_u1) + 2*(a)$)$)--cycle;
        \fill[lightgray!20]
        ($($(upper_v1)+ (a)$)!0.8!($(lower_v1) + 2.5*(a)$)$)--
        ($($(upper_u1)+ (a)$)!0.8!($(lower_u1) + 2.5*(a)$)$)--
        ($(lower_u1) + 2.5*(a) -(0,0.05)$)--
        ($(lower_v1) + 2.5*(a) -(0,0.05)$)--cycle;
        \filldraw[lightgray!50]
        ($(upper_u1) + 2.5*(a)$)--
        ($(upper_v1) + 2.5*(a)$)--
        ($(upper_v1) + 2.5*(a) + 0.5*(a) + (-0.2,0)$) arc (180:90:0.35)--
        ($(upper_u1) + 2.5*(a) + 0.5*(a) + (-0.2,0) + (0.35,0.35)$ ) arc (90:180:0.35)--
        ($(upper_u1) + 2.5*(a) + 0.5*(a) + (-0.2,0)$ )--cycle;
        \draw[line width=1.5,lightgray!50] 
        ($(upper_v1) + 2.5*(a) + 0.5*(a) + (-0.2,0) + (0.35,0.35)$ ) arc (90:0:0.35) [rounded corners]--
        ($(upper_v1) + 2.5*(a) + 0.5*(a) + (-0.2,0) +(0.7,0) + (0, -0.6)$);
        \draw[line width=1.5,dashed,lightgray!50]
        ($(upper_v1) + 2.5*(a) + 0.5*(a) + (-0.2,0) +(0.7,0) + (0, -0.7)$)--
        ($(upper_v1) + 2.5*(a) + 0.5*(a) + (-0.2,0) +(0.7,0) + (0, -2.3)$);
        \draw[line width=1.5,lightgray!50]
        ($(upper_v1) +2.5*(a) + 0.5*(a) + (-0.2,0) +(0.7,0) + (0, -2.4)$)--
        ($(upper_v1) + 2.5*(a) + 0.5*(a) + (-0.2,0) +(0.7,0) + (0, -4)$);
        \draw[line width=1.5,dashed, lightgray!50]
        ($(upper_v1) + 2.5*(a) + 0.5*(a) + (-0.2,0) +(0.7,0) + (0, -4.1)$)--
        ($(upper_v1) +2.5*(a) + 0.5*(a) + (-0.2,0) +(0.7,0) + (0, -6.1)$);
        \draw[line width=1.5,lightgray!50]
        ($(upper_v1) +2.5*(a) + 0.5*(a) + (-0.2,0) +(0.7,0) + (0, -6.15)$)--
        ($(upper_v1) +2.5*(a) + 0.5*(a) + (-0.2,0) +(0.7,0) + (0, -6.6)$);
        \draw[line width=1.5,dashed, lightgray!50]
        ($(upper_v1) + 2.5*(a) + 0.5*(a) + (-0.2,0) +(0.7,0) + (0, -6.7)$)--
        ($(upper_v1) +2.5*(a) + 0.5*(a) + (-0.2,0) +(0.7,0) + (0, -7)$)[rounded corners]
        to [out = -90, in = 0]
        ($(lower_v1)       - 0.2*(a) + (+0.2,0)$);
        \fill[lightgray!20]
        ($(lower_u1)-(0,0.05)$)--
        ($(lower_v1)-(0,0.05)$)--
        ($(lower_v1) - 0.2*(a) + (+0.2,0)$)--
        ($(lower_u1) - 0.2*(a) + (+0.2,0)$)--cycle;
        \draw($($(upper_u1)+ (a)$)!0.5!($(lower_v1)+ 2.5*(a)$)$) node[above=-3mm] {\scalebox{1}{$\vdots $}};
        \draw ($(left_lower_edge_of_plane)+(1,0.5)$) node {\scalebox{1}{\rotatebox{0}{$1$}}};
        \draw ($(left_lower_edge_of_plane)+(a)+(1,0.5)$) node {\scalebox{1}{\rotatebox{0}{$2$}}};
        \draw ($(left_lower_edge_of_plane)+2.5*(a)+(1,0.5)$) node {\scalebox{1}{\rotatebox{0}{$n$}}};
        \draw ($($(left_upper_edge_of_plane)!0.2!(right_upper_edge_of_plane)+2.5*(a)$)!0.3!($(left_lower_edge_of_plane)!0.2!(right_lower_edge_of_plane)+2.5*(a)$)$) node {\scalebox{1}{$\times$}};
        \draw ($($(left_upper_edge_of_plane)!0.2!(right_upper_edge_of_plane)+2.5*(a)$)!0.7!($(left_lower_edge_of_plane)!0.2!(right_lower_edge_of_plane)+2.5*(a)$)$) node {\scalebox{1}{$\times$}};
        \draw ($($(left_upper_edge_of_plane)!0.2!(right_upper_edge_of_plane)+(a)$)!0.3!($(left_lower_edge_of_plane)!0.2!(right_lower_edge_of_plane)+(a)$)$) node {\scalebox{1}{$\times$}};
        \draw ($($(left_upper_edge_of_plane)!0.2!(right_upper_edge_of_plane)+(a)$)!0.7!($(left_lower_edge_of_plane)!0.2!(right_lower_edge_of_plane)+(a)$)$) node {\scalebox{1}{$\times$}};
        \draw ($($(left_upper_edge_of_plane)!0.2!(right_upper_edge_of_plane)$)!0.3!($(left_lower_edge_of_plane)!0.2!(right_lower_edge_of_plane)$)$) node {\scalebox{1}{$\times$}};
        \draw ($($(left_upper_edge_of_plane)!0.2!(right_upper_edge_of_plane)$)!0.7!($(left_lower_edge_of_plane)!0.2!(right_lower_edge_of_plane)$)$) node {\scalebox{1}{$\times$}};
        \draw[line width=1,->,blue!80] 
        ($($($(left_upper_edge_of_plane)!0.2!(right_upper_edge_of_plane)+(a)$)!0.3!($(left_lower_edge_of_plane)!0.2!(right_lower_edge_of_plane)+(a)$)$)+(0,0.4)$)
        arc(-270:118:0.4);
        \draw[line width=1,->,blue!80] 
        ($($($(left_upper_edge_of_plane)!0.2!(right_upper_edge_of_plane)+(a)$)!0.7!($(left_lower_edge_of_plane)!0.2!(right_lower_edge_of_plane)+(a)$)$)+(0,0.4)$)
        arc(-270:118:0.4);
        \draw[line width=1,->,blue!80] 
        ($($($(left_upper_edge_of_plane)!0.2!(right_upper_edge_of_plane)+2.5*(a)$)!0.3!($(left_lower_edge_of_plane)!0.2!(right_lower_edge_of_plane)+2.5*(a)$)$)+(0,0.4)$)
        arc(-270:118:0.4);
        \draw[line width=1,->,blue!80] 
        ($($($(left_upper_edge_of_plane)!0.2!(right_upper_edge_of_plane)+2.5*(a)$)!0.7!($(left_lower_edge_of_plane)!0.2!(right_lower_edge_of_plane)+2.5*(a)$)$)+(0,0.4)$)
        arc(-270:118:0.4);
        \draw[blue!80]
        ($($($(left_upper_edge_of_plane)!0.2!(right_upper_edge_of_plane)+(a)$)!0.3!($(left_lower_edge_of_plane)!0.2!(right_lower_edge_of_plane)+(a)$)$)+(0.2,-0.8)$)
        node {\scalebox{1}{$g_1$}};
        \draw[blue!80]
        ($($($(left_upper_edge_of_plane)!0.2!(right_upper_edge_of_plane)+2.5*(a)$)!0.3!($(left_lower_edge_of_plane)!0.2!(right_lower_edge_of_plane)+2.5*(a)$)$)+(1.2,-1)$)
        node {\scalebox{1}{$g_{n-1}\cdots g_1$}};
    \end{tikzpicture}\nonumber\\\nonumber\\
    &=
    \frac{1}{Z_{\CO}^n Z_n}
    \left\langle
    \CO_1 \CO^{\dagger}_1
    \prod_{i=2}^{n} 
    \left[ R (g_{i-1} \cdots g_{1}) \CO_i \right]
    \left[ \overline{R} (g_{i-1} \cdots g_{1}) \CO^{\dagger}_i \right]
    \right\rangle_{\CR_n}\ .
    \label{eq:Tr_rho_AG^n}
\end{align}
Substituting \eqref{eq:Tr_rho_A^n} and \eqref{eq:Tr_rho_AG^n} into \eqref{eq:charged_moment}, the charged moment can be expressed in terms of correlation functions
\begin{align}
    Z(g_1,\cdots ,g_{n-1})
    &=
    \frac{
    \left\langle
    \CO_1 \CO^{\dagger}_1
    \prod_{i=2}^{n} 
    \left[ R (g_{i-1} \cdots g_{1}) \CO_i \right]
    \left[ \overline{R} (g_{i-1} \cdots g_{1}) \CO^{\dagger}_i \right]
    \right\rangle_{\CR_n}
    }{
    \left\langle
    \prod_{i=1}^{n} 
    \CO_i \CO^{\dagger}_i
    \right\rangle_{\CR_n}
    }\ .
    \label{eq:charged_moment_def}
\end{align}

\subsection{Entanglement asymmetry in $(1+1)$-dimensional CFT}
In the case of a $(1+1)$-dimensional CFT, the charged moments \eqref{eq:charged_moment_def} can be further simplified. The $n$-sheeted manifold $\CR_n$ can be mapped to plane via the following conformal transformation \cite{He:2014mwa, Guo:2015uwa, Benini:2024xjv}:
\begin{align}
    z
    &\to
    \zeta = \left(\frac{z}{z - \ell} \right)^{1/n}\ ,
\end{align}
where we take the interval as $A=[0, \ell]$.
Under this mapping, the correlation functions appearing in the charged moments reduce to those on the plane:
\begin{align}
    Z(g_1,\cdots ,g_{n-1})
    &=
    \frac{
    \left\langle
    \tilde{\CO}_1 \tilde{\CO}^{\dagger}_1
    \prod_{i=2}^{n} 
    \left[ R (g_{i-1} \cdots g_{1}) \tilde{\CO}_i \right]
    \left[ \overline{R} (g_{i-1} \cdots g_{1}) \tilde{\CO}^{\dagger}_i \right]
    \right\rangle_{\text{Plane}}
    }{
    \left\langle
    \prod_{i=1}^{n} 
    \tilde{\CO}_i \tilde{\CO}^{\dagger}_i
    \right\rangle_{\text{Plane}}
    }\ ,
\end{align}
where $\tilde{\CO}_i = \CO \left(\zeta_{1,-}, \overline{\zeta}_{1,-}\right)$, and $\tilde{\CO}^{\dagger}_i = \CO^{\dagger} \left(\zeta_{1,+}, \overline{\zeta}_{1,+}\right)$. The conformal factors cancel in the numerator and the denominator in the charged moments. The coordinates $\zeta_{i,\pm}$ on the plane are given by
\begin{align}
    \zeta_{i,\pm}(t)
    &=
    \left(\frac{z_{\pm}(t)}{z_{\pm}(t) - \ell} \right)^{1/n}
    \exp \left(\frac{2\pi \i }{n} (i-1)\right)\ ,\ i=1,\cdots ,n\ .
    \label{eq:zeta_z_rel}
\end{align}
In the following, we focus on the second R\'enyi entanglement asymmetry ($\text{EA}_2$) by restricting to $n=2$. In this case, the charged moments reduce to the ratio of four-point functions:
\begin{align}
    Z(g)
    =
    \frac{
    \left\langle
    \tilde{\CO}_1 \tilde{\CO}^{\dagger}_1
    \left[ R (g) \tilde{\CO}_2 \right]
    \left[ \overline{R} (g) \tilde{\CO}^{\dagger}_2 \right]
    \right\rangle_{\text{Plane}}
    }{
    \left\langle
    \tilde{\CO}_1 \tilde{\CO}^{\dagger}_1
    \tilde{\CO}_2 \tilde{\CO}^{\dagger}_2
    \right\rangle_{\text{Plane}}
    }\ ,
    \label{eq:charged_moment_4pt_functions}
\end{align}
with the coordinates $\zeta_{i,\pm}(t)$:\footnote{Here we follow the convention used in \cite{He:2014mwa}}
\begin{align}
\begin{aligned}
    \zeta_{1,\pm}(t)
    &=
    - \zeta_{2,\pm}(t)
    =
    \sqrt{\frac{x_0 + t  \pm \i \tau_0 }{x_0 + t -\ell \pm \i \tau_0 } }\ ,
\end{aligned}
\begin{aligned}
    \overline{\zeta_{1,\pm}}(t)
    &=
    - \overline{\zeta_{2,\pm}}(t)
    =
    \sqrt{\frac{ x_0 - t  \mp \i \tau_0 }{ x_0 - t - \ell \mp \i \tau_0 } }\ ,
\end{aligned}
\label{eq:zeta_1_2_value}
\end{align}
where we define the square root of complex function as 
\begin{align}
    \sqrt{\omega} 
    =
    \sqrt{r} e^{\i\theta/2} 
    \quad, \quad \theta \in [-\pi,\pi)
    \quad,\quad \omega = \frac{z}{z-\ell}\ . 
\end{align}
Here, we choose the branch cut along the negative real axis in the $\omega$-plane, which corresponds to $A = [0,\ell]$ in the original $z$-plane.

We further perform the following conformal transformation:
\begin{align}
    \xi
    &=
    \frac{(\zeta - \zeta_{2,+})(\zeta_{1,+}-\zeta_{2,-})}{(\zeta - \zeta_{2,-})(\zeta_{1,+}-\zeta_{2,+})}\ .
\end{align}
This conformal transformation maps the coordinates $\left( \zeta_{1,-}, \zeta_{1,+},\zeta_{2,-}, \zeta_{2,+} \right) \mapsto \left( x,1,\infty,0 \right)$ and the cross-ratios $x$ and $\overline{x}$ is given by\footnote{See also footnote  \ref{foodnote:z}.}
\begin{align}
x(t) 
&= 
\left\{\quad
    \begin{aligned}
        &
        \frac{1}{2} 
        + \frac{1}{2} 
            \left[
            1 + \left( 
                \frac{\tau_0 \ell}{\left(t + x_0 \right) \left( t +x_0 -\ell \right) + \tau_0^2}
                \right)^2
            \right]^{-1/2}
            \quad \text{for} \quad
            \left(t + x_0 \right) \left( t +x_0 -\ell \right) + \tau_0^2 \geq 0 \\
        &
        \frac{1}{2} 
        - \frac{1}{2} 
            \left[
            1 + \left( 
                \frac{\tau_0 \ell}{\left(t + x_0 \right) \left( t +x_0 -\ell \right) + \tau_0^2}
                \right)^2
            \right]^{-1/2}
            \quad \text{for} \quad
            \left(t + x_0 \right) \left( t +x_0 -\ell \right) + \tau_0^2 < 0
    \end{aligned}
\label{eq:cross-ratio}
,\right. \\
\overline{x}(t) 
&= 
\left\{\quad
    \begin{aligned}
        &
        \frac{1}{2} 
        + \frac{1}{2} 
            \left[
            1 + \left( 
                \frac{\tau_0 \ell}{\left(t -x_0 \right) \left( t - x_0 + \ell \right) + \tau_0^2}
                \right)^2
            \right]^{-1/2}
            \quad \text{for} \quad
            \left(t -x_0 \right) \left( t - x_0 + \ell \right) + \tau_0^2 \geq 0 \\
        &
        \frac{1}{2} 
        - \frac{1}{2} 
            \left[
            1 + \left( 
                \frac{\tau_0 \ell}{\left(t -x_0 \right) \left( t - x_0 + \ell \right) + \tau_0^2}
                \right)^2
            \right]^{-1/2}
            \quad \text{for} \quad
            \left(t -x_0 \right) \left( t - x_0 + \ell \right) + \tau_0^2 < 0
    \end{aligned}
,\right.
\label{eq:cross-ratio_bar}
\end{align}
It turns out that the conformal factors are again canceled in the numerator and the denominator in \eqref{eq:charged_moment_4pt_functions}, then the $\text{EA}_2$ reduces to
\begin{align}
    \Delta S_{A}^{(2)} (t)
    =
    -
    \log 
    \left[
    \int_G \text{d} g\
        \frac{
        \left\langle
        \CO \left(x(t), \overline{x}(t)\right) 
        \CO^{\dagger} \left(1,1\right)
        \left[ R (g) \CO \right] \left(\infty, \infty\right) 
        \left[ \overline{R} (g) \CO^{\dagger} \right] \left(0, 0\right)
        \right\rangle
        }{
        \left\langle
        \CO \left(x(t), \overline{x}(t)\right) \CO^{\dagger} \left(1, 1\right)
        \CO \left(\infty, \infty \right) \CO^{\dagger} \left(0, 0\right)
        \right\rangle
        }
    \right] \ .
    \label{eq:REA_charged_moment_n=2}
\end{align}
Here, we omitted the subscript ``Plane'' in the correlation functions, and all operators in this expression are defined on the plane. From this expression, the $\text{EA}_2$ is determined by four-point functions, which are well studied in the context of CFTs.

\section{Quantum Mpemba effect: fundamental case}\label{sec:EA_in_WZW}

In previous section, we expressed $\text{EA}_2$ in terms of the four-point conformal correlators. The main goal of this section is to explore the symmetry restoration structure in $\widehat{su}(N)_k$ WZW model by focusing on the initial state built from primary operator $\Phi_{i}(z, \bar{z})=\Phi_{i}(z)\otimes\overline{\Phi}_{i}(\bar{z})$ in the fundamental representation.\footnote{In principle, one can consider the most general state as $\Phi_{ij}(z, \bar{z})=\Phi_{i}(z)\otimes\overline{\Phi}_{j}(\bar{z})$. In this paper, for simplicity, we restrict to the diagonal case: $i=j$.\label{footnote:operaotor_insertion}} (The index $i$ runs from 1 to $N$.) 

\subsection{Exact results}
In this case, $\text{EA}_2$ can be written in the following decomposed form:\footnote{Remark that, in this expression, the index $i$ is fixed and we do not take the sum over $i$.}
\begin{align}
    \Delta S_{A}^{(2)} (t)
    =
    -
    \log 
    \left[
        \frac{
        \sum_{j,k} \sum_{j', k'}
        \sum_{A,B=1}^{2} 
        \left( M_4 \right)_{iikk'}^{jj'ii}\ 
        \left( I_A\right)_{ij}^{ki}  \ \left( \overline{I}_B \right)_{ij'}^{k'i}    \
        G_{A,B} \left(t\right)
        }{
        \sum_{A,B=1}^{2} 
        G_{A,B} \left(t\right)
        }
    \right] \ ,
    \label{eq:EA_2_WZW_general}
\end{align}
where $M$ is given by the integral \cite{Creutz:1978ub}
\begin{align}
\begin{aligned}
    M_{iikk'}^{jj'ii}
    &=
    \int_{\text{SU}(N)} \text{d} g\
    \left[ R (g) \right]_{i}^{\,\, j}  
    \left[ R (g) \right]_{i}^{\,\, j'}  
    \left[ R^{-1} (g) \right]_{k}^{\,\, i}
    \left[ R^{-1} (g) \right]_{k'}^{\,\, i}\ \\
    &=
    \frac{1}{N(N+1)} 
    \left(
        \delta_k^j \delta_{k'}^{j'} + \delta_{k'}^j \delta_k^{j'}
    \right)\ ,
\end{aligned}
\label{eq:Haar_int_moment}
\end{align}
and $I_A$ and $\overline{I}_B$ are invariant tensors in $\mathbf{N} \otimes \overline{\mathbf{N}} \otimes \overline{\mathbf{N}} \otimes \mathbf{N}$ for holomorphic and anti-holomorphic sectors, respectively. Also, $G_{A,B} \left(t\right)\equiv G_{A,B}\left(x(t), \overline{x}(t) \right)$ is the four-point function associated to the invariant tensors $I_A$ and $\overline{I}_B$, which can be derived by solving the Knizhnik-Zamolodchikov equation\cite{Knizhnik:1984nr}:
\begin{align}
    G_{AB}(t)
    =
    f(t)
    \left[
    F_A^{(-)}(x(t)) F_B^{(-)} (\overline{x}(t))
    +\frac{1-c_{--}^2}{c_{+-}^2}F_A^{(+)}(x(t)) F_B^{(+)}(\overline{x}(t))
    \right]
    \ ,
    \label{eq:G_AB_su_main}
\end{align}
where $f(t)$ is the function which does not depend on the indices $A$ and $B$. Also, $c_{--}$ and $c_{+-}$ are constants given by
\begin{align}
    c_{--}
    =
    N\frac{
    \Gamma\left(\frac{N}{N+k}\right)
    \Gamma\left(-\frac{N}{N+k}\right)
    }{
    \Gamma\left(\frac{1}{N+k}\right)
    \Gamma\left(-\frac{1}{N+k}\right)}
    \ , \quad 
    c_{+-}
    =
    -N\frac{
    \Gamma\left(\frac{N}{N+k}\right)^2
    }{
    \Gamma\left(\frac{N+1}{N+k}\right)
    \Gamma\left(\frac{N-1}{N+k}\right)
    }\ ,
\end{align}
and $F_{A}^{(\pm)}(x)$ are defined by
\begin{align}
    &F_{1}^{(-)}(x)
    =
    F\left(\frac{1}{N+k}, -\frac{1}{N+k}; \frac{k}{N+k};x\right) \ , \\
    &F_{1}^{(+)}(x)
    =
    x^{\frac{N}{N+k}}
    F\left(\frac{N-1}{N+k}, \frac{N+1}{N+k};1+\frac{N}{N+k};x\right) \ , \\
    &F_{2}^{(-)}(x)
    =
    \frac{1}{k} x
    F\left(1+\frac{1}{N+k}, 1-\frac{1}{N+k};2-\frac{N}{N+k};x\right) \ , \\
    &F_{2}^{(+)}(x)
    =
    -N x^{\frac{N}{N+k}} 
    F\left(\frac{N-1}{N+k}, \frac{N+1}{N+k};\frac{N}{N+k};x\right) \ ,
\end{align}
where $F(a,b;c;x)$ is the hypergeometric function. For completeness, we present the review on the four-point function in $\widehat{su}(N)_k$ WZW model in Appendix \ref{sec:4pt_function_in_WZW}. By plugging the above all results into \eqref{eq:EA_2_WZW_general}, we can obtain the exact expression for $\text{EA}_2$:
\begin{align}
    \Delta S_{A}^{(2)} (t)
    &=
    -
    \log 
    \left[
    \frac{
    2G_{1,1}(t) + (N+1) \left( G_{1,2}(t)+G_{2,1}(t) \right) +N(N+1)G_{2,2}(t)
    }{
    N(N+1)\sum_{A,B =1}^{2} G_{A,B} (t)}
    \right]\ .
    \label{eq:EA_2_su_exact}
\end{align}

\subsection{Asymptotic behaviors}
In the previous subsection, we derived the exact result of $\text{EA}_2$ for an initial state build from the primary operator in the fundamental representation. 
However, due to the complexity of the expression, it is quite hard to give physical insights as it is. In this subsection, to get over this complexity, we investigate the behaviors of the exact result by taking various limits. We also give intrinsic interpretations of our results from the view point of the quasi-particle picture. For later convenience, we define dimensionless parameters $\tilde{\tau}_0,\ \tilde{x}_0,\ \tilde{t}$ as follows:
\begin{align}
    \tilde{\tau}_0 = \frac{\tau_0}{\ell},\quad
    \tilde{x}_0 = \frac{x_0}{\ell},\quad
    \tilde{t} = \frac{t}{\ell}\ .
\end{align}

\subsection*{Long time limit: $ \tilde{t}  \to \infty $\ .}
In this limit, the asymptotic form of $\text{EA}_2$ \eqref{eq:EA_2_su_exact} is given by
\begin{align}
    \Delta S_{A}^{(2)} \left(\tilde{t} \right)
    =
    \frac{(N-1)(N^2-2)}{N^2(N+1)} c_{N,k} \epsilon^{\frac{2N}{N+k}}
    +
    \frac{2}{k} \left( 1-\frac{1}{N} \right) \epsilon 
    +
    \mathcal{O}\left( \epsilon^{1+\frac{2N}{N+k}} \right)
    \ ,\label{eq:EA_2_asymptotic_t->infty}
\end{align}
where the infinitesimally small parameter $\epsilon$ is given by
\begin{align}
    \epsilon
    &=
    \frac{1}{4} 
    \left( \frac{\tilde{\tau}_0 }{\tilde{t}^2} \right)^2\ ,
    \label{eq:epsilon_t->infty}
\end{align}
and the coefficient $c_{N,k}$ is defined as
\begin{align}
    c_{N,k}
    &=
    \frac{
    \Gamma\left( \frac{k}{N+k} \right)^2
    \Gamma\left( \frac{N-1}{N+k} \right)
    \Gamma\left( \frac{N+1}{N+k} \right)
    }{
    \Gamma\left( \frac{N}{N+k} \right)^2
    \Gamma\left( \frac{k-1}{N+k} \right)
    \Gamma\left( \frac{k+1}{N+k} \right)
    }
    \ge 0\ .
    \label{eq:c_(N,k)}
\end{align}
The detailed derivation is provided in Appendix \ref{sec:Asymptotic_derivation}. Note that $\text{EA}_2$ is always non-negative and $\text{EA}_2$ decays to zero as $\tilde{t} \to \infty$. This property is in line with our intuition since we start from a symmetry-breaking initial state and evolve it by the symmetry-preserving Hamiltonian. After sufficiently long time, the symmetry has to be restored in subsystem $A$, i.e., $\Delta S_{A}^{(2)} \left(\tilde{t}=\infty \right) = 0$.

\subsection*{Large interval limit: $ \tilde{\tau}_0 \to 0 $ .}
\begin{figure}
    \centering
    \begin{minipage}[b]{0.49\columnwidth}
    \centering
    \begin{tikzpicture}
        \coordinate(o) at (0,0); 
        \coordinate(x0) at (2,0); 
        \coordinate(x1) at (4,0); 
        \coordinate(R) at (6,0); 
        \coordinate(h) at (0,4); 
        \coordinate(h1) at (0,3); 
        \draw[line width=1.5, ->] ($(o)+(-0.5,0)$)--(R) node[right = 2mm] {\scalebox{1.2}{$\tilde{t}$}};
        \draw[line width=1.5, ->] ($(o)+(0,-0.5)$)--(h) node[above = 1mm] {\scalebox{1}{$\Delta S_{A}^{(2)} \left(\tilde{t} \right)$}};
        \draw[line width=1.5,blue!80] (o)--(x0)--($(x0)+(h1)$)--($(x1)+(h1)$)--(x1)--($(R)+(-0.5,0)$);
        \draw[line width=1, dashed] ($(x0)+(h1)$)--(h1) node[left = 2mm] {$\log N$};
        \draw ($(o)+(-0.3,-0.3)$) node {$0$};
        \draw (x0) node[below] {$|\tilde{x}_0|$};
        \draw (x1) node[below] {$|\tilde{x}_0| + 1 $};
    \end{tikzpicture}
    \end{minipage}
    \begin{minipage}[b]{0.49\columnwidth}
    \centering
    \begin{tikzpicture}
        \coordinate(L) at (0,0); 
        \coordinate(u) at (2,0); 
        \coordinate(v) at (5,0); 
        \coordinate(R) at (7,0); 
        \coordinate(p) at (1,0); 
        \coordinate(h) at (0,3); 
        \coordinate(hL) at ($(h)+(0,-2.5)$); 
        \coordinate(hR) at ($(h)+(R)+(-1,0)$); 
        \path[name path = u_line] (u)--($(u)+(h)$);
        \path[name path = v_line] (v)--($(v)+(h)$);
        \path[name path = phL] (p)--(hL);
        \path[name path = phR] (p)--(hR);
        \draw[line width=1.5] (L)--(u) (v)--(R);
        \draw[line width=1.5, red!80] (u)--(v);
        \draw[line width=1.5,red!80, dashed] (u)--($(u)+(h)$);
        \draw[line width=1.5, red!80, dashed] (v)--($(v)+(h)$);
        \draw[line width=1.5,blue!80 ,->, dashed] (p)--(hL);
        \draw[line width=1.5, blue!80,->, dashed, name intersections = {of = u_line and phR, by ={i}}] (p)--(i);
        \draw[line width=1.5,->, blue!80, name intersections = {of = u_line and phR, by ={i}}, name intersections = {of = v_line and phR, by ={j}}] (i)--(j);
        \draw[line width=1.5, blue!80, dashed, ->, name intersections = {of = v_line and phR, by ={i}}] (i)--(hR);
        \draw[->,line width=1.5] ($(L)+(-0.5,0)$)--($(h)+(-0.5,0)$);
        \draw ($($(L)+(-0.5,0)$)!0.5!($(h)+(-0.5,0)$)$) node[left = 2mm] {\scalebox{1.2}{$t$}};
        \fill[red!80] (u) circle (3pt) (v) circle (3pt);
        \draw[blue!80] (p) node {\scalebox{1.3}{$\bigstar$}};
        \draw[red!80] ($(u)!0.5!(v)$) node[below = 2mm] {\scalebox{1.2}{$A$}};
        \draw ($(L)!0.5!(u)$) node[below = 2mm] {\scalebox{1.2}{$B$}};
        \draw ($(v)!0.5!(R)$) node[below = 2mm] {\scalebox{1.2}{$B$}};
    \end{tikzpicture}
    \end{minipage}
\caption{Real-time dependence of $\text{EA}_2$ for $\tilde{x}_0 < 0$ in the large interval limit $\tilde{\tau}_0 \to 0$ (left panel) and its quasi-particle interpretation (right panel). In the right panel, the symbol ``$\star$'' denotes the operator insertion point $(x_0,0)$, and red line indicates the subregion $A$. The trajectories of the quasi-particles are depicted by the blue lines, and the horizontal and vertical axes represent space and real time directions, respectively.}
\label{fig:quasi-particle picture_out}
\end{figure}
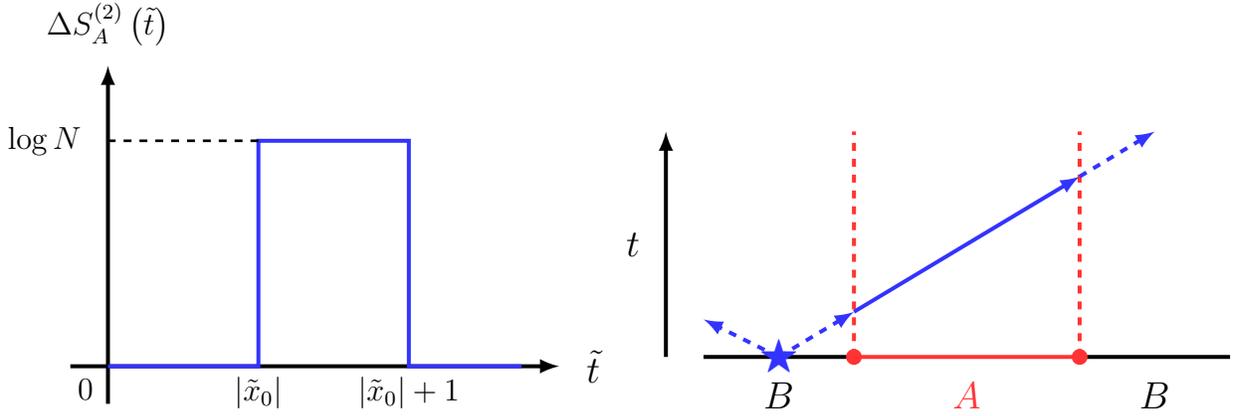

In this limit, the behavior of $\text{EA}_2$ depends on the sign of $\tilde{x}_0$. For $\tilde{x}_0 \le 0$, the asymptotic form of $\text{EA}_2$ is given by
\begin{align}
    \lim_{\tilde{\tau}_0 \to 0}
    \Delta S_{A}^{(2)} \left(\tilde{t} \right)
    &=
    \left\{ \
    \begin{aligned}
        & \log N \quad &, \quad \tilde{t} 
        \in \left[|\tilde{x}_0|,|\tilde{x}_0| + 1 \right] \ ,\\
        & \ 0 \quad &, \quad \tilde{t} 
        \not\in \left[|\tilde{x}_0|, |\tilde{x}_0| + 1\right]\ .
    \end{aligned}
    \right.
    \label{eq:EA_2_tau0->x_0<0}
\end{align}
The above result is physically intuitive: the operator is inserted at $(x_0,0)$, emitting a right-moving quasi-particle (holomorphic part of the inserted operator) and left-moving quasi-particle (anti-holomorphic part) with same constant speed. These quasi-particles propagate in opposite directions to each other under the real-time evolution (see Fig.\,\ref{fig:quasi-particle picture_out}). When $\tilde{t} < |\tilde{x}_0|$, both quasi-particles remain outside the subregion $A$ and thus do not break the symmetry there. At $\tilde{t} = |\tilde{x}_0| $, the right-moving quasi-particle enters subregion $A$ and induces symmetry breaking in subregion $A$, yielding the non-zero value of $\text{EA}_2$. Finally, when $\tilde{t} > |\tilde{x}_0| + 1$, the right-moving quasi-particle does no longer exist in the subregion $A$, and correspondingly $\text{EA}_2$ returns to zero.\footnote{A similar quasi-particle picture has been used in the context of entanglement entropy \cite{Calabrese:2005in,Alba:2017ekd}, although in our case the quasi-particle and anti quasi-particle are not entangled to each other.}\\
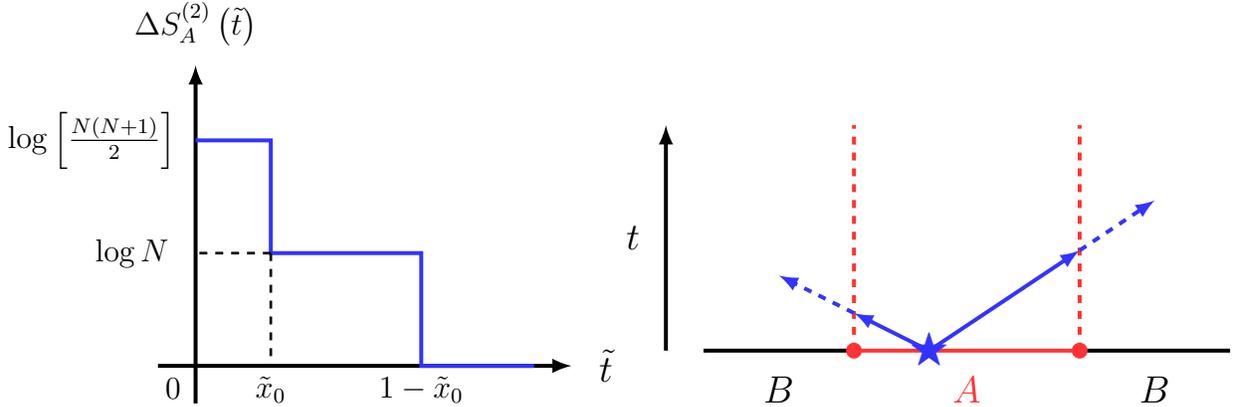
\begin{figure}
    \centering
    \begin{minipage}[b]{0.49\columnwidth}
    \centering
    \begin{tikzpicture}
        \coordinate(o) at (0,0); 
        \coordinate(x0) at (1,0); 
        \coordinate(x1) at (3,0); 
        \coordinate(R) at (5,0); 
        \coordinate(h) at (0,4); 
        \coordinate(h1) at (0,3); 
        \coordinate(h2) at (0,1.5); 
        \draw[line width=1.5, ->] ($(o)+(-0.5,0)$)--(R) node[right = 2mm] {\scalebox{1.2}{$\tilde{t}$}};
        \draw[line width=1.5, ->] ($(o)+(0,-0.5)$)--(h) node[above = 1mm] {\scalebox{1}{$\Delta S_{A}^{(2)} \left( \tilde{t} \right)$}};
        \draw[line width=1.5,blue!80] (h1) node[black, left = 1mm] {$\log \left[\frac{N(N+1)}{2} \right]$}--($(x0)+(h1)$)--($(x0)+(h2)$)--($(x1)+(h2)$)--(x1)--($(R)+(-0.5,0)$);
        \draw[line width=1, dashed] ($(x0)+(h2)$)--(h2) node[left = 2mm] {$\log N$};
        \draw[line width=1, dashed] ($(x0)+(h2)$)--(x0);
        \draw ($(o)+(-0.3,-0.3)$) node {$0$};
        \draw (x0) node[below] {$\tilde{x}_0$};
        \draw (x1) node[below] {$1 - \tilde{x}_0$};
    \end{tikzpicture}
    \end{minipage}
    \begin{minipage}[b]{0.49\columnwidth}
    \centering
    \begin{tikzpicture}
        \coordinate(L) at (0,0); 
        \coordinate(u) at (2,0); 
        \coordinate(v) at (5,0); 
        \coordinate(R) at (7,0); 
        \coordinate(p) at (3,0); 
        \coordinate(h) at (0,3); 
        \coordinate(hL) at ($(h)+(1,-2)$); 
        \coordinate(hR) at ($(h)+(R)+(-1,-1)$); 
        \path[name path = u_line] (u)--($(u)+(h)$);
        \path[name path = v_line] (v)--($(v)+(h)$);
        \path[name path = phL] (p)--(hL);
        \path[name path = phR] (p)--(hR);
        \draw[line width=1.5] (L)--(u) (v)--(R);
        \draw[line width=1.5, red!80] (u)--(v);
        \draw[line width=1.5, dashed, red!80] (u)--($(u)+(h)$);
        \draw[line width=1.5, dashed, red!80] (v)--($(v)+(h)$);
        \draw[line width=1.5,->, blue!80, name intersections = {of = u_line and phL, by ={i}}] (p)--(i);
        \draw[line width=1.5, blue!80, dashed, ->, name intersections = {of = u_line and phL, by ={i}}] (i)--(hL);
        \draw[line width=1.5,->, blue!80, name intersections = {of = v_line and phR, by ={i}}] (p)--(i);
        \draw[line width=1.5, blue!80, dashed, ->, name intersections = {of = v_line and phR, by ={i}}] (i)--(hR);
        \draw[->,line width=1.5] ($(L)+(-0.5,0)$)--($(h)+(-0.5,0)$);
        \draw ($($(L)+(-0.5,0)$)!0.5!($(h)+(-0.5,0)$)$) node[left = 2mm] {\scalebox{1.2}{$t$}};
        \fill[red!80] (u) circle (3pt) (v) circle (3pt);
        \draw[blue!80] (p) node {\scalebox{1.3}{$\bigstar$}};
        \draw[red!80] ($(u)!0.5!(v)$) node[below = 2mm] {\scalebox{1.2}{$A$}};
        \draw ($(L)!0.5!(u)$) node[below = 2mm] {\scalebox{1.2}{$B$}};
        \draw ($(v)!0.5!(R)$) node[below = 2mm] {\scalebox{1.2}{$B$}};
    \end{tikzpicture}
    \end{minipage}
\caption{The real-time dependence of $\text{EA}_2$ for $\tilde{x}_0 > 0$ in the large interval limit $\tilde{\tau}_0 \to 0$ (left panel) and its quasi-particle interpretation (right panel). Here, we assume $\tilde{x}_0 < 1 /2$ for simplicity.}
\label{fig:quasi-particle picture_in}
\end{figure}

On the other hand, for $\tilde{x}_0 > 0$ we obtain
\begin{align}
    \lim_{\tilde{\tau}_0 \to 0}
    \Delta S_{A}^{(2)} \left( \tilde{t} \right)
    &=
    \left\{ \
    \begin{aligned}
        &\log\left[\frac{N(N+1)}{2}\right]
        \quad &, \quad 0 \le \tilde{t} < \tilde{x}_0\ ,\\
        & \log N
        \quad &, \quad \tilde{x}_0 \le \tilde{t} < 1 - \tilde{x}_0\ ,\\
        &\ 0 \quad &, \quad 1 - \tilde{x}_0 < \tilde{t} \ ,
    \end{aligned}
    \right.
    \label{eq:EA_2_tau0->x_0>0}
\end{align}
where we assume $\tilde{x}_0 < 1/2$ for simplicity. 
This result can be also understood from the viewpoint of the quasi-particle picture, as illustrated in Fig.\,\ref{fig:quasi-particle picture_in}. At $\tilde{t} = 0$, both quasi-particles are located at one point within the subregion $A$, and they lead symmetry breaking and the $\text{EA}_2$ takes the value of $\log \left[ N(N+1)/2\right]$. For $\tilde{x}_0 \le \tilde{t} < 1 - \tilde{x}_0$, the left-moving quasi-particle leaves subregion $A$ while the right-moving quasi-particle remains inside. In this case, the quasi-particle leads weaker symmetry breaking and the $\text{EA}_2$ takes the value of $\log N$. Finally, for $1-\tilde{x}_0 < \tilde{t}$, no quasi-particles remain in subregion $A$, and the $\text{EA}_2$ vanish, indicating symmetry preservation. 

We should also comment that our result is consistent with the intuition of the entanglement asymmetry. Roughly speaking, the entanglement asymmetry measures the effective degrees of freedom which break global symmetries in the subregion $A$. This intuitive feature can be explicitly seen in our result. Indeed, by taking the large $N$ limit, the EA$_2$ given in \eqref{eq:EA_2_tau0->x_0>0} is reduced to
\begin{align}
    \lim_{\tilde{\tau}_0 \to 0}
    \Delta S_{A}^{(2)} \left( \tilde{t} \right)
    &\overset{\text{Large }N}{=}
    \left\{ \
    \begin{aligned}
        &2\log N
        \quad , \quad\quad 0 \le \tilde{t} < \tilde{x}_0\ ,\\
        & \log N
        \quad\ , \quad\quad \tilde{x}_0 \le \tilde{t} < 1 -\tilde{x}_0\ ,\\
        &\ 0 \quad\quad \quad \, , \quad\quad 1 - \tilde{x}_0 < \tilde{t} \ .
    \end{aligned}
    \right.
\end{align}
Clearly, the factor of $2$ in front of the logarithm for 
$0 \leq \tilde{t} < \tilde{x}_0$ originates from the contributions of 
both left- and right-moving quasi-particles. In this regime, 
$\text{EA}_2$ is therefore twice as large as in the interval 
$\tilde{x}_0 \leq \tilde{t} < 1 - \tilde{x}_0$, where only the 
right-moving particle contributes. This behavior is fully consistent 
with the general intuition of the entanglement asymmetry.

\begin{figure}
    \centering
    \begin{minipage}[b]{0.49\columnwidth}
    \centering
    \includegraphics[width=0.9\linewidth]{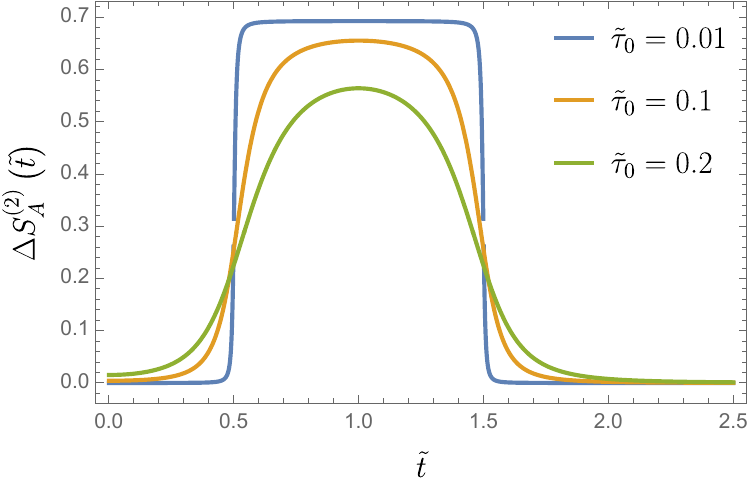}
    \end{minipage}
    \begin{minipage}[b]{0.49\columnwidth}
    \centering
    \includegraphics[width=0.9\linewidth]{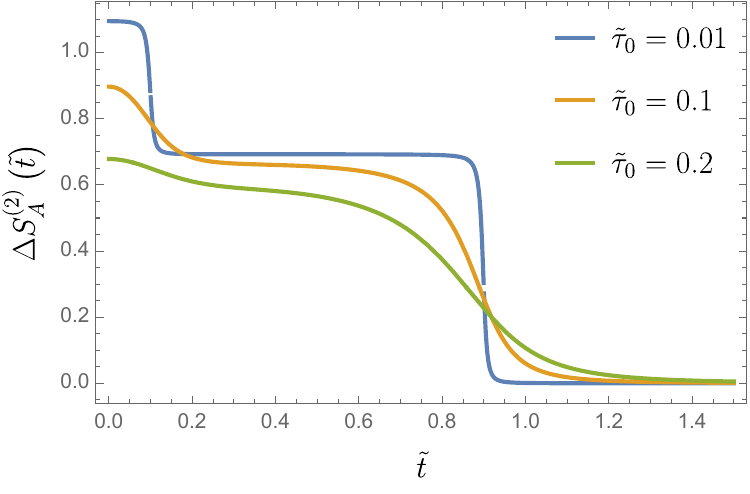}
    \end{minipage}
    \caption{Numerical plot of $\text{EA}_2$ for $(N,k) = (2,1)$ obtained from the analytical expression for $\tilde{\tau}_0 = 0.01, \ 0.1,\ 0.2$, plotted as a function of the dimensionless time $\tilde{t}$. The left panel corresponds to $\tilde{x}_0 = -0.5$ and the right panel to $\tilde{x}_0 = 0.1$. A smaller value of $\tilde{\tau}_0$ preserves the sharp feature predicted by the quasi-particle picture, while larger values smooth out the profiles.}
    \label{fig:quasi_particle_numerical}
\end{figure}

As $\tilde{\tau}_0$ increases, the quasi-particle picture gradually loses its validity due to the finite size effect of the subregion $A$. In Fig.\,\ref{fig:quasi_particle_numerical}, we present the numerical evaluation of \eqref{eq:EA_2_su_exact} for $\tilde{\tau}_0 = 0.01, \ 0.1,\ 0.2$. While the case $\tilde{\tau}_0 = 0.01$ is still well described by the quasi-particle picture, the results for $\tilde{\tau}_0  = 0.1$ and $0.2$ exhibit smoother profiles, signaling the breakdown of this interpretation.

\subsection*{Other limits: $ \tilde{\tau}_0 \to \infty $ or $ |\tilde{x}_0| \to \infty $\ .}
In these limits, the asymptotic form of $\text{EA}_2$ coincides with that in the long time limit \eqref{eq:EA_2_asymptotic_t->infty}, except that the parameter $\epsilon$ is replaced by
\begin{align}
\begin{aligned}
    \epsilon 
    &=
    \frac{1}{4} 
    \left(\frac{1}{\tilde{\tau}_0} \right)^2
    \quad, \quad
    \tilde{\tau}_0 \to \infty\ ,\\
    \epsilon 
    &= \frac{1}{4} 
    \left(\frac{\tilde{\tau}_0 }{\tilde{x}_0^2} \right)^2
    \quad, \quad
    |\tilde{x}_0| \to \infty\ .\\
\end{aligned}
\end{align}
In those cases, the insertion point of the symmetry-breaking operator is sufficiently far from subregion $A$. As a result, the operator insertion does not break the symmetry in subsystem $A$, thereby leading to a vanishing $\text{EA}_2$.

\subsection{Quantum Mpemba effect for $\text{SU}(N)$ symmetry}
\begin{figure}
    \centering
    \begin{minipage}[b]{0.45\columnwidth}
    \centering
    \begin{tikzpicture}[scale=1,baseline={([yshift=-.5ex]current bounding box.center)}]
        \coordinate(left_down)  at (0,-3/2);
        \coordinate(right_down) at (5,-3/2);
        \coordinate(left_up)  at (0,3/2);
        \coordinate(right_up) at (5,3/2);
        \coordinate(epsilon) at (0,0.1);
        \coordinate(u_up) at ($($($(left_down)!0.5!(left_up)$)!0.3!($(right_down)!0.5!(right_up)$)$)+(epsilon)$);
        \coordinate(v_up) at ($($($(left_down)!0.5!(left_up)$)!0.7!($(right_down)!0.5!(right_up)$)$)+(epsilon)$);
        \coordinate(u_down) at ($($($(left_down)!0.5!(left_up)$)!0.3!($(right_down)!0.5!(right_up)$)$)-(epsilon)$);
        \coordinate(v_down) at ($($($(left_down)!0.5!(left_up)$)!0.7!($(right_down)!0.5!(right_up)$)$)-(epsilon)$);
        \coordinate (p_up) at ($($(left_up)!0.5!(right_up)$)!0.3!($(left_down)!0.5!(right_down)$)$);
        \coordinate (p_down) at ($($(left_up)!0.5!(right_up)$)!0.7!($(left_down)!0.5!(right_down)$)$);
        \draw[line width=1] (left_up)--(left_down)--(right_down)--(right_up)--cycle;
        \draw[line width=1,red!80] (u_down)--(v_down);
        \draw[line width=1,red!80] (u_up)--(v_up);
        \draw[line width=0.8, <->] ($($(u_up)!0.5!(v_up)$) + (0.3,0)$)--($(p_up) + (0.3,0)$) node[right = 1mm] {$\tau_0$};
        \draw (p_up) node {\scalebox{1}{$\times$}}; 
        \draw ($(p_up) + (0,0.4)$) node  {$\CO^{\dagger}$};
        \draw (p_down) node {\scalebox{1}{$\times$}}; 
        \draw ($(p_down) + (0,-0.4)$) node {$\CO$}; 
        \draw ($($(left_down)!0.5!(right_down)$)+(0,-0.5)$) node {$x_0 = 1/2$};
    \end{tikzpicture}
    \end{minipage}
    \begin{minipage}[b]{0.45\columnwidth}
    \centering
    \begin{tikzpicture}[scale=1,baseline={([yshift=-.5ex]current bounding box.center)}]
        \coordinate(left_down)  at (0,-3/2);
        \coordinate(right_down) at (5,-3/2);
        \coordinate(left_up)  at (0,3/2);
        \coordinate(right_up) at (5,3/2);
        \coordinate(epsilon) at (0,0.1);
        \coordinate(u_up) at ($($($(left_down)!0.5!(left_up)$)!0.3!($(right_down)!0.5!(right_up)$)$)+(epsilon)$);
        \coordinate(v_up) at ($($($(left_down)!0.5!(left_up)$)!0.7!($(right_down)!0.5!(right_up)$)$)+(epsilon)$);
        \coordinate(u_down) at ($($($(left_down)!0.5!(left_up)$)!0.3!($(right_down)!0.5!(right_up)$)$)-(epsilon)$);
        \coordinate(v_down) at ($($($(left_down)!0.5!(left_up)$)!0.7!($(right_down)!0.5!(right_up)$)$)-(epsilon)$);
        \coordinate (p_up) at ($($(left_up)!0.3!(right_up)$)!0.3!($(left_down)!0.3!(right_down)$)$);
        \coordinate (p_down) at ($($(left_up)!0.3!(right_up)$)!0.7!($(left_down)!0.3!(right_down)$)$);
        \draw[line width=1] (left_up)--(left_down)--(right_down)--(right_up)--cycle;
        \draw[line width=1,red!80] (u_down)--(v_down);
        \draw[line width=1,red!80] (u_up)--(v_up);
        \draw[line width=0.8, <->] ($(u_up) + (0.3,0)$)--($(p_up) + (0.3,0)$) node[right = 1mm] {$\tau_0$};
        \draw (p_up) node {\scalebox{1}{$\times$}}; 
        \draw ($(p_up) + (0,0.4)$) node  {$\CO^{\dagger}$};
        \draw (p_down) node {\scalebox{1}{$\times$}}; 
        \draw ($(p_down) + (0,-0.4)$) node {$\CO$}; 
        \draw ($($(left_down)!0.5!(right_down)$)+(0,-0.5)$) node {$x_0 = 0$};
    \end{tikzpicture}
    \end{minipage}
    \caption{Our setup to discuss the quantum Mpemba effect . The red lines indicate the subregion $A$, while the symbols ``$\times$'' denote the insertion points of the operators $\CO$ and $\CO^{\dagger}$. The spacial coordinate of the operator insertion is set to $\tilde{x}_0 = 1 /2$ in the left panel and $\tilde{x}_0 = 0$ in the right panel. In both cases, the coordinate $\tilde{\tau}_0$ denotes the vertical distance between the operator insertion point and the subregion $A$.}
    \label{fig:operator_insertion_points}
\end{figure}
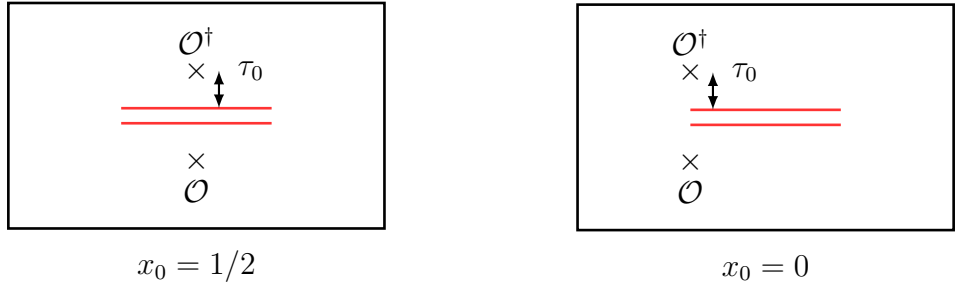

In the previous sections, we have studied various aspects of the $\text{EA}_2$ in $\widehat{su}(N)_{k}$ WZW model. We are now in a position to investigate the quantum Mpemba effect for $\text{SU}(N)$ symmetry via the $\text{EA}_2$.
To discuss the quantum Mpemba effect , we consider the two specific cases: $\tilde{x}_0 = 1 / 2,\ 0$ as illustrated in Fig.\,\ref{fig:operator_insertion_points}. 

\begin{figure}
    \centering
\begin{minipage}[b]{0.49\columnwidth}
    \centering
    \includegraphics[width=0.9\columnwidth]{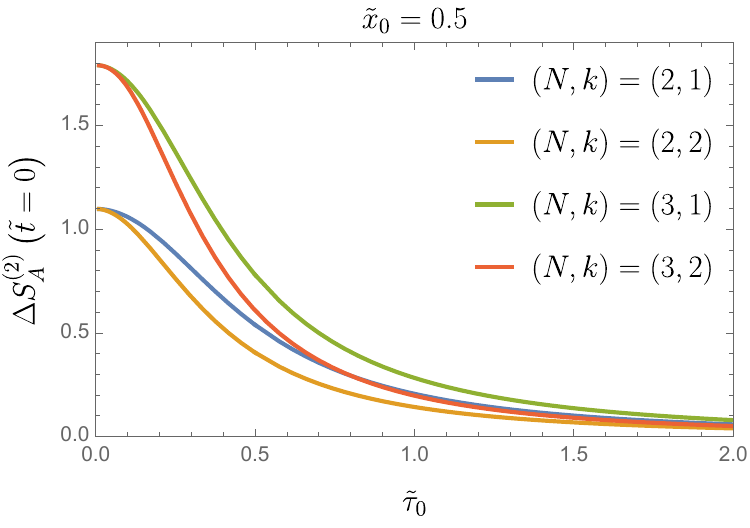}
\end{minipage}
\begin{minipage}[b]{0.49\columnwidth}
    \centering
    \includegraphics[width=0.9\columnwidth]{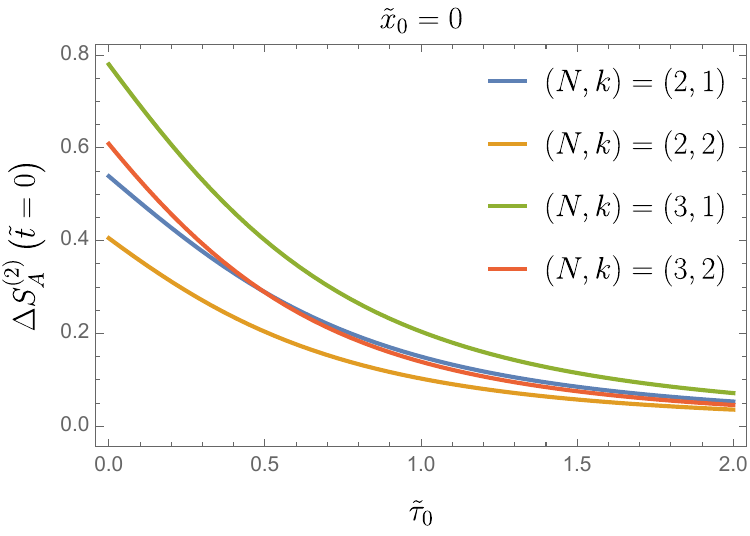}
\end{minipage}
\caption{Numerical evaluation of $\text{EA}_2$ at $\tilde{t}= 0$, plotted as a function of $\tilde{\tau}_0$. The special coordinate of the operator insertion is chosen as $\tilde{x}_0 = 1/2$ in the left panel and $\tilde{x}_0 = 0$ in the right panel.}
\label{fig:EA_2_tau_0}
\end{figure}

The coordinate $\tilde{\tau}_0$ describes the vertical separation between the operator insertion point and subregion $A$, and we treat $\tilde{\tau}_0$ as a parameter controlling the degrees of initial symmetry breaking. The $\text{EA}_2$ also depends on the real time $\tilde{t}$, rank $N$ and level $k$, we thus express the $\text{EA}_2$ as $\Delta S_{A}^{(2)} \left( \tilde{t}; \tilde{x}_0, \tilde{\tau}_0, N, k \right)$.
As shown in Fig.\,\ref{fig:EA_2_tau_0}, the value of $\text{EA}_2$ at $\tilde{t}=0$ decreases monotonically with respect to $\tilde{\tau}_0$ for any $N$ and $k$:
\begin{align}
    \Delta S_{A}^{(2)} \left( \tilde{t} = 0;\tilde{x}_0, \tilde{\tau}_0, N, k \right)
    > 
    \Delta S_{A}^{(2)} \left( \tilde{t} = 0;\tilde{x}_0, \tilde{\tau}_0', N, k \right)
    \quad, \quad
    \tilde{\tau}_0 < \tilde{\tau}_0'\ .
    \label{eq:EA_2_t=0_tau_0}
\end{align}
Increasing $\tilde{\tau}_0$ shifts the operator insertion points further away from subregion $A$, thereby reduces the degree of symmetry breaking. 

\begin{figure}
      \centering
\begin{minipage}[b]{0.49\columnwidth}
    \centering
    \includegraphics[width=0.9\columnwidth]{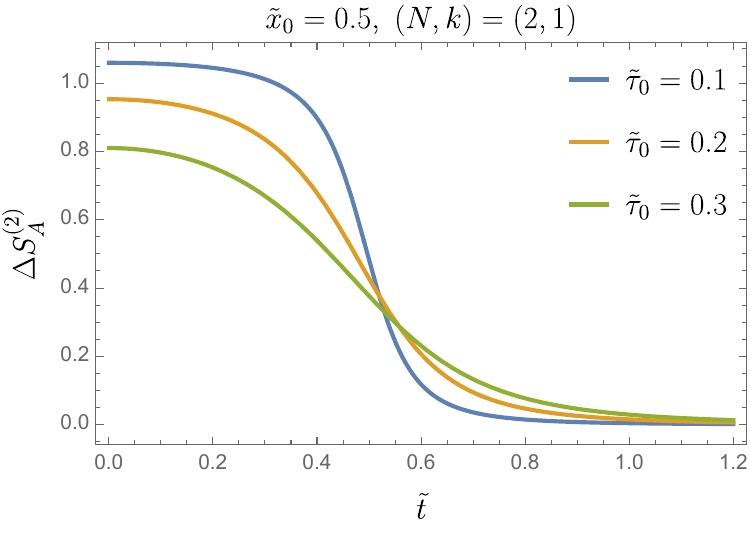}
\end{minipage}
\begin{minipage}[b]{0.49\columnwidth}
    \centering
    \includegraphics[width=0.9\columnwidth]{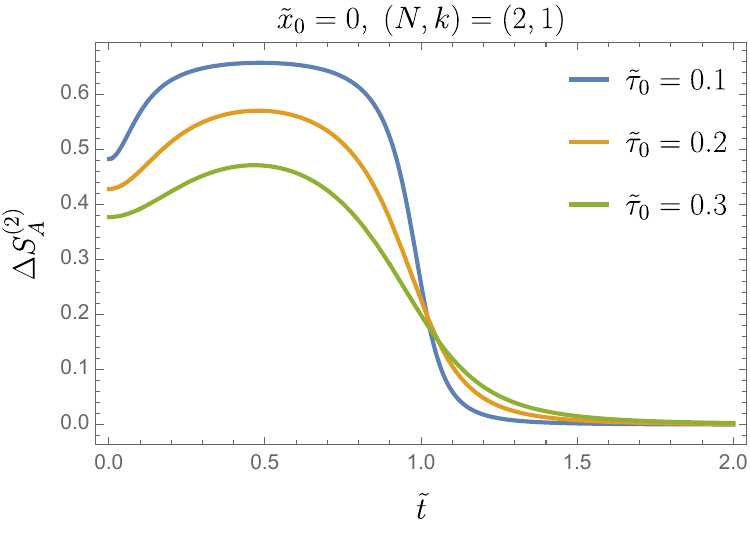}
\end{minipage}
  \caption{Time evolution of $\text{EA}_2$ for $\tilde{x}_0 = 1/2$ (left panel) and $\tilde{x}_0 = 0$ (right panel), with $(N,k) = (2,1)$. In both cases, the curve exhibit crossings, indicating the presence of quantum Mpemba effect.}
  \label{fig:QME_tau0}
\end{figure}

On the other hand, from the long-time behavior of $\text{EA}_2$ in \eqref{eq:EA_2_asymptotic_t->infty}, increasing $\tilde{\tau}_0$ is equivalent to a larger $\epsilon$, and all coefficients in front of $\epsilon$ are positive. As a result, $\text{EA}_2$ increases with $\tilde{\tau}_0$ in the limit $\tilde{t}\to \infty$:
\begin{align}
    \Delta S_{A}^{(2)} \left( \tilde{t} \to \infty;\tilde{x}_0, \tilde{\tau}_0, N, k \right)
    <
    \Delta S_{A}^{(2)} \left( \tilde{t}\to \infty;\tilde{x}_0, \tilde{\tau}_0', N, k \right)
    \quad, \quad
    \tilde{\tau}_0 < \tilde{\tau}_0' <\hspace{-1.5mm}< \tilde{t}^{2} \ .
    \label{eq:EA_2_t=infty_tau_0}
\end{align}
Comparing the inequalities \eqref{eq:EA_2_t=0_tau_0} and \eqref{eq:EA_2_t=infty_tau_0}, we expect that the quantum Mpemba effect always occurs for $\text{SU}(N)$ symmetry. Fig.\,\ref{fig:QME_tau0} shows the time evolution of the exact result \eqref{eq:EA_2_su_exact} setting $(N,k) = (2,1)$ and $\tilde{x}_0 = 1/2, 0$. For $\tilde{x}_0 = 1/2$, the $\text{EA}_2$ decreases monotonically and there is crossings in the plots, which is an evidence for the quantum Mpemba effect. On the other hand, in the case $\tilde{x}_0 = 0$, the $\text{EA}_2$ increases at early time,\footnote{This behavior is consistent with the left panel in Fig.\,\ref{fig:quasi_particle_numerical}.} however it eventually decreases monotonically for large $\tilde{t}$. We can observe crossings again and the quantum Mpemba effect occurs even for $\tilde{x}_0 = 0$. We can see that a similar behavior is found for other values of $(N,k)$ although we do not show the corresponding numerical plots. 

\subsection{New type of quantum Mpemba effect}
In the previous subsection, we analyzed the quantum Mpemba effect for $\text{SU}(N)$ symmetry by varying the parameter $\tilde{\tau}_0$ at fixed $(N, k)$. Here, we investigate the symmetry restoration structures by varying $(N, k)$ while keeping $\tilde{\tau}_0$ fixed. Interestingly, we uncover a qualitatively new type of quantum Mpemba effect: increasing $N$ amplifies the degree of initial symmetry breaking simultaneously accelerates the symmetry restoration, whereas increasing $k$ diminishes the initial symmetry breaking and decelerates the symmetry restoration. 

\subsection*{$N$ dependence with $k$ fixed}
Let us first examine the $N$ dependence of $\text{EA}_2$ with the level $k$ fixed. From the long time behavior in \eqref{eq:EA_2_asymptotic_t->infty}, we obtain the following inequality for $\text{EA}_2$ between different values of $N$:
\begin{align}
\begin{aligned}
    \Delta S_{A}^{(2)} \left(\tilde{t} \to \infty;\tilde{x}_0, \tilde{\tau}_0, N, k\right)
    &>
    \Delta S_{A}^{(2)} \left(\tilde{t}\to \infty;\tilde{x}_0, \tilde{\tau}_0, N', k\right)
    \ 
    &, \ N < N' < k
\end{aligned}
\label{eq:EA_2_t=infty_N}
\end{align}
We next consider the $N$ dependence of $\text{EA}_2$ at $\tilde{t} = 0$. In the long interval limit $\tilde{\tau}_0 \to 0$, \eqref{eq:EA_2_tau0->x_0<0} and \eqref{eq:EA_2_tau0->x_0>0} yield the initial value of $\text{EA}_2$ as\footnote{Strictly speaking, for $\tilde{x}_{0} = 0$, one must slightly shift the time to $\tilde{t} = \tilde{\delta}$ and take $\tilde{\tau}_0 \ll \tilde{\delta}$. For a detailed derivation, see Appendix \ref{sec:Asymptotic_derivation}.}
\begin{align}
    \begin{aligned}
    \lim_{\tilde{\tau}_0 \to 0}
    \Delta S_{A}^{(2)} \left( \tilde{t} = 0;\tilde{x}_0, \tilde{\tau}_0, N, k\right)
    =
    \begin{cases}
        \, \log\left[ N(N+1)/2\right]\ 
        &\ \text{for}\quad \tilde{x}_0 = 1/2\\
        \, \log N\ 
        &\ \text{for}\quad \tilde{x}_0 = 0
    \end{cases}
    \end{aligned} \label{eq:EA_2_initial_tau0->0_N}
\end{align}
which implies that the following inequality holds for arbitrary $k>0$:
\begin{align}
    \Delta S_{A}^{(2)} \left( \tilde{t} = 0;\tilde{x}_0=1/2,0, \tilde{\tau}_0 \to 0, N, k \right)
    &<
    \Delta S_{A}^{(2)} \left( \tilde{t} = 0;\tilde{x}_0=1/2,0, \tilde{\tau}_0 \to 0, N', k\right)
    , \
    N < N'\ .
    \label{eq:EA_2_t=0_tau0->0}
\end{align}
\begin{figure}
\centering
  \begin{subfigure}[t]{0.45\textwidth}
    \centering
    \includegraphics[width=\linewidth]{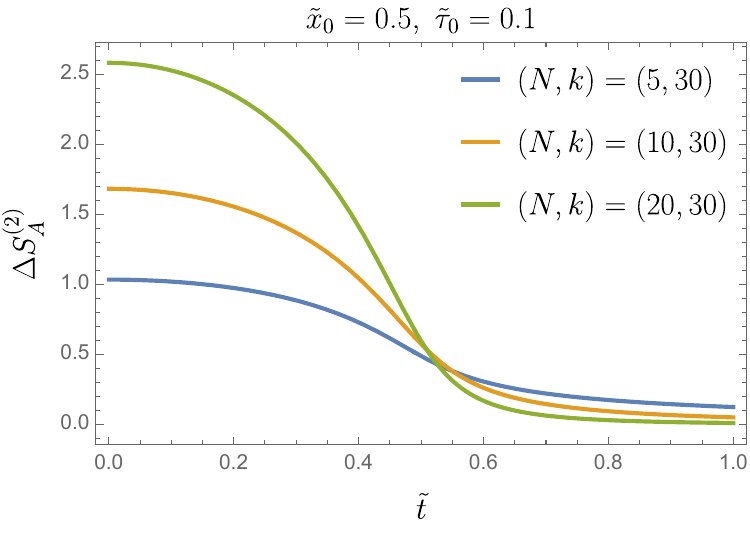}
    \caption{ }
  \end{subfigure}
  \hspace{2mm}
  \begin{subfigure}[t]{0.45\textwidth}
    \centering
    \includegraphics[width=\linewidth]{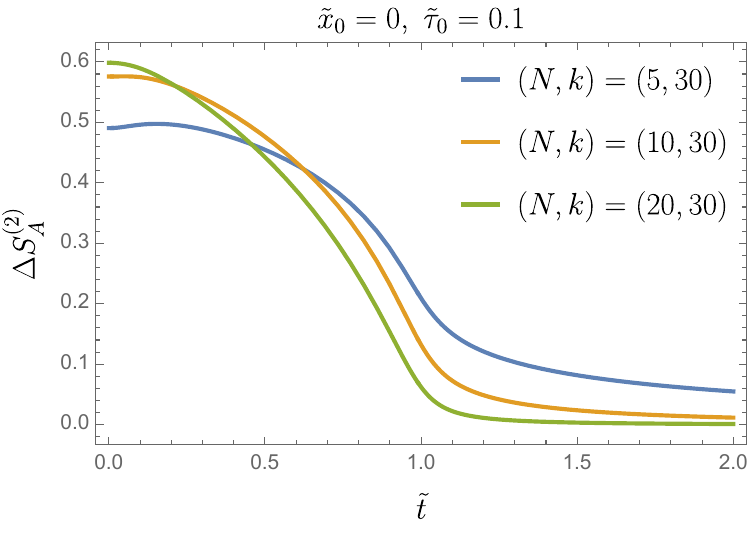}
    \caption{ }
  \end{subfigure}
  \begin{subfigure}[t]{0.45\textwidth}
    \centering
    \includegraphics[width=\linewidth]{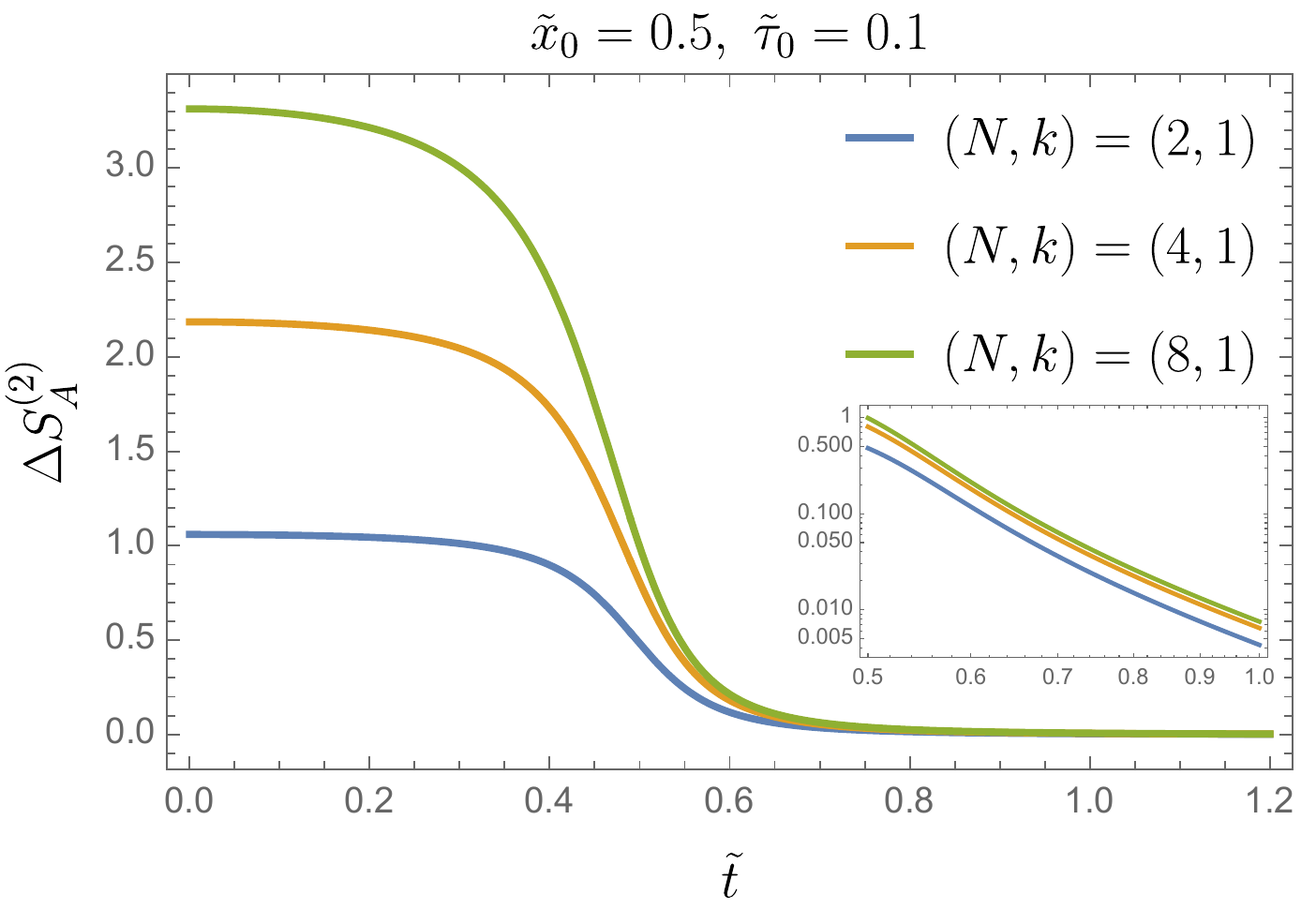}
    \caption{ }
  \end{subfigure}
  \hspace{2mm}
  \begin{subfigure}[t]{0.45\textwidth}
    \centering
    \includegraphics[width=\linewidth]{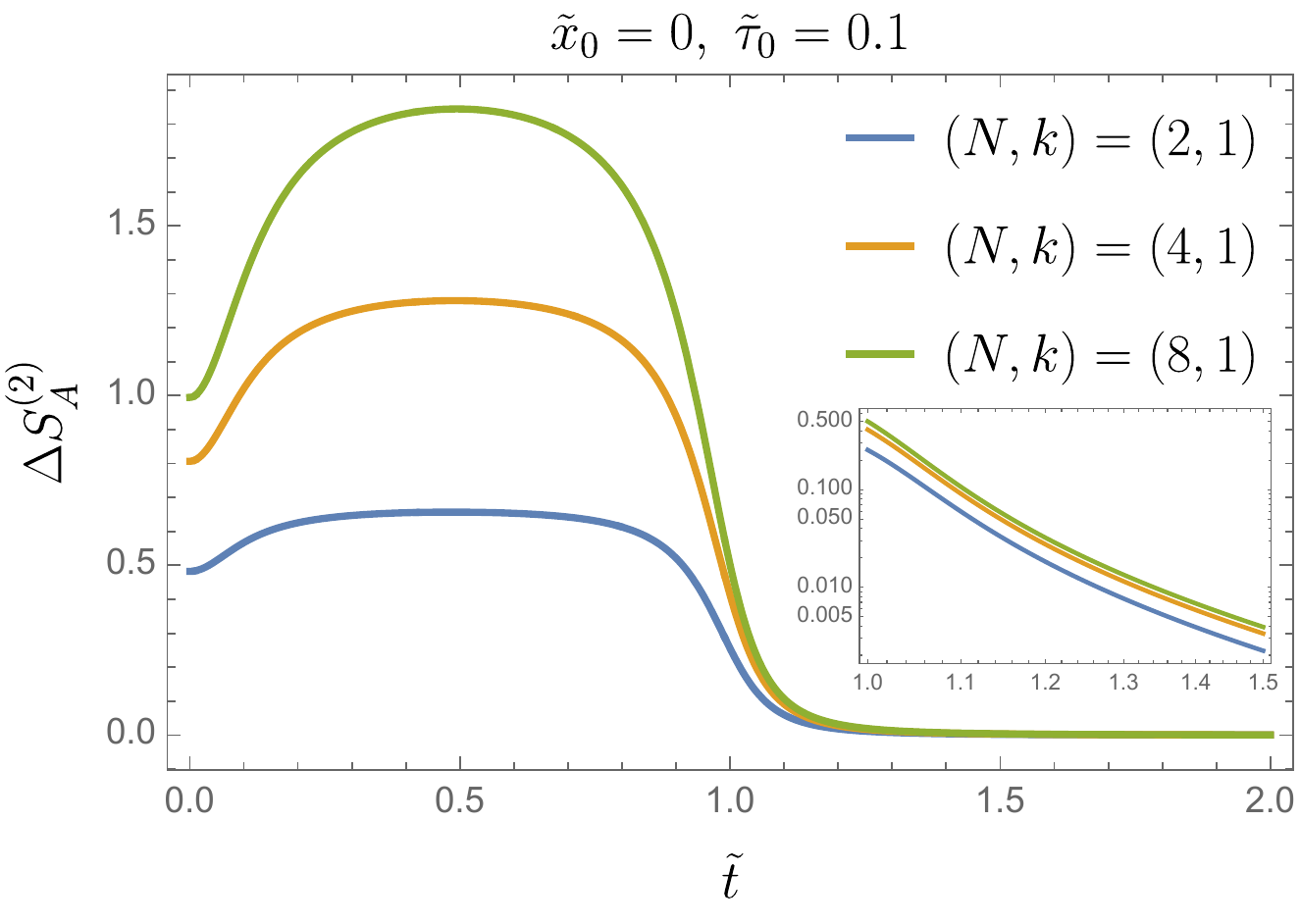}
    \caption{ }
  \end{subfigure}
  \caption{Time evolution of $\text{EA}_2$ for $\tilde{x}_0 = 1/2$ in panels (a), (c), and for $\tilde{x}_0 = 0$ in the panels (b) and (d). Panels (a) and (b) correspond to $N<k$, while panels (c) and (d) correspond to $k<N$. In the panels (c) and (d), we also show the logarithmic plot at intermediate time. For $N<k$, the curves exhibit crossings, indicating the presence of a new type of quantum Mpemba effect. In contrast, for $k<N$, we cannot find any crossings.}
\label{fig:QME_N}
\end{figure}
Comparing the inequalities \eqref{eq:EA_2_t=infty_N} and \eqref{eq:EA_2_t=0_tau0->0}, we expect that the initial value of $\text{EA}_2$ for small $\tilde{\tau}_0$ increases with $N$, while as enough time passes, it decreases with $N$ for $N<k$. This behavior signals a new type of quantum Mpemba effect: increasing $N$ amplifies the degree of initial symmetry breaking while accelerates symmetry restoration. 
In Fig.\,\ref{fig:QME_N}, we show the time evolution of $\text{EA}_2$ for various $N$ with $k$ fixed, in the cases $\tilde{x}_0 = 1/2$ and $\tilde{x}_0 = 0$ with $\tilde{\tau}_0 = 0.1$. As seen in the figures, the curves for $N< k$ exhibit crossings, indicating the new type of quantum Mpemba effect, whereas no crossings appear for $N> k$. In the plots, we present $(N,k) = (5,30),(10,30), (20,30), (2,1), (4,1), (8,1)$, but qualitatively similar behavior is observed for other values of $(N,k)$.

\subsection*{$k$ dependence with $N$ fixed}
\begin{figure}
    \centering
\begin{minipage}[b]{0.49\columnwidth}
    \centering
    \includegraphics[width=0.9\columnwidth]{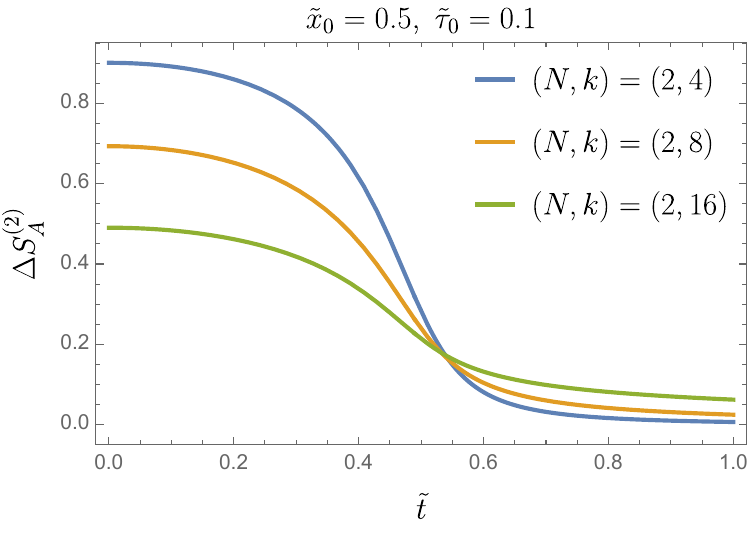}
\end{minipage}
\begin{minipage}[b]{0.49\columnwidth}
    \centering
    \includegraphics[width=0.9\columnwidth]{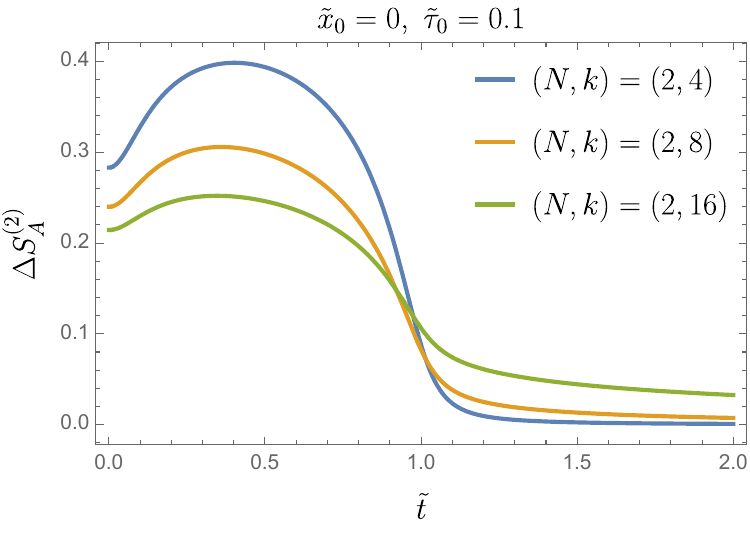}
\end{minipage}
    \caption{Time evolution of $\text{EA}_2$ for $\tilde{x}_0 = 1/2$ (left panel) and $\tilde{x}_0 = 0$ (right panel) with $N<k$ and $\tilde{\tau}_0 = 0.1$. In both cases, the curves exhibit crossings, indicating the presence of a new type of quantum Mpemba effect.}
    \label{fig:QME_k}
\end{figure}

Let us now examine the $k$ dependence of $\text{EA}_2$ with $N$ fixed. From the long time behavior of $\text{EA}_2$ in \eqref{eq:EA_2_asymptotic_t->infty}, we obtain the following inequality between different values of $k$:
\begin{align}
\begin{aligned}
    \Delta S_{A}^{(2)} \left( \tilde{t} \to \infty;\tilde{x}_0, \tilde{\tau}_0, N, k \right)
    &<
    \Delta S_{A}^{(2)} \left( \tilde{t}\to \infty;\tilde{x}_0, \tilde{\tau}_0, N, k'\right)
    \ \ ,\ N < k < k' \ . 
\end{aligned}
\label{eq:EA_2_t=infty_k}
\end{align}
On the other hand, around the initial time $\tilde{t}=0$, the $k$ dependence in the $\text{EA}_2$ appears through the next leading term:
\begin{itemize}
    \item $\tilde{x}_0 = 1/2$ and $N<k$.
        \begin{align}
            &\Delta S_{A}^{(2)} \left( \tilde{t} = 0;\tilde{x}_0, \tilde{\tau}_0, N, k\right)\nonumber\\
            &\quad =
            \log \left[\frac{N(N+1)}{2} \right]
            -\frac{(N-1)(N^2+2N-2)}{2N} c_{N,k}\  \left(4 \tilde{\tau}_0^2 \right)^{\frac{2N}{N+k}}
            +
            \mathcal{O}\left( \tilde{\tau}_0^2 \right)
            \ ,
            \label{eq:EA_2_initial_tau0->0_k_x0=-1/2}
        \end{align}
    where the coefficient $c_{N, k}$ is given in \eqref{eq:c_(N,k)}.
    \item $\tilde{x}_0 = 0$ and $N<k$.
    \begin{align}
        \Delta S_{A}^{(2)} \left( \tilde{t} = \tilde{\delta}; \tilde{x}_0,\tilde{\tau}_0, N, k\right)
        =
        \log N
        - d_{N,k}\  \left( \frac{\tilde{\tau}_0}{2\tilde{\delta}} \right)^{\frac{N}{N+k}}
        + \mathcal{O} \left(\frac{\tilde{\tau}_0}{2\tilde{\delta}} \right)
        \ ,
        \label{eq:EA_2_initial_tau0->0_k_x0=0}
    \end{align}
    where $\tilde{\delta}$ is small parameter such that $\tilde{\tau}_{0} \ll \tilde{\delta}$, and  the coefficient $d_{N,k}$ is defined as
    \begin{align}
        d_{N,k}
        =
        \frac{N(N-1)}{N+1}
        \frac{
        \Gamma\left( \frac{1}{N+k} \right)
        \Gamma\left(1-\frac{1}{N+k} \right)
        \Gamma\left( \frac{k}{N+k} \right)
        }{
        \Gamma\left( \frac{k-1}{N+k} \right)
        \Gamma\left(\frac{k+1}{N+k} \right)
        \Gamma\left( \frac{N}{N+k} \right)
        }
        \ .
    \end{align}
    Note that $d_{N,k}$ is positive for $k \ge 2$ and vanishes for $k=1$. A detailed derivation of these expressions is found in Appendix \ref{sec:Asymptotic_derivation}. 
\end{itemize}
In both cases, we find that $\text{EA}_2$ decreases as $k$ increases for $N < k < k'$:
\begin{align}
    \Delta S_{A}^{(2)} \left( \tilde{t} = +0;\tilde{x}_0=1/2,0, \tilde{\tau}_0 \to 0, N, k \right)
    &>
    \Delta S_{A}^{(2)} \left( \tilde{t} = +0;\tilde{x}_0=1/2,0, \tilde{\tau}_0 \to 0, N, k'\right)
    \ .
    \label{eq:EA_2_t=0_tau0->0_k}
\end{align}
Comparing the inequalities \eqref{eq:EA_2_t=infty_k} and \eqref{eq:EA_2_t=0_tau0->0_k} we expect that increasing $k$ diminishes the degree of initial symmetry breaking while decelerates symmetry restoration, which is another new type of quantum Mpemba effect. In Fig.\,\ref{fig:QME_k}, we present the time evolution of $\text{EA}_2$ for various $k$ with $N$ fixed, for $\tilde{x}_0 = 1/2$ and $\tilde{x}_0 = 0$ with $\tilde{\tau}_0 = 0.1$. As seen in the figures, the curves for $N< k$ exhibit crossings, indicating new type of quantum Mpemba effect. Although we show the plots for $(N,k) = (2,4),(2,8), (2,16)$, qualitatively similar behavior is observed for other $(N,k)$. 

\subsection*{Special case: $k=N$}
\begin{figure}
    \centering
\begin{minipage}[b]{0.49\columnwidth}
    \centering
    \includegraphics[width=0.9\columnwidth]{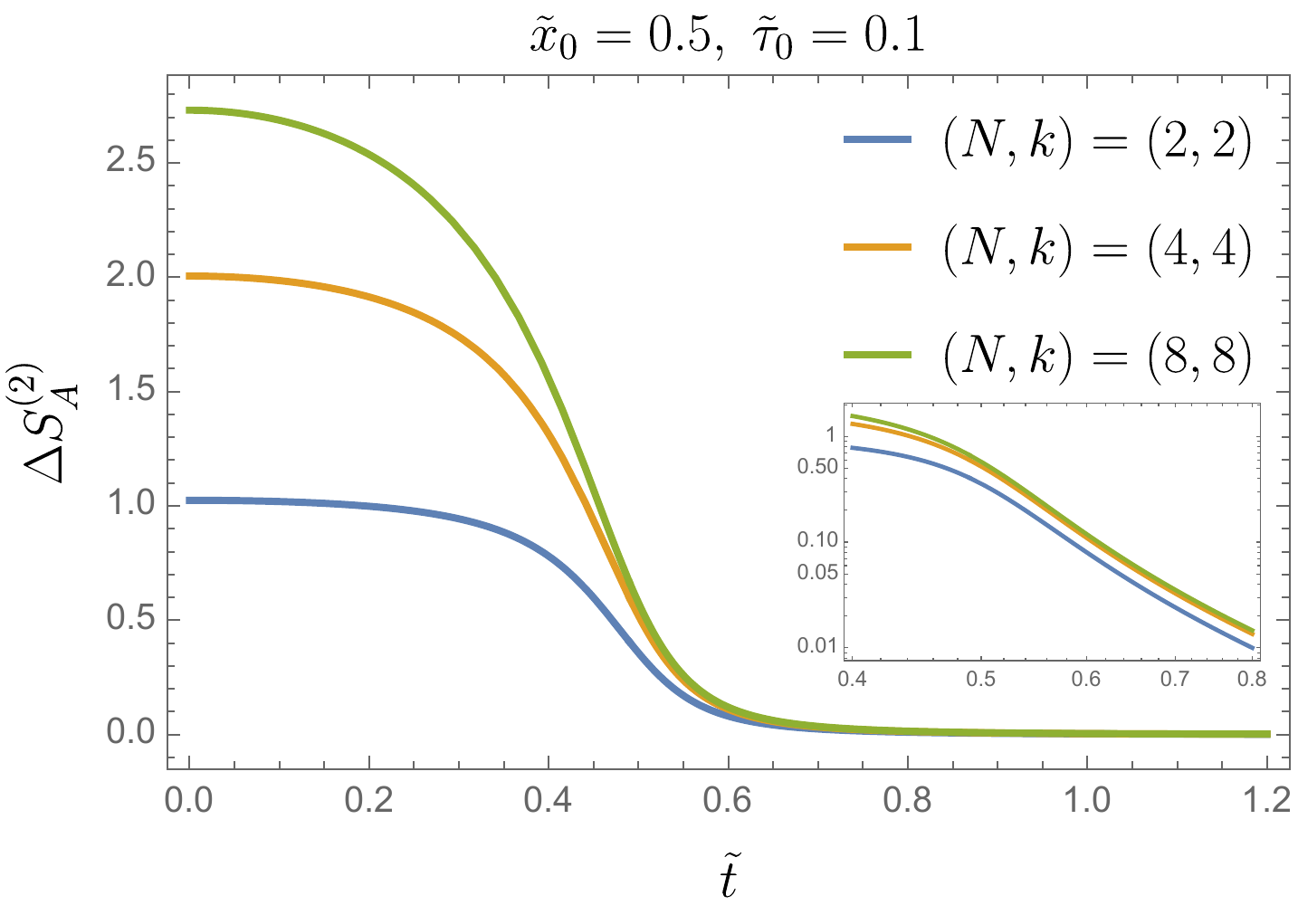}
\end{minipage}
\begin{minipage}[b]{0.49\columnwidth}
    \centering
    \includegraphics[width=0.9\columnwidth]{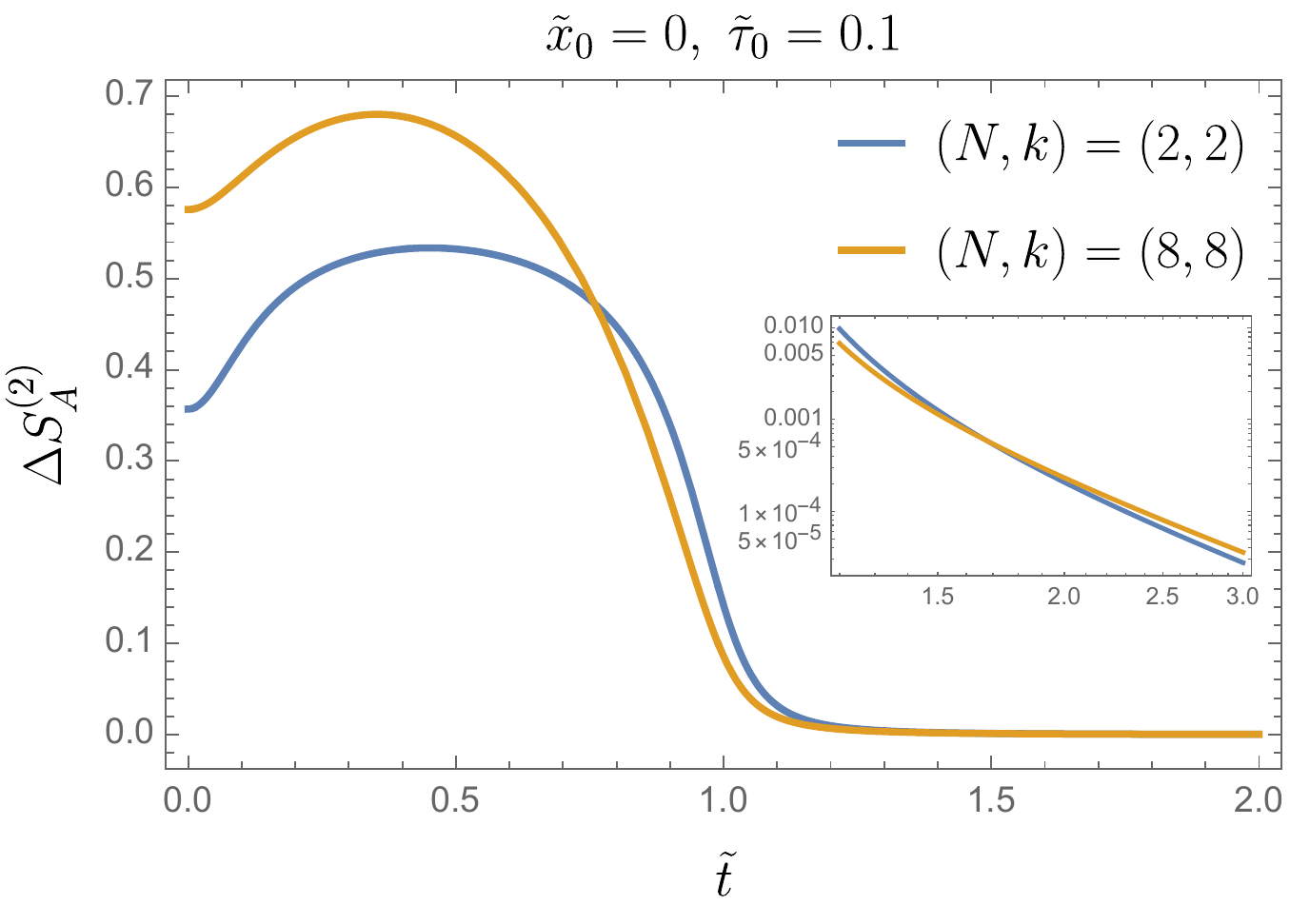}
\end{minipage}
    \caption{Time evolution of $\text{EA}_2$ for $\tilde{x}_0 = 1/2$ (left panel) and $\tilde{x}_0 = 0$ (right panel) with $N=k$ and $\tilde{\tau}_0 = 0.1$. In both panels, we also show the logarithmic plot of the long time behavior. For $\tilde{x}_0 = 1/2$, the curves show no crossings, indicating the absence of the quantum Mpemba effect. In contrast, for $\tilde{x}_0 = 0$, the curves exhibit double crossings.}
    \label{fig:QME_N=k}
\end{figure}

So far, we have investigated the $N$ and $k$ dependencies of $\text{EA}_2$ independently. There, we found new types of quantum Mpemba effects which admit only a single crossing. Here, we consider the case of $N=k$. Interestingly, in this case, we can observe a symmetry restoration structure that allows for double crossings.\footnote{Some examples where such multiple crossings occur are given in \cite{Chalas:2024wjz}.}  In this case, the long time behavior \eqref{eq:EA_2_asymptotic_t->infty} of $\text{EA}_2$ is given by
\begin{align}
    \Delta S_{A}^{(2)} \left(\tilde{t} \to \infty;\tilde{x}_0, \tilde{\tau}_0, N, N\right)
    =
    \left( 1 - \frac{2}{N(N+1)} \right)\cdot \frac{1}{4} 
    \left( \frac{\tilde{\tau}_0 }{\tilde{t}^2} \right)^2
    + \mathcal{O}\left( \left( \frac{\tilde{\tau}_0 }{\tilde{t}^2} \right)^4 \right)
    \ ,
\end{align}
which implies
\begin{align}
\begin{aligned}
    \Delta S_{A}^{(2)} \left(\tilde{t} \to \infty;\tilde{x}_0, \tilde{\tau}_0, N, N\right)
    &<
    \Delta S_{A}^{(2)} \left(\tilde{t}\to \infty;\tilde{x}_0, \tilde{\tau}_0, N', N'\right)
    \ 
    &, \ N < N'
\end{aligned}
\label{eq:EA_2_t=infty_N=k}
\end{align}
From \eqref{eq:EA_2_initial_tau0->0_N}, we also find that the $\text{EA}_2$ increase with $N$ around $\tilde{t}=0$:
\begin{align}
    \Delta S_{A}^{(2)} \left( \tilde{t} = +0;\tilde{x}_0=1/2,0, \tilde{\tau}_0 \to 0, N, N \right)
    &<
    \Delta S_{A}^{(2)} \left( \tilde{t} = +0;\tilde{x}_0=1/2,0, \tilde{\tau}_0 \to 0, N', N'\right)
    \ .
    \label{eq:EA_2_t=0_tau0->0_N=k}
\end{align}
From the inequalities \eqref{eq:EA_2_t=infty_N=k} and \eqref{eq:EA_2_t=0_tau0->0_N=k}, we see that either no crossing or an even number crossings are allowed. In Fig.\,\ref{fig:QME_N=k}, we present the time evolution of $\text{EA}_2$ for $\tilde{x}_0 = 1/2$ and $\tilde{x}_0 = 0$ with $\tilde{\tau}_0 = 0.1$. As shown in the figure, the curves for $\tilde{x}_0 = 1/2$ do not exhibit any crossings, indicating the absence of the quantum Mpemba effect. In contract, the plot for $\tilde{x}_0 = 0$ exhibits the double crossing. We comment that similar behaviors can be found for other values of $k=N$. 

\section{Quantum Mpemba effect: adjoint case}\label{sec:current_EA}
So far, we have explored the symmetry restoration structures focusing on the initial state constructed from the primary operator in the fundamental representation. It is then natural to ask if the phenomena observed in the previous section are modified or not when the initial state is built from primaries in other representations. To answer this question, in what follows, we consider the initial state constructed from the WZW current $J^{a}(z, \bar{z})$:\footnote{We can straightforwardly generalize the discussion to the non-diagonal case. See also footnote \ref{footnote:operaotor_insertion}.}
\begin{align}
    \ket{\psi} 
    =
    \frac{1}{\sqrt{Z_0}} J^{a}\left(z_{-}, \overline{z}_{-}\right) \ket{0} \ , \quad J^{a}\left(z_{-}, \overline{z}_{-}\right)\equiv J^{a}\left(z_{-}\right)\otimes \overline{J}^{\,a}\left(\overline{z}_{-}\right)\ . 
\end{align} 
To derive $\text{EA}_{2}$ for the above initial state, we need the following four-point function of the WZW currents:
\begin{align}
    \begin{aligned}
        F^{abcd}(x)
        \equiv
        \langle\, J^{a}(x)J^{b}(0)J^{c}(1)J^{d}(\infty)\, \rangle
        \ .
    \end{aligned}
\end{align}
Using the operator product expansion between the WZW currents, one can compute this four-point function \cite{Ketov:1995yd}:
\begin{align}\label{eq:F(x)}
    F^{abcd}(x)
    =
    k^{2}
    \left[
         \frac{\delta^{ab}\delta^{cd}}{x^2}
        +\frac{\delta^{ac}\delta^{bd}}{(1-x)^2}
        +\delta^{ad}\delta^{bc}
    \right]
    +
    \frac{k}{3}
    \left[
         \frac{C_{1}^{abcd}}{x(1-x)}
        +\frac{C_{2}^{abcd}}{x}
        +\frac{C_{3}^{abcd}}{1-x}
    \right]\ . 
\end{align}
where $C_{i}^{abcd}$ is defined in terms of the structure constant $f_{abc}$ of $\text{SU}(N)$
\begin{align}
    \begin{aligned}
    C_{1}^{abcd}\equiv f^{abe}f^{cde}+f^{ace}f^{bde}\ , \\
    C_{2}^{abcd}\equiv f^{abe}f^{cde}+f^{ade}f^{cbe}\ , \\ 
    C_{3}^{abcd}\equiv f^{ace}f^{bde}+f^{ade}f^{bce}\ .
    \end{aligned}
\end{align}
Moreover, the WZW current $J^{a}$ transforms under $g \in \text{SU}(N)$ as 
\begin{align}\label{eq:D}
    J'^{a}(z)=\sum_{b}[D(g)]^{ba}J^{b}\ , \quad D(g)^{ab}=2 \text{Tr}(g^{\dagger}t^{a}gt^{b})\ , 
\end{align}
where $t^{a}$ is the generator of $\text{SU}(N)$, and the trace is normalized to $\text{Tr}(t^{a}t^{b})=\frac{1}{2}\delta^{ab}$.  Repeating a similar analysis to that in Section \ref{sec:EA_in_CFT}, we obtain the following expression for the $\text{EA}_2$:
\begin{align}
    \Delta S_{A}^{(2)}(t)
    =
    -
    \log 
    \left[
    \sum_{b, b' , c, c'}\int_{\text{SU}(N)} \text{d} g
    D(g)^{ba}D(g)^{ca}D(g)^{b' a}D(g)^{c'a}
    \frac{
    F^{abac}(t) \overline{F}^{ab'ac'}\left(t\right)
    }{
    F^{aaaa}(t)\overline{F}^{aaaa} \left(t\right)}
    \right]\ ,
\end{align}
where $F^{abcd}(t)\equiv F^{abcd}(x(t))$, $\overline{F}^{abcd}(t)\equiv F^{abcd}(\overline{x}(t))$ and the time-dependence of cross-ratios are given in \eqref{eq:cross-ratio} and \eqref{eq:cross-ratio_bar}.
Employing the Haar integral formula\cite{Collins_2009, Weingarten:1977ya}:\footnote{The easiest way to understand this integral formula is to use the fact that the matrix $D(g)$ furnishes the fundamental representation of $\text{SO}(N^{2}-1)$.}
\begin{align}
    \int_{\text{SU}(N)} \text{d} g D(g)^{ba}D(g)^{ca}D(g)^{b' a}D(g)^{c'a}
    =
    \frac{1}{N^4 -1}(\delta_{bc}\delta_{b'c'}+\delta_{bb'}\delta_{cc'}+\delta_{bc'}\delta_{cb'})\ , 
\end{align}
we obtain the following expression.
\begin{align}\label{eq:current_EA_F}
    \Delta S_{A}^{(2)}(t)
    =
    -
    \log 
    \left[
        \frac{
            \left( \sum_b F^{abab}(t) \right) 
            \left( \sum_b \overline{F}^{abab}\left(t\right) \right)
            +
            2\sum_{b,c} F^{abac}(t) \overline{F}^{abac}\left(t\right)
        }{
            (N^4-1) F^{aaaa}(t)\overline{F}^{aaaa}\left(t\right)
        }
    \right]\ .
\end{align}
One can readily check that the $\text{EA}_2$ exactly vanishes in the limit $t\to \infty$:
\begin{align}
    \begin{aligned}
\lim_{t\to\infty}\Delta S_{A}^{(2)}(t)=0\ ,
    \end{aligned}
\end{align}
which means that, at the level of subsystem $A$, the global symmetry $\text{SU}(N)$ is preserved at late times. For simplicity, in the remainder of this section, we restrict ourselves to the large $k$ limit, where the second term in the RHS of \eqref{eq:F(x)} can be dropped. In this limit, the $\text{EA}_{2}$ does not depend on $k$, and has the following simplified form:
\begin{align}\label{eq:current_EA_exact}
    \Delta S_{A}^{(2)}\left( \tilde{t} ;\tilde{x}_0, \tilde{\tau}_0 , N\right)
    =
    -\log 
    \left[
    \frac{(N^4-1)f_1\left(x, \overline{x} \right)
    +(N^2+1) f_2\left(x, \overline{x} \right)
    +3 f_3\left(x, \overline{x} \right)
    }{
    (N^4-1)(1-x+x^2)^2 \left(1-\overline{x} + \overline{x}^2 \right)
    }
    \right]\ ,
\end{align}
where $x=x(t)$, $\overline{x}=\overline{x}(t)$, and $f_i \left(x, \overline{x} \right)$ are defined as
\begin{align}
\begin{aligned}
    f_1 \left(x, \overline{x} \right)
    &=
    x^2 \overline{x}^2\ ,\\
    f_2 \left(x, \overline{x} \right)
    &=
    x^2  \left(1-\overline{x} \right)^2 \left(1+\overline{x}^2 \right)
    +
    \overline{x}^2\left(1-x \right)^2 (1 + x^2)\ ,\\
    f_3 \left(x, \overline{x} \right)
    &=
    (1-x)^2 \left(1-\overline{x} \right)^2
    \left(1 + x^2 + \overline{x}^2 + x^2 \overline{x}^2 \right)\ .
\end{aligned}
\end{align}
Also, the time dependence of the cross-ratios are explicitly given in \eqref{eq:cross-ratio} and \eqref{eq:cross-ratio_bar}. As in the previous section, we summarize the asymptotic behaviors of $\text{EA}_2$ in the long time limit and large interval limit:
\begin{itemize}
    \item \textbf{Long time limit: $ \tilde{t}  \to \infty $\ .}\\
    \begin{align}\label{eq:current_EA_t->infty}
        \Delta S_{A}^{(2)}\left( \tilde{t} ;\tilde{x}_0, \tilde{\tau}_0 , N\right)
        &=
        \frac{1}{4}\left( 1 - \frac{1}{N^2-1}\right) 
        \left( \frac{\tilde{\tau}_0 }{\tilde{t}^2} \right)^4
        + \mathcal{O}\left(\left( \frac{\tilde{\tau}_0 }{\tilde{t}^2} \right)^6 \right)\ .
    \end{align}
    \item \textbf{Large interval limit: $ \tilde{\tau}_0  \to 0 $\ .}
    \begin{itemize}
        \item[(i)] $\tilde{x}_0 = 1/2\ .$
        \begin{align}\label{eq:current_EA_tau0->0_x0=-1/2}
            \Delta S_{A}^{(2)}\left( \tilde{t} = 0 ;\tilde{x}_0, \tilde{\tau}_0 , N\right)
            &=
            \log \left[\frac{N^4-1}{3} \right]
            -\frac{32}{3} \left( N^2 - 1\right) \tilde{\tau}_0^4
            + \mathcal{O}\left(\tilde{\tau}_0^6 \right)
            \ ,
        \end{align}
        \item[(ii)] $\tilde{x}_0 = 0\ .$
        \begin{align}\label{eq:current_EA_tau0->0_x0=0}
            \Delta S_{A}^{(2)} \left( \tilde{t} = \tilde{\delta}; \tilde{x}_0,\tilde{\tau}_0, N\right)
            &=
            \log \left[N^2 - 1\right]
            -\frac{(N^2 - 2)(N^2-1)}{N^2+1} 
            \left(\frac{\tilde{\tau}_0}{2\tilde{\delta}} \right)^4 
            + \mathcal{O}\left(\left(\frac{\tilde{\tau}_0}{2\tilde{\delta}} \right)^6 \right)
            \ .
        \end{align}
    \end{itemize}
\end{itemize}
We make a comment on the asymptotic behavior \eqref{eq:current_EA_tau0->0_x0=0}. The leading term in \eqref{eq:current_EA_tau0->0_x0=0} is equivalent to the logarithm of the dimension of the adjoint representation of $\text{SU}(N)$. Together with the fundamental case, where the leading contribution is $\log N$ as \eqref{eq:EA_2_initial_tau0->0_k_x0=0}, this strongly suggests that, in general, the leading term is given by the logarithm of the dimension of the representation associated with the operator that prepares the initial state.

\begin{figure}
      \centering
\begin{minipage}[b]{0.49\columnwidth}
    \centering
    \includegraphics[width=0.9\columnwidth]{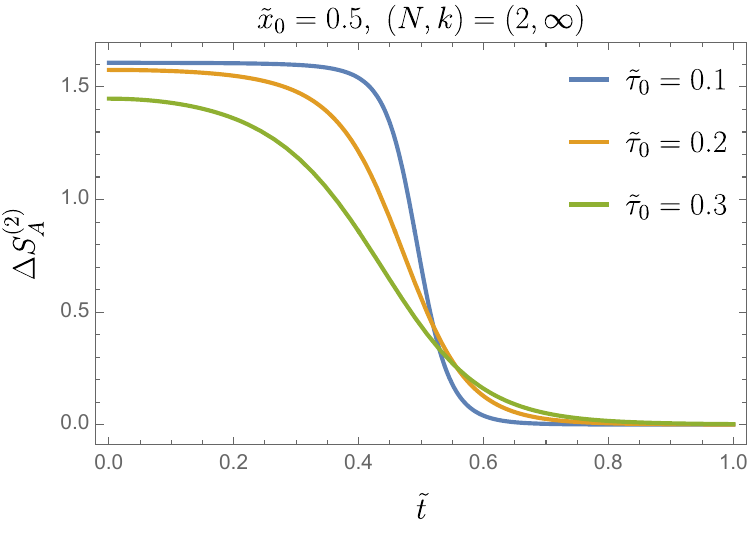}
\end{minipage}
\begin{minipage}[b]{0.49\columnwidth}
    \centering
    \includegraphics[width=0.9\columnwidth]{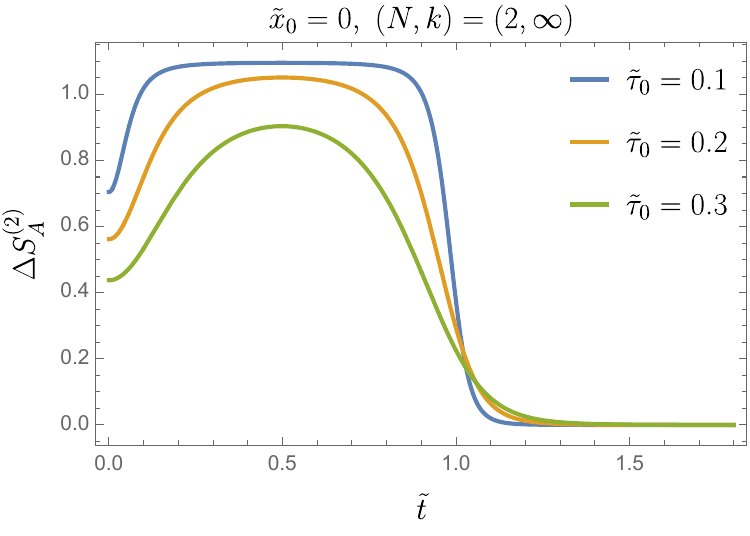}
\end{minipage}
  \caption{Time evolution of $\text{EA}_2$ constructed from currents for $\tilde{x}_0 = 1/2$ (left panel) and $\tilde{x}_0 = 0$ (right panel), with $(N,k) = (2,\infty)$. In both cases, the curves exhibit crossings, indicating the presence of quantum Mpemba effect.}
  \label{fig:QME_current_tau0}
\end{figure}

We now investigate the quantum Mpemba effect using \eqref{eq:current_EA_exact}. Firstly, fixing the rank $N$ and varying the parameter $\tilde{\tau}_0$, we obtain the following inequalities from the above asymptotic behaviors:
\begin{align}
    \left\{\ 
    \begin{aligned}
        \Delta S_{A}^{(2)} \left( \tilde{t} =+0;\tilde{x}_0=1/2,0, \tilde{\tau}_0 \to 0, N\right)
        &>
        \Delta S_{A}^{(2)} \left( \tilde{t} =+0;\tilde{x}_0=1/2,0, \tilde{\tau}_0' \to 0, N\right)\\
        \Delta S_{A}^{(2)} \left( \tilde{t} \to \infty;\tilde{x}_0, \tilde{\tau}_0, N\right)
        &<
        \Delta S_{A}^{(2)} \left( \tilde{t} \to \infty;\tilde{x}_0, \tilde{\tau}_0', N\right)
    \end{aligned}
    \right.
    \ ,\ \tilde{\tau}_0 < \tilde{\tau}_0'\ .
\end{align}
Note that we have already seen the same inequalities as these ones in   \eqref{eq:EA_2_t=0_tau_0} and \eqref{eq:EA_2_t=infty_tau_0}.
We therefore expect the quantum Mpemba effect to occur here as well. Fig.\,\ref{fig:QME_current_tau0} presents the time evolution of the result \eqref{eq:current_EA_exact} for $N=2$ and $\tilde{x}_0 = 1/2,0$. Indeed, the curves clearly exhibit crossings, providing evidence of the quantum Mpemba effect. Similar quantitative behavior is observed for other values of $N$.

\begin{figure}
    \centering
\begin{minipage}[b]{0.49\columnwidth}
    \centering
    \includegraphics[width=0.9\columnwidth]{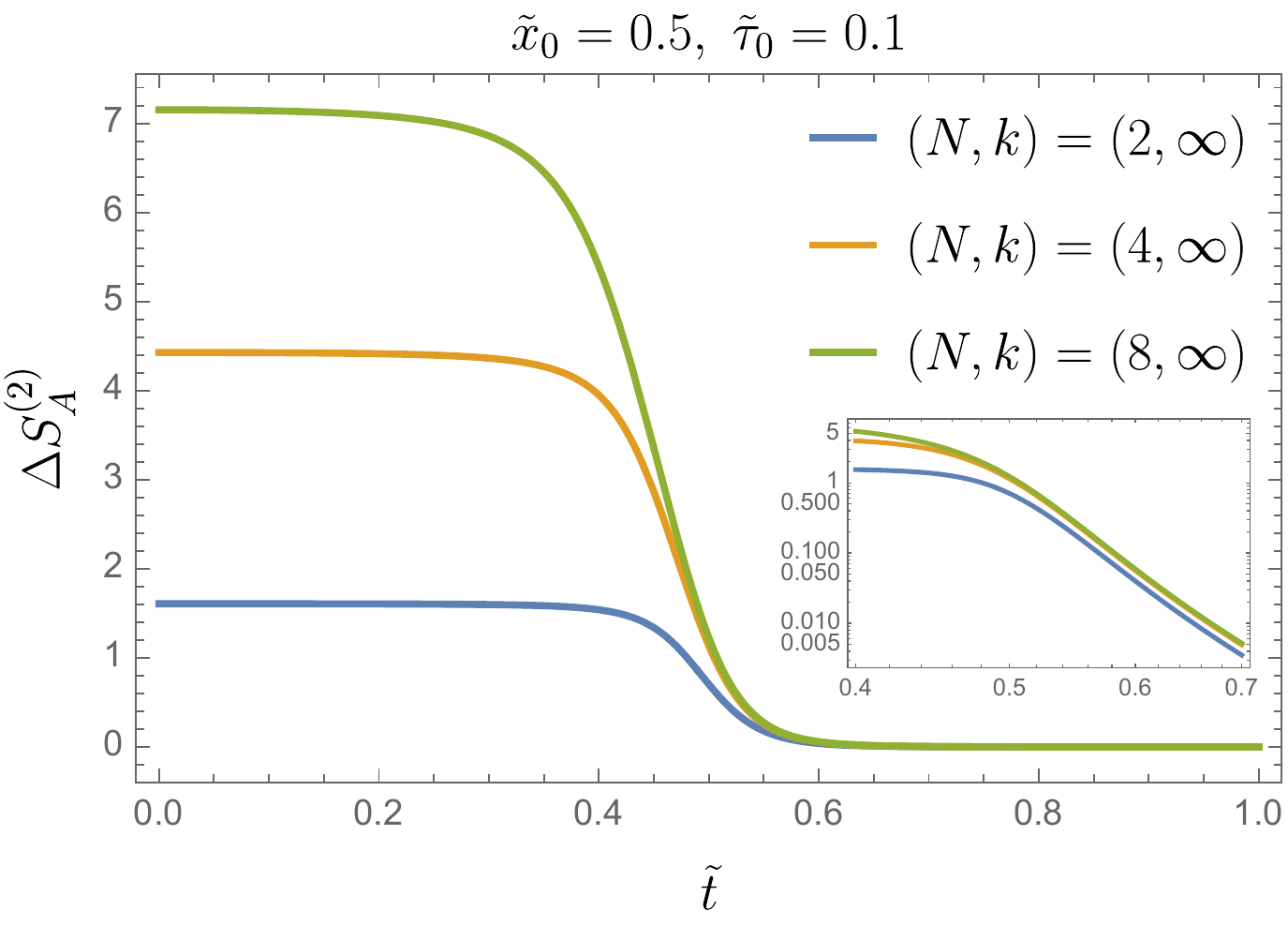}
\end{minipage}
\begin{minipage}[b]{0.49\columnwidth}
    \centering
    \includegraphics[width=0.9\columnwidth]{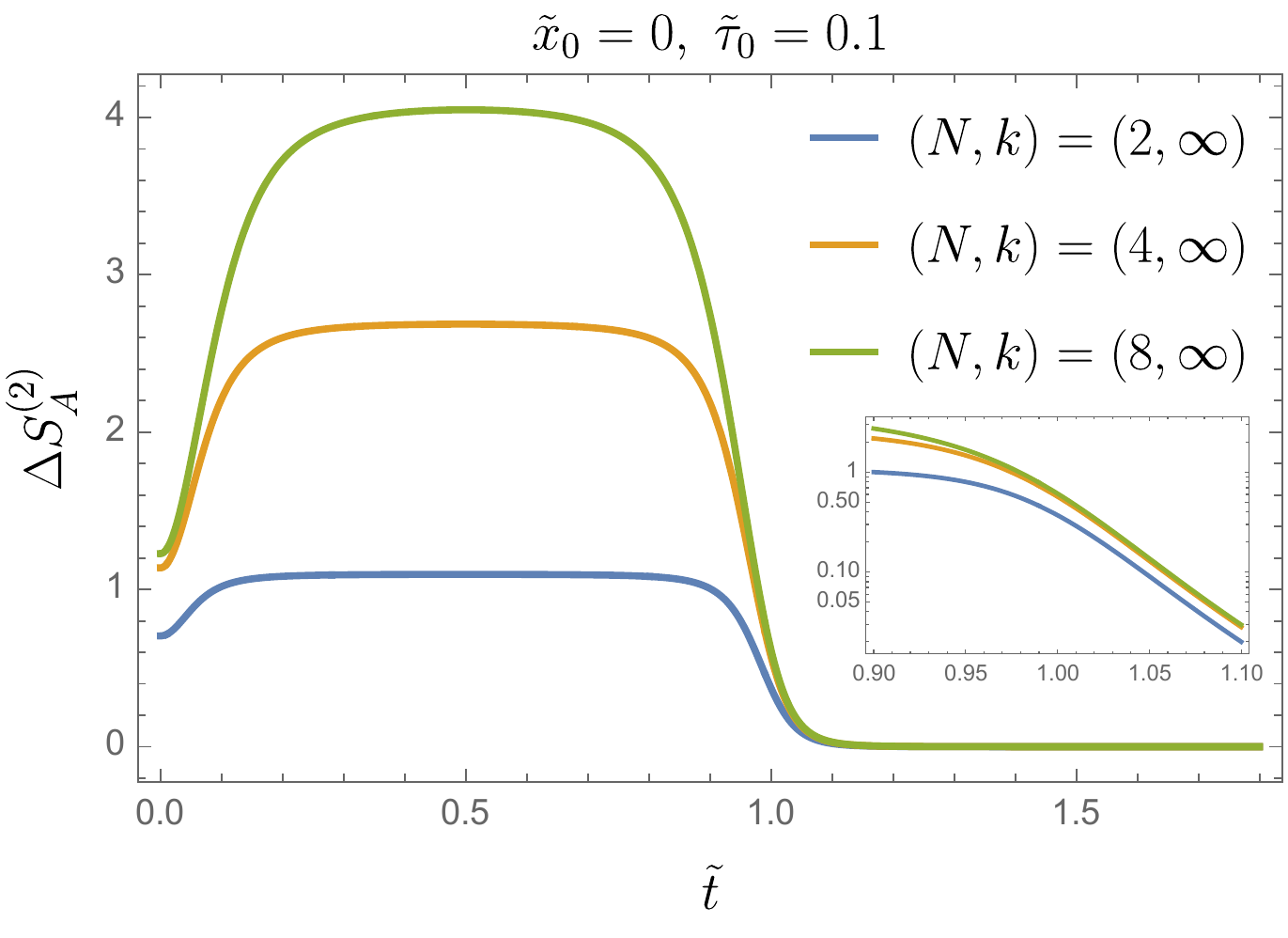}
\end{minipage}
    \caption{Time evolution of $\text{EA}_2$ constructed from currents for $\tilde{x}_0 = 1/2$ (left panel) and $\tilde{x}_0 = 0$ (right panel) with $\tilde{\tau}_0 = 0.1$. In both cases, the curves exhibit no crossings, indicating the absence of quantum Mpemba effect.}
    \label{fig:QME_current_N}
\end{figure}

On the other hand, we find no evidence of the new type of quantum Mpemba effect in the current case. Indeed, the asymptotic behaviors \eqref{eq:current_EA_t->infty}, \eqref{eq:current_EA_tau0->0_x0=-1/2} and \eqref{eq:EA_2_initial_tau0->0_k_x0=0} imply that for $\tilde{x}_{0}=1/2, 0$, 
\begin{align}
    \left\{\ 
    \begin{aligned}
        \Delta S_{A}^{(2)} \left( \tilde{t} =+0;\tilde{x}_0, \tilde{\tau}_0 \to 0, N\right)
        &<
        \Delta S_{A}^{(2)} \left( \tilde{t} =+0;\tilde{x}_0, \tilde{\tau}_0 \to 0, N'\right)\\
        \Delta S_{A}^{(2)} \left( \tilde{t} \to \infty;\tilde{x}_0, \tilde{\tau}_0, N\right)
        &<
        \Delta S_{A}^{(2)} \left( \tilde{t} \to \infty;\tilde{x}_0, \tilde{\tau}_0, N'\right)
    \end{aligned}
    \right.
    \ ,\ N < N'\ .
\end{align}
which forbids an odd number of crossings in the time evolution of $\text{EA}_2$. Fig.\,\ref{fig:QME_current_N} shows the time evolution of $\text{EA}_2$ for various $N$ with $\tilde{x}_0 = 1/2, 0$ and $\tilde{\tau}_0 = 0.1$. As evident from the figures, no crossings occur, indicating the absence of the quantum Mpemba effect. Similar behavior is observed for other values of $N$.

In summary, we analyzed the $\text{EA}_2$ constructed from currents in the adjoint representation. While the standard quantum Mpemba effect is observed, we are not able to find the new type of quantum Mpemba effect in this case.

\section{Conclusion and discussion}\label{sec:conclusion_and_discussion}
In this paper, we presented the first systematic study of symmetry restoration of non-Abelian symmetries by using the $\widehat{su}(N)_k$ WZW model. 
Due to the Coleman-Mermin-Wagner theorem, spontaneous breaking of continuous global symmetries is forbidden in $1+1$ dimensions. We therefore prepared excited initial states constructed from the primary operator in fundamental representation and the WZW current, both of which explicitly break the global symmetry $\text{SU}(N)$. In Section \ref{sec:EA_in_CFT}, we derived the general form of $\text{EA}_2$ for arbitrary initial states, which is given in \eqref{eq:REA_charged_moment_n=2}. In the derivation, we used the conformal symmetry to express $\text{EA}_2$ in terms of four-point function. In Section \ref{sec:EA_in_WZW}, we applied the general formula for $\text{EA}_{2}$ to the case where the initial state is created by the primary operator in the fundamental representation, and investigated its time evolution. We confirmed that the quantum Mpemba effect, which is the phenomenon that the state with stronger initially symmetry breaking relaxes more rapidly, also occurs for the global symmetry $\text{SU}(N)$. 
Furthermore, we uncovered a new type of quantum Mpemba effect: increasing the rank $N$ amplifies the degree of initial symmetry breaking, while accelerating the symmetry restoration. Conversely, increasing the level $k$ weakens the degree of initial symmetry breaking while decelerating the symmetry restoration. In Section~\ref{sec:current_EA}, as another representation, we also studied how the initially broken global symmetry is restored by employing the initial state built from the WZW current. We take the large $k$ limit, and see that this limit simplifies the expression of $\text{EA}_{2}$. In this case, we find that the quantum Mpemba effect always occurs at fixed rank $N$, but no evidence for the new type of quantum Mpemba effect was observed. 

This work opens several interesting directions for future studies, and some of them are listed as follows. 
\begin{itemize}
    \item 
    Firstly, it is interesting to extend our analysis to the other representations and study whether a new type of quantum Mpemba effects occur or not. This leads to more well-understanding of symmetry restoration for non-Abelian global symmetries. Although our analysis focused on $\widehat{su}(N)_k$ symmetry, it can be straightforwardly extended to other affine Lie algebras. 
    For instance, the explicit forms of four-point correlations of two-fundamental and two anti-fundamental primaries are known for $\widehat{so}(N)_k$ and $\widehat{sp}(N)_k$ symmetries \cite{Fuchs:1986ew,Fuchs:1987rf}, and thus our approach can be directly applied to these cases.  

    \item 
    In this work, we focus on the second R\'enyi entanglement asymmetry ($n=2$) for simplicity. A natural future direction is to explore the entanglement asymmetry as the relative entropy ($n\to 1$). As discussed in Section 2, the $n$-th R\'enyi entanglement asymmetry is expressed in terms of $2n$-point correlation function. As a result, it is difficult to obtain an analytical expression for general $n$ and to subsequently take the limit $n\to 1$.
    On the other hand, in \cite{Benini:2024xjv}, the authors analyzed the entanglement asymmetry ($n=1$)  for a weakly symmetry-broken initial state, where one can control the degrees of symmetry breaking in perturbative way. In particular, the authors computed the leading perturbative contribution to the entanglement asymmetry in terms of two-point function. This perturbative analysis can in principle be applied to our setup as well. Since the analytical expression for the four-point function in WZW model is known, it is also possible to derive the subleading contributions beyond the leading perturbative order. We leave this point to a future work.
    
    
    \item 
    Another interesting direction is to explore finite temperature systems. 
    It is known that the R\'enyi entanglement entropy at low temperature can be expressed in terms of correlation functions \cite{Cardy:2014jwa, Ghasemi:2022jxg, Chen:2014unl,Guo:2015uwa}. 
    This analysis can be extended to the R\'enyi entanglement asymmetry, making it intriguing to investigate how thermal effects contribute to the quantum Mpemba effect though symmetry restoration.
    
    \item 
    It would be valuable to interpret our result from the perspective of three-dimensional Chern-Simons theory, which is dual to the WZW model\cite{Witten:1988hf}. In our analysis, we saw that the new type of quantum Mpemba effects occurs only for fundamental primary, but not for adjoint primary. It is interesting to explore some implications of our results from the three-dimensional perspective.
\end{itemize}

\section*{Acknowledgments}
The work of H.F. and S.S. was supported by Grant-in-Aid for JSPS fellows, Grant Nos. 24KJ1637 and 23KJ1533, respectively.

\appendix

\section{Four-point function in Wess-Zumino-Witten models}
\label{sec:4pt_function_in_WZW}
In this appendix, we present the review on how we can determine the four-point functions in the $\widehat{su}(N)_k$ WZW model by solving the Knizhnik-Zamolodchikov (KZ) equation. We follow the notation used in \cite{DiFrancesco:1997nk}. 

\subsection{Preliminary}
\label{subsec:preliminary}
Before delving into the details, for completeness, it is convenient to recap the basics of WZW models whose affine Lie algebra is given by $\hat{\mathfrak{g}}_{k}$. Let $\Phi_{\Lambda}^{\alpha}$ is the primary field in the irreducible module of the highest weight $\Lambda$ with weight~$\alpha$, and its conformal dimension $h_{\Lambda}$ is given by
\begin{align}\label{eq:conf}
    h_{\Lambda}=\frac{(\Lambda, \Lambda+2\rho)}{2(k+g)}\ .
\end{align}
Here, $\rho$ is the Weyl vector which is defined in terms of the fundamental weights $\{w_{i}\}$ as
\begin{align}\label{eq:Weyl}
    \rho\equiv\sum_{i}w_{i}\ . 
\end{align}
Also, $g$ is the dual Coxeter number of the Lie algebra $\mathfrak{g}$. In general, all representations can be expanded in terms of the fundamental weights:
\begin{align}\label{eq:weight_expansion}
    \Lambda=\sum_{i}\lambda_{i}w_{i}\ , \quad \lambda_{i}\in \mathbb{Z}\ ,
\end{align}
where the set of the coefficients is referred to as the Dynkin label. It is convenient to introduce the quadratic form matrix $F_{ij}$ which is the inner product between fundamental weights:
    \begin{align}
        F_{ij}\equiv(w_{i} , w_{j})\ . 
    \end{align}
Then, by plugging \eqref{eq:Weyl} and \eqref{eq:weight_expansion} into \eqref{eq:conf}, we can rewrite the conformal dimension $h_{\Lambda}$ in terms of the Dynkin label and quadratic form matrix. 

\subsection{Knizhnik-Zamolodchikov equation}
Let us consider the following four-point function:\footnote{ Here, we omit the anti-holomorphic part for simplicity.}
\begin{align}\label{eq:four_pt}
    \mathcal{F}_{\alpha_{1} \alpha_{4}}^{\alpha_{2}\alpha_{3}} (z_{i})
    \coloneqq
    \left\langle\Phi_{\Lambda_{1},\alpha_{1}}(z_{1})
    \Phi_{\Lambda_{2}}^{\alpha_{2}}(z_{2})
    \Phi_{\Lambda_{3}}^{\alpha_{3}}(z_{3})
    \Phi_{\Lambda_{4},\alpha_{4}}(z_{4}) \right\rangle\ , 
\end{align}
where $\Phi_{\Lambda_{i}, \alpha_i}$ is the primary operator in the representation $\Lambda_{i}$, and $\alpha_{i}=1, 2, \cdots, \dim \Lambda_{i}$.  
By using the conformal transformation, we can always map the four holomorphic coordinates $(z_{1}, z_{2}, z_{3}, z_{4})$ to $(x, 0, 1, \infty)$ with
\begin{align}
    x
    \coloneqq
    \frac{z_{12}z_{34}}{z_{14}z_{32}}\ ,
\end{align}
where $z_{ij}\coloneqq z_{1}-z_{2}$.
Then, the above four-point function \eqref{eq:four_pt} can be rewritten as
\begin{align}
    \mathcal{F}_{\alpha_{1} \alpha_{4}}^{\alpha_{2}\alpha_{3}} (z_{i})
    =
    \left(
    z_{14} z_{23}
    \right)^{-2h}
    F_{\alpha_{1} \alpha_{4}}^{\alpha_{2}\alpha_{3}} (x)\ ,
\end{align}
where $F_{\alpha_{1} \alpha_{4}}^{\alpha_{2}\alpha_{3}} (x)$ is defined by
\begin{align}\label{eq:four_cross}
    F_{\alpha_{1} \alpha_{4}}^{\alpha_{2}\alpha_{3}} (x)
    \equiv
    \left\langle\Phi_{\Lambda_{1}, \alpha_{1}}(x)
    \Phi_{\Lambda_{2}}^{\alpha_{2}}(0)
    \Phi_{\Lambda_{3}}^{\alpha_{3}}(1)
    \Phi_{\Lambda_{4}, \alpha_{4}}(\infty) \right\rangle\ .
\end{align}
In order for the four-point function not to vanish, the tensor product of the four affine integrable representations of primary fields must contain at least one identity. Suppose that this tensor product takes the following form: 
\begin{align}\label{eq:id}
\Lambda_{1}\otimes\Lambda_{2}\otimes\Lambda_{3}\otimes\Lambda_{4}\supset m\,\mathbf{1}\ , 
\end{align}
where $m$ is some positive integer. This implies that there are $m$ invariant tensors $\{I_{A}\}$, and the four-point function can be decomposed as
\begin{align}
    F_{\alpha_{1} \alpha_{4}}^{\alpha_{2}\alpha_{3}} (x)
    =
    \sum_{A=1}^{m} F_{A}(x)(I_{A})_{\alpha_{1} \alpha_{4}}^{\alpha_{2}\alpha_{3}} \ . 
\end{align}
Then, the four-point function \eqref{eq:four_pt} satisfies the following KZ equation \cite{Knizhnik:1984nr}:
\begin{align}\label{eq:KZ_eq}
    \left[
    \partial_{z_{\alpha}} 
    - \frac{1}{k+g} \sum_{\beta\not=\alpha} 
    \frac{
    \sum_{a}R_{\Lambda_{\alpha}}(t^{a})\otimes R_{\Lambda_{\beta}}(t^{a})
    }{
    z_{\alpha} - z_{\beta}}
    \right]
    (z_{14}z_{23})^{-2h} \sum_{A=1}^{m}I_{A} F_{A}(x)
    =0\ ,
\end{align}
where $R_{\Lambda_{\alpha}}(t^{a})$ is the representation matrix of the generator $t^a$ in the representation $\Lambda_{\alpha}$. 
We should remark that we omit the indices for the representation space, and the generators~$R_{\Lambda_{\alpha}}(t^{a})$ act on the indices of the invariant tensor $I_{A}$ properly. In particular, we set $\alpha=1$, and the above KZ equation turns into
\begin{align}
    \left[
    \frac{-2h}{z_{14}}
    +\left(\frac{x}{z_{12}}-\frac{x}{z_{14}}\right)\partial_{x}
    -\frac{1}{k+g} \sum_{\beta\not=1}
    \frac{\sum_{a}R_{\Lambda_{1}}(t^{1})\otimes R_{\Lambda_{\beta}}(t^{a})
    }{
    z_{1}-z_{\beta}}
    \right]\sum_{A=1}^{m}I_{A} F_{A}(x)
    =0\ . 
\end{align}
At this stage, we set the four space-time coordinates as 
\begin{align}
    z_{1}=x\ , \quad z_{2}=0\ , \quad z_{3}=1\ , \quad z_{4}=\infty\ ,  
\end{align}
then the above differential equation is reduced to
\begin{align}\label{eq:KZ_eq}
    \left[
    \partial_{x}
    -\frac{1}{k+g}
    \frac{\sum_{a}R_{\Lambda_{1}}(t^{a})\otimes R_{\Lambda_{2}}(t^{a})
    }{
    x}
    -\frac{1}{k+g} 
    \frac{
    \sum_{a}R_{\Lambda_{1}}(t^{a})\otimes R_{\Lambda_{3}}(t^{a})
    }{
    x-1}
    \right]\sum_{A=1}^{m}I_{A} F_{A}(x)
    =0\ .
\end{align}

\subsection{Computing the four-point function for $\widehat{su}(N)_k$}
In the following, we focus on the case $\hat{\mathfrak{g}}_{k} = \widehat{su}(N)_k$ and derive the four-point function. In \eqref{eq:four_pt}, we specify the representations of the primary operators as follows. We take $\Lambda_{1}$ and $\Lambda_{4}$ to be in the fundamental representation $\mathbf{N}$, whose affine Dynkin labels are
\begin{align}\label{eq:Dynkin_label_A}
\Lambda_{1}=\Lambda_{4}=\mathbf{N}\ : \ [\lambda_{0};\lambda_{1}, \lambda_{2}, \cdots, \lambda_{N-1}]=[k-1;1, 0, 0, \cdots , 0]\ .
\end{align}
The conformal dimension $h$ of this representation is
\begin{align}
h = \frac{N^{2}-1}{2N(k+N)}\ .
\end{align}
This expression follows directly from \eqref{eq:conf}, using the fact that the dual Coxeter number is $g=N$ and that the quadratic form matrix is given by
\begin{align}
    \begin{aligned}
        F_{1 j} = \frac{N-j}{N}\ .
    \end{aligned}
\end{align}
Similarly, we choose $\Lambda_{2}$ and $\Lambda_{3}$ to be in the anti-fundamental representation $\overline{\mathbf{N}}$, with affine Dynkin labels
\begin{align}
\Lambda_{2}=\Lambda_{3}=\overline{\mathbf{N}}\ : \ [\lambda_{0};\lambda_{1}, \lambda_{2}, \cdots, \lambda_{N-1}]=[k-1;0, 0, 0, \cdots , 1]\ .
\end{align}
Let $\Phi_{i}$ denote a primary operator in the fundamental representation ($i = 1, \dots , N$). The corresponding four-point function \eqref{eq:four_cross} is written as
\begin{align}
    F_{ik}^{j \ell}(x)
    \coloneqq
    \left\langle
    \Phi_{i}(x)\Phi^{\dagger j}(0)\Phi^{\dagger \ell}(1)\Phi_{k}(\infty)
    \right\rangle\ .
\end{align}
In our case, we have two invariant tensors\footnote{Precisely speaking, when $k=1$, we have only one invariant tensor. This can be verified by using the representation theory of the affine Lie algebra. We will come back to this issue later again.} (i.e., $m=2$ in \eqref{eq:id}) and the invariant tensors are explicitly given by
\begin{align}
    (I_{1})_{ik}^{j \ell}=\delta_{i}^{j}\delta_{k}^{\ell}\ , \quad (I_{2})_{ik}^{j \ell}=\delta_{i}^{\ell}\delta_{k}^{j}\ . 
    \label{eq:inv_tensor_su(N)}
\end{align}
Then, the KZ equation \eqref{eq:KZ_eq} reads
\begin{align}
        \sum_{A=1}^{2}
    \left[
    (I_{A})_{ik}^{j\ell} \partial_{x}
    +\frac{1}{k+g}
    \frac{T_{i p}^{q j} (I_{A})_{q k}^{p\ell}   
    }{x}
    +\frac{1}{k+g}
    \frac{T_{i p}^{q\ell} (I_{A})_{q k}^{jp}
    }{x-1}
    \right] F_{A}(x)
    =0\ ,
\end{align}
where we restored the full indices for representations, and the tensor $T_{ik}^{j\ell}$ is given by
\begin{align}
    T_{ik}^{j\ell}=\delta_{i}^{\ell}\delta_{k}^{j}-\frac{1}{N}\delta_{i}^{j}\delta_{k}^{\ell} \ .
\end{align}
Then, the KZ equations are equivalent to the following two differential equations:
\begin{align}
    \partial_{x}F_{1}+\frac{1}{\kappa}\left(\frac{N^2 -1}{N}\frac{F_1}{x}+\frac{F_2}{x}-\frac{1}{N}\frac{F_1}{x-1}\right)&=0\ , \\
    \partial_{x}F_{2}+\frac{1}{\kappa}\left(\frac{N^2 -1}{N}\frac{F_2}{x-1}+\frac{F_1}{x-1}-\frac{1}{N}\frac{F_2}{x}\right)&=0\ , 
\end{align}
where $\kappa\equiv k+N$. To solve these equations, we firstly express $F_{2}$ in terms of $F_1$ by using the first equation:
\begin{align}\label{eq:F2}
    F_{2}=-\kappa x \partial_{x}F_1 -\frac{N^2 -1}{N}F_1 +\frac{1}{N}\frac{x}{x-1}F_{1}\ . 
\end{align}
By substituting this into the second equation, one can obtain the second-order differential equation for $F_{1}$. By rewriting $F_{1}$ as
\begin{align}
    F_{1}=x^{r}(1-x)^{s} f_1(x)\ , 
\end{align}
this differential equation can be written as
\begin{align}
    x(1-x)\partial_{x}^{2}f_{1}+A(x)\partial_{x}f_{1}+B(x)f_{1}=0\ , 
\end{align}
where $A(x)$ and $B(x)$ are given by
\begin{align}
    \begin{aligned}
    \kappa N \cdot A(x)&\equiv (1-x)(2r\kappa N +N^{2} -2+\kappa N)-x(2s\kappa N +N^2 -2)\ , \\
    (\kappa N)^{2} \cdot B(x)&\equiv\frac{1-x}{x}\left(\kappa^2 N^2 r(r-1)+r \kappa N (N^2 -2)+r\kappa^2 N^2 -(N^2 -1)\right) \\
    &\quad +\frac{x}{1-x}\left(\kappa^2 N^2 s(s-1)+ s \kappa N (N^2 -2)+\kappa N -(N^2 -1)\right) \\
    &\quad -2rs\kappa^{2}N^{2}-(s+r)\kappa N (N^2 -1)-s\kappa^2 N^2+\kappa N -1 -(N^{2}-1)^2 +N^{2}\ . 
    \end{aligned}
\end{align}
Note that if $B(x)$ do not depend on the cross-ratio $x$, the solution $f_{1}$ can be written in terms of the hypergeometric function. This condition gives rise to non-trivial constraints on $r$ and $s$. Indeed, by requiring that the terms which are proportional $(x/(1-x))^{\pm 1}$ in the function $B(x)$, $r$ and $s$ are fixed to be
\begin{align}
    &r=r_{+}\equiv \frac{1}{\kappa N}=h_{\hat{\theta}}-2h \qquad  \text{or} \qquad r=r_{-}\equiv -\frac{N^{2}-1}{\kappa N}=-2h\ , \\
    &s=s_{+}\equiv \frac{1}{\kappa N}=h_{\hat{\theta}}-2h\qquad  \text{or} \qquad s=s_{-}\equiv 1-\frac{N^{2}-1}{\kappa N}=1-2h\ ,
\end{align}
where $h$ and $h_{\hat{\theta}}$ are the conformal dimensions of the chiral WZW primary fields in fundamental and adjoint representations.
At first glance, we have four solutions, but the true number of independent solutions are two. Following the yellow book, we fix $s=s_{+}$, and the two independent solutions are written as
\begin{align}
    \begin{aligned}
        &r=r_{-}\ : \ f_{1}^{(-)}=F\left(\frac{1}{\kappa}, -\frac{1}{\kappa};1-\frac{N}{\kappa};x\right)\ , \\
        &r=r_{+}\ : \ f_{1}^{(+)}=F\left(\frac{N-1}{\kappa}, \frac{N+1}{\kappa};1+\frac{N}{\kappa};x\right)\ .
    \end{aligned}
\end{align}
Then, the two solutions of $F_{1}$ become
\begin{align}
    &F_{1}^{(-)}(x)=x^{-2h}(1-x)^{h_{\hat{\theta}}-2h}F\left(\frac{1}{\kappa}, -\frac{1}{\kappa};1-\frac{N}{\kappa};x\right) \ , 
    \label{eq:F_1^(-)}\\
    &F_{1}^{(+)}(x)=x^{h_{\hat{\theta}}-2h}(1-x)^{h_{\hat{\theta}}-2h}F\left(\frac{N-1}{\kappa}, \frac{N+1}{\kappa};1+\frac{N}{\kappa};x\right) \ .
    \label{eq:F_1^(+)}
\end{align}
By plugging these solutions to \eqref{eq:F2}, we obtain the solutions for $F_{2}$:\footnote{Remark that there is a typo in \cite[equation (15.170)]{DiFrancesco:1997nk}.}
\begin{align}
    &F_{2}^{(-)}(x)
    =
    \frac{1}{k}x^{1-2h}(1-x)^{h_{\hat{\theta}}-2h}
    F\left(1+\frac{1}{\kappa}, 1-\frac{1}{\kappa};2-\frac{N}{\kappa};x\right) \ , 
    \label{eq:F_2^(-)}\\
    &F_{2}^{(+)}(x)
    =
    -N x^{h_{\hat{\theta}}-2h}(1-x)^{h_{\hat{\theta}}-2h}
    F\left(\frac{N-1}{\kappa}, \frac{N+1}{\kappa};\frac{N}{\kappa};x\right) \ .
    \label{eq:F_2^(+)}
\end{align}
These results imply that $F_{A}^{(-)} (A=1, 2)$ are conformal blocks for the identity channel, while $F_{A}^{(+)} (A=1, 2)$ are conformal blocks for the adjoint field channel. 

By restoring the dependence on the anti-holomorphic coordinate, the four-point function can be written as
\begin{align}
F_{ik}^{j \ell}(x, \bar{x}):= \left\langle\Phi_{i}(x, \bar{x})\Phi^{\dagger j}(0, 0)\Phi^{\dagger \ell}(1, 1)\Phi_{k}(\infty, \infty)\right\rangle=\sum_{A, B=1, 2}(I_{A})_{ik}^{j\ell}(\bar{I}_{B})_{ik}^{j\ell}G_{AB}(x, \bar{x})\ , 
\end{align}
where $G_{A, B}$ is given by
\begin{align}
        G_{A,B} \left( x, \overline{x} \right)=
\sum_{\alpha, \beta=\pm} X_{\alpha, \beta} F_A^{(\alpha)}(x) F_B^{(\beta)}\left( \bar{x} \right) \ ,
\end{align}
with some coefficient $X_{\alpha, \beta}$. Physical correlation functions must satisfy the locality condition and the crossing-symmetry, which can be used to fix coefficients $X_{\alpha, \beta}$. The locality condition is given by
\begin{align}
    G_{AB}(e^{2\pi \i}x, e^{-2\pi \i}\bar{x})=G_{AB}(x, \bar{x})\ .
\end{align}
This clearly prohibits the off-diagonal contributions:
\begin{align}\label{eq:constraint_locality}
    X_{-+}=X_{+-}=0\ . 
\end{align}
The second condition is the crossing-symmetry:
\begin{align}
    F_{ik}^{j \ell}(x, \bar{x})=F_{i k}^{\ell j }(1-x, 1-\bar{x})\ , 
\end{align}
which is equivalent to
\begin{align}\label{eq:crossing}
    G_{A,B} \left( x, \overline{x} \right)=G_{3-A, 3-B} \left( 1-x, 1-\overline{x} \right)\ . 
\end{align}
This crossing symmetry can be used to fix the remaining coefficients $X_{++}$ and $X_{--}$ as seen below. Thanks to the following identity:
\small
\begin{align}
    F(a, b; c;x)=A_{1}F(a, b;a+b-c+1;1-x)+A_{2}(1-x)^{c-a-b}F(c-a, c-b;c-a-b+1;1-x)\ , 
\end{align}
\normalsize
where $A_1$ and $A_2$ are given by
\begin{align}
    A_1 = \frac{\Gamma(c)\Gamma(c-a-b)}{\Gamma(c-a)\Gamma(c-b)}\ , \quad A_1 = \frac{\Gamma(c)\Gamma(a+b-c)}{\Gamma(a)\Gamma(b)}\ . 
\end{align}
one can write down the conformal blocks as the power series of $1-x$. This implies that we can express the conformal blocks $F_{A}^{(\alpha)}(x)$ as follows:\footnote{Recall that the crossing symmetry flips $I_{1}$ and $I_{2}$.}
\begin{align}\label{eq:F}
    F_{A}^{(\alpha)}(x)=\sum_{\beta}c_{\alpha\beta}F_{3-A}^{(\beta)}(1-x)\ . 
\end{align}
Clearly, the matrix $c_{\alpha\beta}$ must satisfy the idempotent condition:
\begin{align}\label{eq:c}
    c_{++}=-c_{--}\ , \quad c_{++}^2 + c_{+-}c_{-+}=1\ . 
\end{align}
After some computations, one can fix $c_{--}$ and $c_{+-}$ as
\begin{align}
    c_{--}=N\frac{\Gamma(N/\kappa)\Gamma(-N/\kappa)}{\Gamma(1/\kappa)\Gamma(-1/\kappa)}\ , \quad c_{+-}=-N\frac{\Gamma(N/\kappa)^2}{\Gamma((N+1)/\kappa)\Gamma((N-1)/\kappa)}\ . 
\end{align}
By substituting \eqref{eq:F} into \eqref{eq:crossing}, the crossing relation \eqref{eq:crossing} is equivalent to
\begin{align}
    \sum_{\gamma, \delta}X_{\gamma \delta}c_{\gamma\alpha}c_{\delta\beta}=X_{\alpha\beta}\ . 
\end{align}
More explicitly, these can be written as
\begin{align}
    X_{++}c_{++}c_{+-}+X_{--}c_{-+}c_{--}&=0\ , \\
    X_{--}c^{2}_{--}+X_{++}c^{2}_{+-}&=X_{--}\ , \\
    X_{--}c^{2}_{-+}+X_{++}c^{2}_{++}&=X_{++}\ , 
\end{align}
where we used \eqref{eq:constraint_locality}. By using \eqref{eq:c}, these constraints are reduced to
\begin{align}
    X_{++}=\frac{1-c_{--}^2}{c_{+-}^2}X_{--}\ . 
\end{align}
By choosing the normalization $X_{--}=1$, one can obtain the physical four-point function satisfying both the locality condition and the crossing symmetry:
\begin{align}
    G_{AB}(x, \bar{x})=F_A^{(-)}(x) F_B^{(-)} (\bar{x})+\frac{1-c_{--}^2}{c_{+-}^2}F_A^{(+)}(x) F_B^{(+)}(\bar{x})\ . 
    \label{eq:G_AB_su}
\end{align}
Note that only when $k=1$, the second term vanishes due to $c_{--}=1$. This means that the only identity block contributes to the four-point function. This is consistent with the fusion rule of WZW primaries.

\section{Asymptotic behavior of $\text{EA}_2$ in $\widehat{su}(N)_k$ WZW model}
\label{sec:Asymptotic_derivation}
In this appendix, we derive the asymptotic behavior of $\text{EA}_2$ in $\widehat{su}(N)_k$ WZW model. For later convenience, let us consider the case in which the following parameters $\epsilon_{\pm}$ are infinitesimally small:
\begin{align}
    \epsilon_{\pm}
    =
    \frac{1}{4}
    \left(
    \frac{
    \tilde{\tau}_0
    }{
    \left( \tilde{t} \pm \tilde{x}_0 \right) \left(\tilde{t} \pm \tilde{x}_0 \mp 1 \right)+\tilde{\tau}_0^2
    }
    \right)^2
    \ ,
    \label{eq:epsilon_pm}
\end{align}
where $\tilde{\tau}_0 = \frac{\tau_0}{\ell},$ $\tilde{x}_0 = \frac{x_0}{\ell}$ and $\tilde{t} = \frac{t}{\ell}$ are dimensionless parameters. In this case, the cross-ratios $x$ in \eqref{eq:cross-ratio} and $\overline{x}$ in \eqref{eq:cross-ratio_bar} can be approximated as
\begin{align}
    x
    =
    \left\{\, 
    \begin{aligned}
        &1 - \epsilon_{+}
        \quad &, \quad  
        \left(\tilde{t} + \tilde{x}_0 \right) \left(\tilde{t} + \tilde{x}_0 -1 \right) + \tilde{\tau}_0^2 \geq 0\\
        &\epsilon_{+}   \quad &, \quad  
        \left(\tilde{t} + \tilde{x}_0 \right) \left(\tilde{t} + \tilde{x}_0 -1 \right) + \tilde{\tau}_0^2 < 0
    \end{aligned}
    \right.\nonumber\\
    \label{eq:cross-ratio_asymmptotic}\\
    \overline{x}
    =
    \left\{\, 
    \begin{aligned}
        &1 - \epsilon_{-}
        \quad &, \quad  
        \left(\tilde{t} - \tilde{x}_0 \right) \left(\tilde{t} - \tilde{x}_0 +1 \right) + \tilde{\tau}_0^2 \geq 0\\
        &\epsilon_{-}   \quad &, \quad  
        \left(\tilde{t} - \tilde{x}_0 \right) \left(\tilde{t} - \tilde{x}_0 +1 \right) + \tilde{\tau}_0^2 < 0
    \end{aligned}
    \right.\nonumber
\end{align}
Since $\text{EA}_2$ essentially depends on $x$, $\overline{x}$, $N$ and $k$, we denote it as $\Delta S_{A}^{(2)} \left(x,\overline{x} ; N, k \right)$.
Note that this notation differs from that used in the main text. In the following, we derive the asymptotic behaviors of $\text{EA}_2$ in various limits.

\subsection*{Long time limit: $ \tilde{t}  \to \infty $\ .}
For $\tilde{t} \gg \tilde{\tau}_0$, the infinitesimally small parameter $\epsilon_{\pm}$ take the form:
\begin{align}
    \epsilon_{\pm}
    =
    \epsilon
    =
    \frac{1}{4}
    \left( \frac{\tilde{\tau}_0}{\tilde{t}^2} \right)^2\ ,
\end{align}
and the cross-ratios in \eqref{eq:cross-ratio_asymmptotic} are given by
\begin{align}
    \left(x,\overline{x} \right)
    =
    \left(1-\epsilon, 1-\epsilon \right)\ .
\end{align}
In the limit $\left(x,\overline{x} \right) \to (1,1)$, the 4pt function $G_{2,2}$ dominates in \eqref{eq:EA_2_su_exact}, and $\text{EA}_2$ converges to zero. The asymptotic behavior is obtained by expanding $\text{EA}_2$ with respect to $\epsilon$:
\begin{align}
    \Delta S_{A}^{(2)} \left(1-\epsilon,1-\epsilon ; N, k \right)
    =
    \frac{(N-1)(N^2-2)}{N^2(N+1)} c_{N,k} \epsilon^{\frac{2N}{N+k}} 
    +
    \frac{2}{k} \left( 1-\frac{1}{N} \right) \epsilon
    +
    \mathcal{O}\left( \epsilon^{1+\frac{2N}{N+k}} \right)
\end{align}
where the coefficient $c_{N,k}$ is given by
\begin{align}
    c_{N,k}
    &=
    \frac{
    \Gamma\left( \frac{k}{N+k} \right)^2
    \Gamma\left( \frac{N-1}{N+k} \right)
    \Gamma\left( \frac{N+1}{N+k} \right)
    }{
    \Gamma\left( \frac{N}{N+k} \right)^2
    \Gamma\left( \frac{k-1}{N+k} \right)
    \Gamma\left( \frac{k+1}{N+k} \right)
    }
    \ge 0\ .
\end{align}
Note that $\text{EA}_2$ is positive and decays to zero as $\epsilon \to 0$. In this expansion, the first term $\mathcal{O}\left( \epsilon^{2N/(N+k)} \right)$ dominates for $N<k$, while the second term $\mathcal{O}(\epsilon)$ dominates for $k<N$.

For $N=k$, the coefficient reduces to $c_{N,N} = 1$, and the two contributions merge into:
\begin{align}
    \Delta S_{A}^{(2)} \left(1-\epsilon,1-\epsilon ; N, N \right)
    =
    \left( 1 - \frac{2}{N(N+1)} \right)\epsilon 
    + \mathcal{O}\left( \epsilon^{2} \right)\ .
\end{align}

A same analysis applies to other limits, such as $\tilde{\tau}_0 \to \infty$ and $\tilde{x}_0 \to 0$. In those cases, the infinitesimal parameter $\epsilon$ is replaced by
\begin{align}
\begin{aligned}
    \epsilon 
    &=
    \frac{1}{4} 
    \left(\frac{1}{\tilde{\tau}_0} \right)^2
    \quad, \quad
    \tilde{\tau}_0 \to \infty\\
    \epsilon 
    &= \frac{1}{4} 
    \left(\frac{\tilde{\tau}_0 }{\tilde{x}_0^2} \right)^2
    \quad, \quad
    |\tilde{x}_0| \to \infty
\end{aligned}\ .
\end{align}

\subsection*{Large interval limit: $ \tilde{\tau}_0 \to 0 $ .}
Let us now consider the case $\tilde{\tau}_0 \ll \left| \tilde{t} \pm \tilde{x}_0 \right|$. For $\tilde{x}_0 < 0$, the time dependence of the cross-ratios \eqref{eq:cross-ratio_asymmptotic} is given by
\begin{align}
    \lim_{\tilde{\tau}_0 \to 0}
    \left(x(t), \overline{x}(t) \right)
    =
    \left\{\ 
    \begin{aligned}
        &\left(1, 1 \right)\ 
        &,\ 0\le \tilde{t} \le |\tilde{x}_0|,\  |\tilde{x}_0| + 1 \le \tilde{t}\\
        &\left(0, 1 \right)\ 
        &,\ |\tilde{x}_0| < \tilde{t} < |\tilde{x}_0| + 1 
    \end{aligned}
    \right.\ ,
\end{align}
where we have neglected the $\mathcal{O}\left(\epsilon_{\pm} \right)$ terms for simplicity. When $\left(x, \overline{x} \right) \to (1,1)$, the contribution from $G_{22}$ dominates over the other therms $G_{11},\ G_{12}$ and $G_{21}$, due to the explicit form of $F_{A}^{\pm}(x)$ in \eqref{eq:F_1^(-)}-\eqref{eq:F_2^(+)} together with the crossing symmetry \eqref{eq:crossing}. As a result, $\text{EA}_2$ in \eqref{eq:EA_2_su_exact} converges to zero. In contrast, when $\left(x, \overline{x} \right) \to (0,1)$, $G_{12}$ dominates over $G_{11},\ G_{21}$ and $G_{22}$, owing to the relation \eqref{eq:F}, and $\text{EA}_2$ converges to $\log N$. We may summarize this behavior as
\begin{align}
    \lim_{\tilde{\tau}_0 \to 0}
    \Delta S_{A}^{(2)}
    =
    \left\{\ 
    \begin{aligned}
        &0\ 
        &,\ 0\le \tilde{t} \le |\tilde{x}_0|,\  |\tilde{x}_0| + 1 \le \tilde{t}\\
        &\log N\ 
        &,\ |\tilde{x}_0| < \tilde{t} < |\tilde{x}_0| + 1 
    \end{aligned}
    \right.\ ,
\end{align}
A qualitatively similar behavior holds for $1<\tilde{x}_0$.

For $0 < \tilde{x}_0 < 1$, the time dependence of the cross-ratios \eqref{eq:cross-ratio_asymmptotic} changes as
\begin{align}
    \lim_{\tilde{\tau}_0 \to 0}
    \left(x(t), \overline{x}(t) \right)
    =
    \left\{\ 
    \begin{aligned}
        &\left(0, 0 \right)\ 
        &,\ 0\le \tilde{t} \le \tilde{x}_0\\
        &\left(0, 1 \right)\ 
        &,\ \tilde{x}_0  < \tilde{t} < 1-\tilde{x}_0 \\
        &(1,1)
        &,\ 1-\tilde{x}_0 \le \tilde{t}
    \end{aligned}
    \right.\ ,
\end{align}
where we have assumed $0<\tilde{x}_0 < 1/2$ (a similar argument applies to for $1/2 < \tilde{x}_0 < 1$). In this regime, when $\left(x, \overline{x} \right) \to (0,0)$, the $G_{11}$ dominates over $G_{12},\ G_{21}$ and $G_{22}$, and $\text{EA}_2$ converges to $\log[N(N+1)/2]$. Thus, the time evolution of $\text{EA}_2$ can be summarized as
\begin{align}
    \lim_{\tilde{\tau}_0 \to 0}
    \Delta S_{A}^{(2)}
    =
    \left\{\ 
    \begin{aligned}
        &\log \left[\frac{N(N+1)}{2} \right]\ 
        &,\ 0\le \tilde{t} \le \tilde{x}_0  \\
        &\log N\ 
        &,\ \tilde{x}_0 < \tilde{t} < 1-\tilde{x}_0  \\
        &0
        &,\ 1-\tilde{x}_0 \le \tilde{t}
    \end{aligned}
    \right.\ ,
\end{align}
Here, we have omitted the infinitesimally small corrections of order $\mathcal{O} \left(\epsilon_{\pm} \right)$. In the following, we derive the explicit form of those terms in the special cases relevant to the quantum Mpemba effect.
\begin{itemize}
    \item Case 1: $\tilde{\tau}_0 \to 0 ,\ \tilde{t} = 0,\ 0< \tilde{x}_0 < 1$\\
    This corresponds to the left panel of Fig.\,\ref{fig:operator_insertion_points} in the limit $\tilde{\tau}_0 \to 0$. The cross-ratios take the form
    \begin{align}
        \left(x, \overline{x} \right)
        =
        (\epsilon, \epsilon)\ ,
    \end{align}
    where the infinitesimally small parameter $\epsilon$ is given by
    \begin{align}
        \epsilon
        =
        \frac{1}{4} \left( \frac{\tilde{\tau}_0}{\tilde{x}_0(\tilde{x}_0-1)} \right)^2\ .
    \end{align}
    Expanding $\text{EA}_2$ with respect to $\epsilon$, we obtain the next leading correction:
    \begin{align}
        \Delta S_{A}^{(2)}
        &= 
        \log \left[\frac{N(N+1)}{2} \right]
        -\frac{(N-1)(N^2+2N-2)}{2N} c_{N,k}\  \epsilon^{\frac{2N}{N+k}} 
        - \frac{N-1}{k} \epsilon 
        + \mathcal{O}\left( \epsilon^{1+\frac{2N}{N+k}} \right)
        \ .
    \end{align}
    In this expansion, the first term $\mathcal{O}\left( \epsilon^{2N/(N+k)} \right)$ dominates for $N<k$, while the second term $\mathcal{O}(\epsilon)$ dominates for $k<N$.
    For $N=k$, the two terms marge and $\text{EA}_2$ becomes
    \begin{align}
        \Delta S_{A}^{(2)}
        &= 
        \log \left[\frac{N(N+1)}{2} \right]
        - \frac{1}{2}(N-1)(N+2) \epsilon 
        + \mathcal{O}\left( \epsilon^{2} \right)\ .
    \end{align}
    
    \item Case 2: $\tilde{\tau}_0 \ll \tilde{\delta}  ,\ \tilde{t} = +\tilde{\delta},\ \tilde{x}_0 = 0$\\
    This corresponds to the right panel of Fig.\,\ref{fig:operator_insertion_points} in the limit $\tilde{\tau}_0 \to 0$. The cross-ratios take the form
    \begin{align}
        \left(x, \overline{x} \right)
        =
        (\epsilon, 1-\epsilon)\ ,
    \end{align}
    where the infinitesimally small parameter $\epsilon$ is given by
    \begin{align}
        \epsilon
        =
        \frac{1}{4} \left( \frac{\tilde{\tau}_0}{\tilde{\delta}} \right)^2\ .
    \end{align}
    Expanding $\text{EA}_2$ with respect to $\epsilon$, we obtain the next leading correction:
    \begin{align}
        \Delta S_{A}^{(2)}
        &=
        \log N
        - d_{N,k}\  \epsilon^{\frac{N}{N+k}}
        -\left( N-2 + \frac{2}{N+1} \right)\epsilon
        +\mathcal{O}\left( \epsilon^{1+\frac{N}{N+k}} \right)
    \end{align}
    with the coefficient $d_{N,k}$ defined by
    \begin{align}
        d_{N,k}
        =
        \frac{N(N-1)}{N+1}
        \frac{
        \Gamma\left( \frac{1}{N+k} \right)
        \Gamma\left(1-\frac{1}{N+k} \right)
        \Gamma\left( \frac{k}{N+k} \right)
        }{
        \Gamma\left( \frac{k-1}{N+k} \right)
        \Gamma\left(\frac{k+1}{N+k} \right)
        \Gamma\left( \frac{N}{N+k} \right)
        }
        \ .
    \end{align}
    Note that $d_{N,k}$ is positive for $k \ge 2$ and vanishes for $k=1$. Consequently, the next leading contribution is the second term $\mathcal{O}\left( \epsilon^{\frac{N}{N+k}} \right)$ for $k\ge2$, while for $k=1$ it is the $\mathcal{O}(\epsilon)$ term.
\end{itemize}


\bibliographystyle{utphys}
\bibliography{EA} 

\end{document}